\documentclass[useAMS]{mn2e}
	\usepackage{longtable}
        \usepackage{graphicx}

\def\oversim#1#2{\lower0.5pt\vbox{\baselineskip0pt \lineskip-0.5pt
     \ialign{$\mathsurround0pt #1\hfil##\hfil$\crcr#2\crcr\sim\crcr}}}

      \let\leq=\leqslant 
      \let\geq=\geqslant 

\title[Near-Infrared Photometry of Carbon Stars]
{Near-Infrared Photometry of Carbon Stars\thanks{This
paper is based on observations made at the South African
Astronomical Observatory.}}

\author[P.A. Whitelock et al.]
{Patricia A. Whitelock$^{1,2,3}$\thanks{e-mail: \tt paw@saao.ac.za}, 
    Michael W. Feast$^2$, Freddy Marang$^1$ and \newauthor  M.A.T. 
Groenewegen$^4$\\
      $^1$ South African Astronomical Observatory, P.O.Box 9, 7935
           Observatory, South Africa\\
       $^2$ Astronomy Department, University of Cape Town, 7701 Rondebosch,
           South Africa\\
       $^3$ National Astrophysics and Space Science Programme,
            University of Cape Town, 7701 Rondebosch, South Africa\\
       $^4$ Instituut voor Sterrenkunde, KU Leuven, Celestijnenlaan 200B,
	B-3001 Leuven, Belgium  
}
\date{Received date; accepted date}

\pubyear{2006}
\begin{document}
\maketitle

\begin{abstract} Near-infrared, $JHKL$, photometry of 239 Galactic
carbon-rich variable stars is presented and discussed. From these and
published data the stars were classified as Mira or non-Mira variables and
amplitudes and pulsation periods, ranging from 222 to 948 days for the
Miras, were determined for most of them. A comparison of the colour and
period relations with those of similar stars in the Large Magellanic Cloud
indicates minor differences, which may be the consequence of sample
selection effects. Apparent bolometric magnitudes were determined by
combining the mean $JHKL$ fluxes with mid-infrared photometry from IRAS and
MSX. Then, using the Mira period luminosity relation to set the absolute
magnitudes, distances were determined -- to greater accuracy than has
hitherto been possible for this type of star. Bolometric corrections to the
$K$ magnitude were calculated and prescriptions derived for calculating
these from various colours. Mass-loss rates were also calculated and
compared to values in the literature.

Approximately one third of the C-rich Miras and an unknown fraction of the
non-Miras exhibit apparently random obscuration events that are
reminiscent of the phenomena exhibited by the hydrogen deficient RCB stars.
The underlying cause of this is unclear, but it may be that mass loss, and
consequently dust formation, is very easily triggered from these very
extended atmospheres.

\end{abstract}


\section{Introduction}
 Intrinsic carbon stars are produced when asymptotic giant branch (AGB)
stars experience sufficient dredge-up to change their surface carbon to
oxygen ratios from $\rm C/O<1$ to $\rm C/O>1$. Exactly how and when this
occurs depend on the initial mass and the abundances of the star in
question, but the details remain controversial. It has been known for a long
while that the majority of the C stars in the Magellanic Clouds were low
mass objects (e.g. Iben 1981), but it has only recently been possible to
model the production of C stars among relatively low mass stars (e.g.
Stancliffe et al. 2005).

In attempting to find an appropriate group of local C stars to study it is
natural to examine the large amplitude variables. First, the amplitude of
their variability identifies them as AGB stars and therefore one can be
reasonably certain that their carbon enrichment is intrinsic rather than the
result of binary mass transfer. Secondly, observations of extragalactic C
Miras suggest that they obey a well defined period luminosity relation
(Feast et al. 1989, Whitelock et al. 2003) and it should therefore be
possible to establish their distances, provided their periods and
apparent luminosities can be measured.

This particular project arose from the requirement to identify a significant
local population of Galactic C stars with distances and radial
velocities which could be used for kinematic studies. From these it should
be possible to gain insight into the nature, in particular the ages and
masses, of the local C star population which will be invaluable for
comparison with theory and with similar populations in Local Group
galaxies (e.g. Menzies et al. 2002).

In this paper, the first of three, we discuss infrared photometry of C
variables observable from the southern hemisphere. Subsequent papers will
deal with radial velocities and northern C-rich Miras (Menzies et al. 2006,
Paper II) and the kinematics of the full sample (Feast et al. 2006, Paper
III).

\section{Source Selection}\label{sources}
 Much of the $JHKL$ photometry reported here comes from two specific
programmes; first, stars selected from the catalogue of Aaronson et al.
(1989) and secondly, stars selected from the IRAS Point Source Catalogue
(IRAS Science Team 1985), henceforth referred to as the `Aaronson' and `IRAS'
samples, respectively. The stars from each group were monitored for several
years to determine their pulsation periods and to establish the
characteristics of their variability. These data were supplemented by
observations of C stars that had been obtained from SAAO over the years
as part of other programmes. In general stars were selected to be
south of declination $+30^{\rm o}$ for ease of monitoring from Sutherland.

Aaronson et al. (1989) published velocities and $JHK$ photometry for C
stars. The stars selected for monitoring were chosen on the basis of their
colours  ($(H-K)_0>0.8$) (using a very rough estimate of the reddening
correction) or their $K$ variability as potential Mira variables which could
be used to establish the kinematic properties of the C Miras (the
colour criterion may have resulted in the omission of some C Miras, but the
objective was to find stars with a high probability of being C Miras rather
than to be complete). These stars are identified with an `A' in column 10
(G) of Table~\ref{names}. Data are reported here for 60 such stars,
including three which are probably not C stars.
 
The IRAS sample was selected from the Point Source Catalogue (PSC) using the
following 3 criteria: spectral types 4n, indicating SiC features in the LRS
spectra; 25 to 12 $\mu$m flux ratio, $\rm F_{25}/F_{12}>0.4$\rm; 12
$\mu$m flux, $F_{12}>40$\,Jy.  The flux ratio criterion was intended to isolate
stars with dust shells; while the limit to the 12 $\mu$m flux was intended
to ensure that the stars would be observable, at least at $K$ and
$L$, on the 0.75m at Sutherland.  Note that the spectral-type criterion
will have resulted in the omission of C Miras without SiC shells, but again
the objective was to find stars with a high probability of being C-rich Miras
rather than to be complete. Some of the stars selected in this way were
already being observed as part of other programmes; nevertheless, they will
be identified as part of the `IRAS' sample for the purpose of the discussion
below. These stars are identified with an `I' in column 10 (G) of Table
\ref{names}. Data are presented for 96 such IRAS sources, including
five which are unlikely to be C stars.

A further 32 C-stars with IRAS photometry have $\rm F_{25}/F_{12}>0.4$, but
with $F_{12}<40$. These were not chosen on the basis of their IRAS
characteristics but are identified here with an `i' in column 10 (G) as
an interesting group to compare with their brighter counterparts. Five of the
stars selected in this way are probably not C stars.

A number of sources from the Aaronson and IRAS samples were identified by the
selection criteria mentioned above, but observations proved impossible
because of severe crowding.

There are only two stars in common between the Aaronson and the IRAS samples,
although another 8 are found in the faint IRAS sample. This might suggest 
that the fainter IRAS sources are not simply more distant but that they 
actually have thinner shells than their brighter counterparts.

The names of the C stars are listed in Table~\ref{names} in one of the forms
recognizable by the SIMBAD data base, with preference being given to a
variable star designation (from Samus et al. 2004, henceforth GCVS) where
one exists. IRAS names are also listed, and are used throughout this paper
without the `IRAS' prefix. The C numbers listed in the table are from the
updated version of Stephenson's Catalogue of Galactic Carbon Stars (Alksnis
et al. 2001). Most of the coordinates were taken from Cutri et al. (2003,
henceforth 2MASS).

At the end of Table~\ref{names} 18 stars are listed which were observed as
part of the programme, but which are either peculiar in some way or are
probably not bona fide C stars. Peculiarities include transitional objects,
i.e. SC stars.

\section{Infrared Photometry}\label{ir_phot}
 The detailed SAAO photometry is reported in Table~\ref{irdata} for the
first star only, the full dataset of photometry is available electronically.
The table gives times of the observations as Julian Date (JD) followed by
the $JHKL$ mags measured for that date. It is organized in order of right
ascension with the dubious C stars (the 18 listed at the end of Table~1)
mixed among the definite ones. Where measurements were not made, usually at
$J$ or $L$ because the star was too faint, the column is left blank. Most of
the $JHKL$ measurements were made with the MkII infrared photometer on the
0.75-m telescope at SAAO, Sutherland. They are on the current SAAO system as
defined by Carter (1990). A few measurements were made on the SAAO 1.9-m and
these have been transformed to the SAAO system\footnote{Transformation from
the 1.9-m natural system to the SAAO system defined by Carter (1990) assumes
$K_{1.9}=K$ and $(J-H)_{1.9}=0.95(J-H)$ or $(J-K)_{1.9}=0.955(J-K)$.}. These
are marked `1.9m' in the last column of the table. 

The post-1979 photometry is accurate to better than $\pm 0.03$ mag at $JHK$,
and to better than $\pm 0.05$ mag at $L$; observations with $J>13.0$,
measured using the 1.9m telescope, or $J>10.4$ measured using the 0.75m, are
good to $\pm 0.06$ at $J$. Measurements marked with a colon are accurate to
better than $0.1$ mag.  Some older (pre-1979) measurements were reported by
Catchpole et al. (1979). The measurements listed here differ slightly from
those in Catchpole et al., because they have been corrected to Carter's
improved values for the standard stars. These measurements are slightly less
accurate than the more modern values, but see Catchpole et al.\ for details.

%

\begin{center}
\onecolumn
\begin{longtable}{lcrrrrrccrrc}
\caption[Stars observed in this survey.]{Stars observed in this survey.}\label{names} \\
\hline
 name & IRAS &\multicolumn{3}{c}{RA}&\multicolumn{3}{c}{Dec} & MSX & G 
& C & comment\\
& & \multicolumn{6}{c}{Equinox 2000}\\
\hline
\endfirsthead

\hline
 name & IRAS &\multicolumn{3}{c}{RA}&\multicolumn{3}{c}{Dec} & MSX & G
& C & comment \\
& & \multicolumn{6}{c}{Equinox 2000}\\
\hline
\endhead
  \multicolumn{12}{l}{{Continued on Next Page\ldots}} \\
\endfoot    

  \\ \hline 
\endlastfoot

R Scl           & 01246--3248 &  1& 26& 58& --32& 32& 36& & I&  234\\
YY Tri          & 02152+2822 &  2& 18&  6& +28& 36& 45& & I& 6028\\
R For           & 02270--2619 &  2& 29& 15& --26&  5& 56& &  &  361\\
EV Eri          & 04067--0922 &  4&  9&  6& -- 9& 14& 12& & i& 6070\\
$[$TI98$]$0418+0122 & 04188+0122 &  4& 21& 27& + 1& 29& 14& &  & 6075\\
V718 Tau        & 04284+1732 &  4& 31& 22& +17& 39& 10& &  &  714\\
TT Tau          & 04483+2826 &  4& 51& 31& +28& 31& 37& &  &  794& \\
R Lep           & 04573--1452 &  4& 59& 36& --14& 48& 23& &  &  833\\
TU Tau          & 05421+2424 &  5& 45& 14& +24& 25& 12&X&  & 1038\\
Y Tau           & 05426+2040 &  5& 45& 39& +20& 41& 42& &  & 1042\\
QS Ori          & 05428+1215 &  5& 45& 37& +12& 16& 15& &  &  395\\
05418--3224     & 05418--3224 &  5& 43& 43& --32& 23& 29& & I& 1045\\
V1259 Ori       & 06012+0726 &  6&  4& 00& + 7& 25& 52& & I& 6113\\
06088+1909      & 06088+1909 &  6& 11& 48& +19&  8& 20&X& I& 1187&1\\
BN Mon          & 06192+0722 &  6& 21& 58& + 7& 20& 58&X&  & 1246\\
ZZ Gem          & 06209+2503 &  6& 24&  01& +25&  1& 53& &  & 1251\\
V617 Mon        & 06210+0831 &  6& 23& 48& + 8& 29& 51&X&  & 1254\\
V636 Mon        & 06226--0905 &  6& 25&  1& -- 9&  7& 16& &  & 6146\\
V477 Mon        & 06268+0849 &  6& 29& 35& + 8& 47& 16&X& I& 1287\\
CR Gem          & 06315+1606 &  6& 34& 24& +16&  4& 30&X& i& 1309\\
GM CMa          & 06391--2213 &  6& 41& 15& --22& 16& 43& & i& 1357\\
V503 Mon        & 06422+0953 &  6& 44& 57& + 9& 50& 48&X&  & 1377\\
RT Gem          & 06436+1840 &  6& 46& 35& +18& 36& 54& &  & 1389\\
06487+0551      & 06487+0551 &  6& 51& 24& + 5& 47& 34&X& I& 1430\\
CG Mon          & 06487+0517 &  6& 51& 27& + 5& 13& 23&X&  & 1431\\
CL Mon          & 06529+0626 &  6& 55& 37& + 6& 22& 43& &  & 1465\\
06531--0216     & 06531--0216 &  6& 55& 40& -- 2& 20& 16&X&  & 1471\\
NP Pup          & 06528--4218 &  6& 54& 27& --42& 21& 56& &  & 1478\\
06564+0342      & 06564+0342 &  6& 59&  6& + 3& 37& 56&X& I& 1494\\
W CMa           & 07057--1150 &  7&  8&  3& --11& 55& 24&X& i& 1565\\
07080--0106     & 07080--0106 &  7& 10& 35& -- 1& 11& 26& & I& 1580\\
VX Gem          & 07099+1441 &  7& 12& 49& +14& 36&  4& &  & 1595\\
07097--1011     & 07097--1011 &  7& 12&  8& --10& 16&39&X&A & 1597\\
R Vol           & 07065--7256 &  7&  5& 36& --73&  0& 52& &  & 1599\\
HX CMa          & 07098--2012 &  7& 12&  4& --20& 17& 23& & I& 1601\\
07136--1512     & 07136--1512 &  7& 15& 57& --15& 18& 08&X&A & 1630\\
07161--0111     & 07161--0111 &  7& 18& 39& -- 1& 16& 52& & I& 1642\\
07217--1246     & 07217--1246 &  7& 24&  3& --12& 52& 28&X&AI& 1696\\
07220--2324      & 07220--2324 &  7& 24&  7& --23& 30& 46& & I& 1699\\
07223--1553      & 07223--1553 &  7& 24& 35& --15& 59& 52&X&A & 1701\\
07293--1832      & 07293--1832 &  7& 31& 31& --18& 39&  4&X&A & 1751\\
07319--1940      & 07319--1940 &  7& 34&  6& --19& 46& 56&X&A & 1775\\
$[$W71b$]$W007--02   & 07348--1926 &  7& 37&  2& --19& 32& 54&X&A & 1798\\
07373--4021      & 07373--4021 &  7& 39&  4& --40& 28& 47& &  & 1825\\
$[$W71b$]$W008--03   &            &  7& 40& 55& --26&  1& 31&X&A & 1831\\
V471 Pup        & 07390--2618 &  7& 41&  6& --26& 25& 19&X&A & 1834\\
$[$ABC89$]$Pup 3    & 07403--2943 &  7& 42& 17& --29& 51&  4&X&Ai& 1847\\
$[$ABC89$]$Pup 17   &            &  7& 49& 32& --27& 23&  36&X&A & 1897\\
07454--7112      & 07454--7112 &  7& 45&  2& --71& 19& 46& & I& 1901\\
$[$ABC89$]$Pup 21   & 07506--2819 &  7& 52& 43& --28& 26& 52&X&A & 1924\\
V831 Mon        & 07551--0032 &  7& 57& 43& -- 0& 41&  6& &  & 1960\\
07576--4054      & 07576--4054 &  7& 59& 24& --41&  3& 16& & I& 1992\\
07582--1933      & 07582--1933 &  8&  0& 25& --19& 42& 11& & I& 1993\\
V509 Pup        & 08004--3023 &  8&  2& 26& --30& 32& 16&X&A & 2010\\
$[$ABC89$]$Pup 38   & 08010--2626 &  8&  3&  7& --26& 34& 31&X&A & 6268\\
FF Pup          & 08014--2356 &  8&  3& 35& --24&  4& 35& &  & 2022\\
$[$ABC89$]$Pup 42   & 08029--2942 &  8&  4& 58& --29& 51& 26&X&A & 2043\\
V518 Pup        & 08045--1524 &  8&  6& 51& --15& 33& 23& & I& 2056\\
08050--2838      & 08050--2838 &  8&  7&  6& --28& 47& 40&X& I& 2062\\
RU Pup          & 08053--2246 &  8&  7& 30& --22& 54& 45& &  & 2064\\
FK Pup          & 08073--3608 &  8&  9& 11& --36& 17&  7&X&  & 2086\\
08074--3615      & 08074--3615 &  8&  9& 20& --36& 24& 27&X& I& 6267\\
$[$ABC89$]$Ppx19    & 08080--3259 &  8& 10&  2& --33&  8& 29&X&A & 2091\\
$[$W71b$]$021--05    & 08083--3145 &  8& 10& 18& --31& 54& 22&X&A & 2095\\
$[$ABC89$]$Ppx22    & 08085--3351 &  8& 10& 29& --34&  0& 36&X&A & 2099\\
V346 Pup        & 08088--3243 &  8& 10& 49& --32& 52&  6&X&AI& 2101\\
$[$W71b$]$026--01    & 08160--3822 &  8& 17& 52& --38& 32& 16&X&A & 2146\\
RY Hya          & 08174+0255 &  8& 20&  6& + 2& 45& 56& &  & 2150\\
$[$ABC89$]$Ppx40    & 08197--3447 &  8& 21& 41& --34& 57& 24&X&A & 2173\\
$[$W71b$]$029--02    & 08233--4110 &  8& 25& 10& --41& 20&  2&X&A & 2203\\
$[$W71b$]$029--04    & 08266--4110 &  8& 28& 26& --41& 20& 43&X&A & 2224\\
08340--3357      & 08340--3357 &  8& 36&  3& --34&  7& 34& & I& 2260\\
R Pyx           & 08434--2801 &  8& 45& 31& --28& 12&  3& &  & 2326\\
UW Pyx          & 08450--3407 &  8& 47& 00& --34& 18& 59& &  & 2334\\
T Cnc           & 08538+2002 &  8& 56& 40& +19& 50& 57& &  & 2384\\
08535--4724      & 08535--4724 &  8& 55& 11& --47& 35& 56&X& I& 2389\\
08534--5055      & 08534--5055 &  8& 55&  2& --51&  7& 20&X& I& 2390\\
IQ Hya          & 09112--2311 &  9& 13& 32& --23& 23& 31& &  & 2450\\
CQ Pyx          & 09116--2439 &  9& 13& 54& --24& 51& 25& & I& 6325\\
09164--5349      & 09164--5349 &  9& 18&  2& --54&  2& 27&X& I& 2473\\
09176--5147      & 09176--5147 &  9& 19& 17& --52&  0& 28&X& I& 2476\\
$[$ABC89$]$Vel19    &            &  9& 26& 19& --52&  6&  4&X&A & 2508\\
$[$W71b$]$046--02    & 09249--4909 &  9& 26& 45& --49& 22& 25&X&A & 2512\\
$[$ABC89$]$Vel44    & 09331--5010 &  9& 34& 57& --50& 24& 30&X&A & 2563\\
09433--6233      & 09433--6233 &  9& 44& 41& --62& 47& 32& & i& 6339\\
CW Leo          & 09452+1330 &  9& 47& 57& +13& 16& 43& & I& 2619\\
W Sex           & 09484--0147 &  9& 50& 58& -- 2&  1& 43& &  & 2635\\
09484--6242      & 09484--6242 &  9& 49& 49& --62& 56&  9& &  & 2645\\
09513--5324      & 09513--5324 &  9& 53&  7& --53& 38& 54&X& I& 2653\\
09529--5506      & 09529--5506 &  9& 54& 41& --55& 20& 16&X& I& 2660\\
09533--6021      & 09533--6021 &  9& 54& 52& --60& 35& 26& & i& 2663\\
09521--7508      & 09521--7508 &  9& 52& 30& --75& 22& 28& & I& 2664\\
09586--6150      & 09586--6150 & 10&  0&  9& --62&  5& 19& & i& 6344\\
10019--6156      & 10019--6156 & 10&  3& 29& --62& 10& 37& &  & 2691\\
10023--5946      & 10023--5946 & 10&  3& 58& --60&  0& 37&X&  & 2692\\
10026--5849      & 10026--5849 & 10&  4& 20& --59&  4&  0&X&  & 6347\\
10052--5906      & 10052--5906 & 10&  6& 57& --59& 21& 25&X&  & 2703\\
10098--5742      & 10098--5742 & 10& 11& 35& --57& 57& 53&X& i& 6352\\
10109--5958      & 10109--5958 & 10& 12& 40& --60& 13& 30&X&  & 2720\\
RW LMi           & 10131+3049 & 10& 16&  2& +30& 34& 19& &  & 2724\\
10130--5703      & 10130--5703 & 10& 14& 49& --57& 18& 45&X& i&     \\
10136--5743      & 10136--5743 & 10& 15& 27& --57& 58& 11&X&  & 2729\\
10145--6046      & 10145--6046 & 10& 16& 13& --61&  1& 43& &  & 2734\\
10149--5919      & 10149--5919 & 10& 16& 43& --59& 34& 52&X&  & 2735\\
10151--6008      & 10151--6008 & 10& 16& 50& --60& 23& 55&X& i& 6354\\
10175--5957      & 10175--5957 & 10& 19& 17& --60& 12& 52&X&  & 2745\\
10199--5801      & 10199--5801 & 10& 21& 44& --58& 16& 35&X& i& 6363\\
10220--5858      & 10220--5858 & 10& 23& 49& --59& 13& 54&X&  & 6366\\
CPD--58 2175     & 10231--5823 & 10& 24& 58& --58& 39& 17&X& i& 2760\\
CZ Hya          & 10249--2517 & 10& 27& 18& --25& 32& 56& &  & 2764\\
$[$ABC89$]$Car5    &            & 10& 29& 44& --62& 28& 29& &A & 2776\\
$[$ABC89$]$Car11   &            & 10& 32& 22& --60& 42& 29&X&A & 2784\\
TV Vel          & 10324--5358 & 10& 34& 28& --54& 14& 28& &  & 2790\\
U Ant           & 10329--3918 & 10& 35& 13& --39& 33& 45& &  & 2793\\
$[$ABC89$]$Car28    &            & 10& 37&  9& --60& 59& 34&X&A & 6391\\
$[$ABC89$]$Car32    & 10366--5950 & 10& 38& 29& --60&  5& 57&X&A & 2817\\
FU Car          & 10390--5907 & 10& 41&  0& --59& 23& 13&X&Ai& 2832\\
$[$ABC89$]$Car54    & 10442--5809 & 10& 46& 16& --58& 25& 21&X&A & 2850\\
$[$ABC89$]$Car59    &            & 10& 48& 30& --60& 11& 32&X&A & 2862\\
V Hya           & 10491--2059 & 10& 51& 37& --21& 15& 00& & I& 2877\\
$[$ABC89$]$Car73    & 10509--6036 & 10& 52& 55& --60& 52& 10&X&A & 6426\\
$[$ABC89$]$Car81    &            & 10& 54& 27& --60& 19&  50&X&A & 2897\\
$[$ABC89$]$Car84    &            & 10& 56& 45& --60&  3& 37&X&A & 2907\\
$[$ABC89$]$Car87    & 10558--6203 & 10& 57& 47& --62& 19& 16&X&A & 2911\\
$[$ABC89$]$Car93    &            & 10& 59&  5& --60& 31& 49& &A & 2917\\
$[$ABC89$]$Car105   & 11009--6117 & 11&  3&  1& --61& 33& 28&X&Ai& 2941\\
11145--6534      & 11145--6534 & 11& 16& 39& --65& 50& 56& & I& 2987\\
$[$W65$]$ c1        &            & 11& 20& 34& --59& 30& 51&X&  & 2997\\
$[$W65$]$ c2        &            & 11& 22&  5& --59& 38& 45&X&A & 3003\\
$[$W65$]$ c13       & 11299--6103 & 11& 32& 19& --61& 20& 34&X&A & 3051\\
$[$TI98$]$1130--1020 & 11308--1020 & 11& 33& 25& --10& 36& 59& &  & 3052\\
11318--7256      & 11318--7256 & 11& 33& 58& --73& 13& 19& &  & 3062\\
$[$ABC89$]$Cen3     &            & 11& 35& 54& --60& 33& 41&X&A & 3068\\
$[$ABC89$]$Cen4     & 11339--6012 & 11& 36& 17& --60& 29& 18&X&A & 3071\\
11463--6320      & 11463--6320 & 11& 48& 48& --63& 37& 28&X& I& 6455\\
$[$ABC89$]$Cen32    & 11468--5950 & 11& 49& 21& --60&  7& 5&X&Ai& 3108\\
$[$ABC89$]$Cen43    & 11510--6046 & 11& 53& 31& --61&  3& 33&X&A & 3120\\
$[$ABC89$]$Cen60    & 11556--6357 & 11& 58&  8& --64& 14& 54&X&A & 3139\\
$[$ABC89$]$Cen78    &            & 12&  4& 10& --62& 42& 26&X&A & 6464\\
CF Cru          & 12023--6230 & 12&  4& 55& --62& 47& 39&X&A & 3165\\
$[$ABC89$]$Cen97    & 12100--6122 & 12& 12& 44& --61& 39& 01&X&A & 6473\\
12194--6007      & 12194--6007 & 12& 22& 10& --60& 24& 15&X& I& 3220\\
SS Vir          & 12226+0102 & 12& 25& 14& + 0& 46& 11& &  & 3236\\
12298--5754      & 12298--5754 & 12& 32& 41& --58& 11& 29&X& I& 3251\\
CGCS3268        & 12374--5706 & 12& 40& 15& --57& 22&  46& &A & 3268\\
12394--4338      & 12394--4338 & 12& 42& 10& --43& 55& 03& & I& 3275\\
12421--6217      & 12421--6217 & 12& 45&  7& --62& 33& 38&X& I& 6489\\
RU Vir          & 12447+0425 & 12& 47& 18& + 4&  8& 41& &  & 3286\\
V Cru           & 12536--5737 & 12& 56& 36& --57& 53& 57&X&  & 3310\\
12540--6845      & 12540--6845 & 12& 57& 16& --69&  1& 51& & I& 3311\\
$[$ABC89$]$Cru17    & 13022--6400 & 13&  5& 26& --64& 16& 11&X&Ai& 3327\\
$[$ABC89$]$Cir1     & 13342--6232 & 13& 37& 44& --62& 48& 28&X&Ai& 3410\\
13343--5807      & 13343--5807 & 13& 37& 41& --58& 23& 10&X& I& 3411\\
13477--6532      & 13477--6532 & 13& 51& 29& --65& 46& 56&X& I& 3439\\
13482--6716      & 13482--6716 & 13& 52&  4& --67& 30& 56& & I& 3441\\
13509--6348      & 13509--6348 & 13& 54& 34& --64&  3& 23&X& I& 3446\\
$[$ABC89$]$Cir18    &            & 13& 55& 26& --59& 22& 24&X&A & 6547\\
$[$ABC89$]$Cir26    & 14004--6047 & 14&  4&  5& --61& 01& 50&X&Ai& 6549\\
$[$ABC89$]$Cir27    & 14010--5927 & 14&  4& 33& --59& 41& 22&X&Ai& 3470\\
$[$W71b$]$093--02& 14192--6327 & 14& 23&  8& --63& 41&  9&X&A & 3487\\
14395--5656      & 14395--5656 & 14& 43& 14& --57&  8& 45&X&A & 3523\\
14404--6320      & 14404--6320 & 14& 44& 26& --63& 33& 28&X& I& 3525\\
14443--5708      & 14443--5708 & 14& 48&  4& --57& 20& 37&X& I& 6565\\
15082--4808      & 15082--4808 & 15& 11& 41& --48& 19& 59& & I& 3570\\
15084--5702      & 15084--5702 & 15& 12& 15& --57& 13& 28&X& I& 6572\\
II Lup          & 15194--5115 & 15& 23&  5& --51& 25& 59&X& I& 3592\\
15261--5702      & 15261--5702 & 15& 30&  2& --57& 12& 46&X& I&     \\
15471--5644      & 15471--5644 & 15& 51&  6& --56& 53& 24&X& I& 6600\\
CGCS3660        &            & 16&  2& 44& --41& 21& 32& &  & 3660\\
16079--4812      & 16079--4812 & 16& 11& 34& --48& 19& 51&X& I& 3670\\
NP Her          & 16150+2558 & 16& 17&  9& +25& 51& 02& &  & 3679\\
16171--4759      & 16171--4759 & 16& 20& 50& --48&  6& 53&X& I& 3681\\
V Oph           & 16239--1218 & 16& 26& 44& --12& 25& 36& &  & 3698\\
SU Sco          & 16374--3217 & 16& 40& 39& --32& 22& 48& &  & 3720\\
CGCS3721        & 16387--5401 & 16& 42& 45& --54&  7& 10& &  & 3721\\
16406--1406      & 16406--1406 & 16& 43& 27& --14& 12& 00& & i&     \\
16538--4633      & 16538--4633 & 16& 57& 32& --46& 37& 47&X& I& 3747\\
16545--4214      & 16545--4214 & 16& 58&  6& --42& 19& 24&X& I& 3748\\
T Ara           & 16584--5459 & 17&  2& 33& --55&  4& 16& & i& 3756 &2\\
V901 Sco        & 16595--3239 & 17&  2& 46& --32& 43& 31&X&  & 3762\\
17047--2848      & 17047--2848 & 17&  7& 56& --28& 52& 06& & I& 3772\\
V2548 Oph       & 17049--2440 & 17&  7& 58& --24& 44&  31& & I& 6661\\
SZ Ara          & 17065--6153 & 17& 11&  7& --61& 57& 15& &  & 3774\\
V617 Sco        & 17103--3551 & 17& 13& 41& --35& 55& 21&X& I& 3786\\
17105--3746      & 17105--3746 & 17& 13& 59& --37& 50&  8&X& I& 6670& 3\\
17130--3907      & 17130--3907 & 17& 16& 33& --39& 10& 46&X& I& 3794\\
17209--3318      & 17209--3318 & 17& 24& 15& --33& 21& 20&X& I& 6674 & 4\\
17217--3916      & 17217--3916 & 17& 25& 13& --39& 19& 22&X& I& 3823\\
17222--2328      & 17222--2328 & 17& 25& 18& --23& 30& 46& & I& 3825\\
V833 Her        & 17297+1747 & 17& 31& 55& +17& 45& 21& & I& 6677\\
V Pav           & 17389--5742 & 17& 43& 19& --57& 43& 26& &  & 3861\\
17446--4048      & 17446--4048 & 17& 48& 12& --40& 49& 36& & I& 6685\\
17446--7809      & 17446--7809 & 17& 52& 35& --78& 10& 42& & I& 6687\\
17463--4007      & 17463--4007 & 17& 49& 50& --40&  7& 58& &  &  & 5\\
V348 Sco        & 17478--4315 & 17& 51& 30& --43& 16& 23& & i& 3886\\
17581--1744      & 17581--1744 & 18&  1&  6& --17& 44& 23&X& I& 3925\\
18036--2344      & 18036--2344 & 18&  6& 42& --23& 44& 22&X& I& 6709\\
FX Ser          & 18040--0941 & 18&  6& 50& -- 9& 41& 16& & I& 6711\\
V1280 Sgr       & 18073--2652 & 18& 10& 28& --26& 51& 58&X&  & 3960\\
18119--2244      & 18119--2244 & 18& 15&  1& --22& 43& 58&X& I& 6729\\
18147--2215      & 18147--2215 & 18& 17& 43& --22& 14& 39&X& I& 6733\\
V5104 Sgr       & 18194--2708 & 18& 22& 35& --27&  6& 29& &  & 6738\\
V2548 Sgr       & 18234--2206 & 18& 26& 29& --22&  4& 15&X& I& 4007\\
18239--0655      & 18239--0655 & 18& 26& 39& -- 6& 54& 4&X& I& 6743\\
18244--0815      & 18244--0815 & 18& 27&  7& -- 8& 13& 10&X& i&     \\
V1076 Her       & 18240+2326 & 18& 25&  6& +23& 28& 47& & I&     \\
18248--0839      & 18248--0839 & 18& 27& 34& -- 8& 37& 23&X& I& 4014\\
18269--1257      & 18269--1257 & 18& 29& 47& --12& 54& 58&X& I& 4024\\
18320--0352      & 18320--0352 & 18& 34& 40& -- 3& 50& 14&X& I& 6750\\
V627 Oph        & 18321+0910 & 18& 34& 34& + 9& 12& 42& &  & 4045\\
18367--0452      & 18367--0452 & 18& 39& 22& -- 4& 48& 45&X& I& 6754& 3\\
V1417 Aql       & 18398--0220 & 18& 42& 25& -- 2& 17& 27&X& I& 4077\\
18400--0704      & 18400--0704 & 18& 42& 45& -- 7&  1& 10&X& i& 6757\\
V821 Her        & 18397+1738 & 18& 41& 55& +17& 41&  8& & I& 4078\\
18424+0346      & 18424+0346 & 18& 44& 59& + 3& 49& 35&X& I& 4093\\
V874 Aql        &            & 18& 45& 41& + 9& 38& 39& &  & 4099\\
V2045 Sgr       & 18463--1706 & 18& 49& 15& --17&  3& 25& &  & 4117\\
S Sct           & 18476--0758 & 18& 50& 20& -- 7& 54& 28&X&  & 4121\\
18475+0926      & 18475+0926 & 18& 49& 55& + 9& 30& 07&X& I& 6761\\
AI Sct          & 18481--0647 & 18& 50& 52& -- 6& 44& 23&X&  & 4124\\
V1418 Aql       & 19008+0726 & 19&  3& 18& + 7& 30& 45&X& I& 4162\\
19029+2017      & 19029+2017 & 19&  5&  7& +20& 22&  4& & I& 6767\\
19068+0544      & 19068+0544 & 19&  9& 16& + 5& 49& 10&X& I& 6772\\
V1420 Aql       & 19175--0807 & 19& 20& 18& -- 8&  2& 12& & I& 6780\\
V374 Aql        & 19276--0056 & 19& 30& 15& -- 0& 50&  9& & I& 4301\\
V1965 Cyg       & 19321+2757 & 19& 34& 10& +28&  4&  8&X& I& 4347\\
19358+0917      & 19358+0917 & 19& 38& 13& + 9& 24& 9& & I& 4378\\
19455+0920      & 19455+0920 & 19& 47& 56& + 9& 28&  9& & I& 4475\\
R Cap           & 20084--1425 & 20& 11& 18& --14& 16&  3& &  & 4701\\
RT Cap          & 20141--2128 & 20& 17&  7& --21& 19&  4& &  & 4774\\
BD Vul          & 20351+2618 & 20& 37& 18& +26& 29& 13& &  & 4915\\
V442 Vul        & 20570+2714 & 20& 59& 10& +27& 26& 39& & I& 5063\\
RV Aqr          & 21032--0024 & 21&  5& 52& -- 0& 12& 42& &  & 5120\\
Y Pav           & 21197--6956 & 21& 24& 17& --69& 44&  2& &  & 5239\\
$[$TI98$]$2223+2548 & 22239+2548 & 22& 26& 19& +26&  3& 38& &  &     \\
$[$TI98$]$2259+1249 & 22592+1249 & 23&  1& 47& +13&  5& 14& &  &     \\
LL Peg          & 23166+1655 & 23& 19& 13& +17& 11& 33& & I& 6913\\
RU Aqr          & 23217--1735 & 23& 24& 24& --17& 19&  9& & i&     \\
IZ Peg          & 23257+1038 & 23& 28& 17& +10& 54& 37& & I& 6916\\
\\
\multicolumn{12}{l}{CS stars, peculiar and uncertain C stars}\\
R Ori           & 04562+0803 &  4& 58& 59& + 8&  7& 49& &  &  828& 6\\
R CMi           & 07059+1006 &  7&  8& 42& +10&  1& 26& &  & 1561& 6\\
08276--5125      & 08276--5125 &  8& 29&  8& --51& 35&  5& & i& & C?\\
08439--2734      & 08439--2734 &  8& 46& 6& --27& 45& 49& & I& 2329 & 7\\
UX Pyx          & 09075--2758 &  9&  9& 41& --28& 10& 21& & i&  & 8\\
MU Vel          & 09450--4716 &  9& 46& 54& --47& 29& 51& & i& & C?\\
10226--5229     & 10226--5229 & 10& 24& 34& --52& 43& 29& & I&  & C?\\
$[$ABC89$]$Car10& 10293--5912 & 10& 31&  9& --59& 28& 16&X&Ai& 6382& 9\\
TU Car          & 10331--6027 & 10& 34& 55& --60& 42& 35&X&A & 2795&10\\
V354 Cen        &            & 11& 50& 59& --47& 55& 21& &  &  &C?  \\
$[$ABC89$]$Cen50    & 11529--6350 & 11& 55& 29& --64&  7& 23&X&A && 9\\
BH Cru          & 12135--5600 & 12& 16& 16& --56& 17&  7& &  & & 6 \\
TT Cen          & 13163--6031 & 13& 19& 34& --60& 46& 44&X& i& 3367& 6\\
RV Cen          & 13343--5613 & 13& 37& 35& --56& 28& 33& &  & 3412& 11\\
16316--5026      & 16316--5026 & 16& 35& 30& --50& 32& 10& & I&  & 8 \\
VX Aql          & 18575--0139 & 19&  0&  9& -- 1& 34& 56&X&  &  &6 \\
18595--3947     &18595--3947 & 19 & 3& 2& --39 & 42& 56&  & I& &  12\\
V1293 Aql       & 19306+0455 & 19& 33&  6& + 5&  1& 45& & I&  & 13\\
\end{longtable}
\end{center}
Notes:\\
1. Complex flattened circumstellar shell (Richichi et al. 1998).\\
2. Super lithium rich star (Catchpole \& Feast 1976).\\
3. No 2MASS photometry; image blended.\\ 
4. C star $35''$ east of OH353.81+1.45; IRAS probably blend of two.\\
5. 17463--4007 is the only known cool C star in the Bulge as noted from an
objective prism survey by S. Hughes (private communication). \\
6. SC star (Keenan \& Boeshaar 1980).\\
7. SC star (Lloyd Evans 1991).\\ 
8. SIMBAD quotes spectral type other than C.\\
9. S star (Lloyd Evans \& Catchpole 1989).\\
10. Not a C star (Aaronson et al. 1989).\\
11. C star with silicate shell (Skinner et al. 1990).\\
12. Not a C star (Chen et al. 2003).\\
13. Not a C star, e.g. Groenewegen (1994) and references therein.\\
C? These have been suggested as C rich on the basis of their infrared colours 
or IRAS spectra, but the evidence is inconclusive.\\ 


%
%
\begin{center}
\begin{longtable}{lrrrrrrrrlllr}
\caption[Near-infrared data.]{Near-infrared data.} \label{jhkl} \\

\hline  \\
Name           &  $J$&  $H$  &    $K$ & $L$  &
$\Delta J$ &$\Delta H$ &$\Delta K$ & $\Delta L$ &${\rm P}_K$ &P$_{lit}$ &
Var& no.  \\
& \multicolumn{8}{c}{(mag)} & \multicolumn{2}{c}{(day)}\\
\hline\\
\endfirsthead

\hline  \\
Name           &  $J$&  $H$  &    $K$ & $L$  &
$\Delta J$ &$\Delta H$ &$\Delta K$ & $\Delta L$ &${\rm P}_K$ &P$_{lit}$ &
Var& no.  \\
\hline\\
\endhead

  \multicolumn{13}{l}{{Continued on Next Page\ldots}} \\
\endfoot

  \\ \hline 
\endlastfoot
R Scl               & 2.02&   0.66& --0.08& --0.73& 0.86& 0.64& 0.36& 0.28& 375 &363&20 &157  	\\ 
YY Tri              &     &   9.82&   6.81&   3.06&     & 1.80& 1.74& 1.40& 624 &    &10 & 74    	\\ 
R For               & 4.08&   2.41&   1.21& --0.07& 1.10& 0.92& 0.66& 0.58& 385 &389 &11$^\dag$ &100   	\\ 
EV Eri              & 5.55&   4.30&   3.60&   3.00&     &     &     &     & 226 &    &21$^\dag$ & 59	\\ 
$[$TI98$]$0418+0122 & 9.23&   7.36&   6.01&   4.53& 1.64& 1.42& 1.08& 0.78& 422 &   &10 & 31	\\ 
V718 Tau            & 5.92&   4.07&   2.84&   1.63& 1.46& 1.18& 0.78& 0.80& 388 &405&13$^*$ & 14	\\ 
TT Tau              & 2.78&   1.45&   0.97&   0.54&     &     &     &     &     &166 &20 &  1	\\ 
R Lep               & 2.49&   1.03&   0.07& --0.94& 0.94& 0.78& 0.52& 0.50& 438 &427 &11$^\dag$ & 71	\\ 
TU Tau              & 3.62&   2.25&   1.68&   1.09&     &     &     &     &     &190:&20 &  1	\\ 
Y Tau               & 2.13&   0.83&   0.27& --0.29&     &     &     &     &     &242 &20 &  6	\\ 
QS Ori              & 6.40&   4.82&   3.80&   2.73& 1.58& 1.44& 1.02& 1.00& 483 &476 &10 & 12	\\ 
05418--3224         & 9.15&   6.68&   4.86&   2.75& 1.70& 1.56& 1.36& 1.30& 483&&11:$^*$ & 15	\\ 
V1259 Ori           &     &  11.29&   7.73&   3.44&     &     &     &     &     &696 &10 &  2	\\ 
06088+1909          & 8.57&   6.07&   4.26&   2.25& 1.38& 1.20& 1.00& 0.86& 493 &    &10 & 13	\\ 
BN Mon              & 4.44&   2.99&   2.24&   1.59&     &     &     &     &     &600:&20 &  3	\\ 
ZZ Gem              & 5.41&   4.02&   3.21&   2.60&$>$0.8&$>$0.6&$>$0.4&$>$0.2&316:&317&10&9 	\\ 
V617 Mon            & 7.24&   5.45&   4.17&   2.83& 0.88& 0.84& 0.72& 0.62& 444 &375:&14$^*$ & 13\\ 
V636 Mon            & 5.01&   3.13&   1.82&   0.41& 1.62& 1.26& 0.86& 0.56& 543 &    &10 & 15	\\ 
V477 Mon            & 8.87&   6.45&   4.55&   2.26& 1.66& 1.48& 1.22& 0.96& 619 &820:&10 & 12	\\ 
CR Gem              & 3.36&   1.91&   1.38&   0.92&     &     &     &     &     &250 &20 &  1	\\ 
GM CMa              & 4.52&   3.10&   2.22&   1.39& 0.76& 0.54& 0.30& 0.22& 403 &    &21 & 94	\\ 
V503 Mon            & 8.11&   6.59&   5.69&   4.9:& 0.60& 0.40& 0.16& 0.16& 357 &355 &20 & 11	\\ 
RT Gem              & 6.51&   5.21&   4.62&   4.21& 0.76& 0.74& 0.58& 0.72& 350 &350 &10 &  9	\\ 
06487+0551          & 9.14&   6.58&   4.62&   2.24& 1.34& 1.20& 1.02& 0.74& 536 &    &10 & 12	\\ 
CG Mon              & 5.33&   3.96&   3.30&   2.75& 0.82& 0.74& 0.60& 0.70& 424 &419 &10 &  9	\\ 
CL Mon              & 4.78&   3.06&   1.88&   0.59& 1.28& 1.02& 0.74& 0.52& 511 &497 &10 & 10	\\ 
06531--0216         & 6.95&   4.89&   3.35&   1.61& 1.32& 1.18& 0.98& 0.64& 595:&    &14$^*$ & 13	\\ 
NP Pup              & 2.49&   1.38&   1.03&   0.64&     &     &     &     &     &    &20 & 3	\\ 
06564+0342          &10.26&   7.59&   5.43&   2.84& 1.50& 1.32& 1.16& 1.02& 584 &    &10 & 12	\\ 
W CMa               & 2.59&   1.40&   0.96&   0.50&     &     &     &     &     &    &20 & 13	\\ 
07080--0106         &11.5 &   8.68&   6.23&   3.49&     & 2.54& 1.60& 1.04& 594 &    &10 & 12	\\ 
VX Gem              & 4.96&   3.74&   3.13&   2.78& 1.14& 1.06& 0.66& 0.78& 391 &379 &10 &  9	\\ 
07097--1011         & 8.91&   7.04&   5.92&   4.95& 0.50& 0.36& 0.22& 0.24& 437 &    &20 & 13	\\ 
R Vol               & 5.08&   3.14&   1.71&   0.08& 1.46& 1.26& 0.98& 0.80& 452 &454&11$^\dag$ & 88	\\ 
HX CMa              & 7.92&   5.46&   3.62&   1.42&     &     &     &     &     &725 &10 &  1	\\ 
07136--1512         & 8.12&   6.50&   5.57&   4.74& 0.36& 0.28& 0.16& 0.10& 486:&    &20 & 12	\\ 
07161--0111         & 7.40&   5.26&   3.58&   1.64&     &     &     &     &     &    &23 & 12	\\ 
07217--1246         & 8.88&   6.32&   4.30&   1.91& 1.42& 1.32& 1.12& 0.88& 620:&    &10 & 12	\\ 
07220--2324         & 9.51&   6.80&   4.77&   2.46& 1.58& 1.24& 1.06& 0.88& 560 &    &10 & 13	\\ 
07223--1553         & 7.98&   6.23&   5.23&   4.38& 0.56& 0.44& 0.24& 0.18& 457 &    &20 & 13	\\ 
07293--1832         & 8.80&   6.99&   6.02&   5.17&     &     &     &     &     &    &20 & 12	\\ 
07319--1940         & 7.91&   6.36&   5.64&   4.99&     &     &     &     &     &    &20 &  6	\\ 
$[$W71b$]$007--02   & 7.85&   6.08&   4.93&   3.79& 1.12& 0.94& 0.70& 0.40& 460:&    &13 & 15	\\ 
07373--4021         & 5.39&   3.52&   2.23&   0.77& 1.40& 1.20& 0.90& 0.78& 459 &471 &13 & 15	\\ 
$[$W71b$]$008--03   & 9.50&   8.01&   7.30&       &     &     &     &     &     &    &20 & 6	\\ 
V471 Pup            & 7.08&   5.78&   5.13&   4.63& 1.04& 0.98& 0.76& 0.92& 390 &    &10 & 16	\\ 
$[$ABC89$]$Pup3     & 7.79&   6.31&   5.37&   4.61& 0.24& 0.22& 0.10&     & 309 &    &20 &  9	\\ 
$[$ABC89$]$Pup17    & 9.42&   7.86&   7.12&   6.54&     &     &     &     &     &    &20 &  6	\\ 
07454--7112         & 8.03&   5.08&   2.82&   0.07& 2.08& 1.98& 1.72& 1.46& 511 &    &10 & 11	\\ 
$[$ABC89$]$Pup21    & 9.27&   7.55&   6.38&   5.34&     &     &     &     &     &    &23 & 17	\\ 
V831 Mon            & 6.53&   4.99&   3.89&   2.76& 1.38& 1.04& 0.74& 0.58& 331 &319 &10 & 13	\\ 
07576--4054         &     &   9.29&   6.62&   3.39&     & 1.98& 1.80& 1.40& 519 &    &10 & 16	\\ 
07582--1933         &10.70&   7.74&   5.39&   2.69& 1.84& 1.82& 1.62& 1.42& 541 &    &10 & 14	\\ 
V509 Pup            & 5.95&   4.48&   3.72&   3.01&     &     &     &     &     &    &20 &  8	\\ 
$[$ABC89$]$Pup38    &10.65&   8.79&   7.43&   6.06& 1.26& 1.08& 0.88& 1.00& 431 &    &10 & 15	\\ 
FF Pup              & 6.64&   5.38&   4.69&   4.00& 1.10& 1.16& 0.94& 1.06& 431 &436 &10 &  8	\\ 
$[$ABC89$]$Pup42    & 8.52&   6.97&   6.16&   5.45& 0.28& 0.22& 0.14& 0.06& 199 &    &20 & 15	\\ 
V518 Pup            & 6.96&   4.90&   3.52&   2.07& 1.06& 0.94& 0.76& 0.80& 228 &448 &13 & 13	\\ 
08050--2838         &10.40&   7.28&   5.08&   2.57& 2.10& 1.80& 1.48& 1.26& 555 &    &13 & 11	\\ 
RU Pup              & 3.65&   2.52&   2.04&   1.48&     &     &     &     &     &425 &20 &  2	\\ 
FK Pup              & 3.55&   2.16&   1.46&   0.78&     &     &     &     &     &502 &20 &  2	\\ 
08074--3615         &     &  13.13&   9.09&   4.50&     & 2.14& 1.74& 1.46& 832 &    &10 & 13	\\ 
$[$ABC89$]$Ppx19    & 9.98&   7.77&   6.11&   4.23& 1.78& 1.30& 0.88& 0.46& 474 &    &13 & 16	\\ 
$[$W71b$]$021--05   & 6.66&   5.19&   4.40&   3.71&     &     &     &     &     &    &20:&  4	\\ 
$[$ABC89$]$Ppx22    & 6.59&   5.10&   4.34&   3.62&     &     &     &     &     &    &20:&  4	\\ 
V346 Pup            & 8.67&   5.92&   3.75&   1.16& 1.78& 1.60& 1.38& 1.14& 568 &571 &10 & 51	\\ 
$[$W71b$]$026--01   & 8.27&   6.51&   5.58&   4.72&     &     &     &     &     &    &20:&  4	\\ 
RY Hya              & 4.36&   2.98&   2.24&   1.46& 0.46& 0.30& 0.14& 0.12& 516 &529 &20 & 61	\\ 
$[$ABC89$]$Ppx40    & 7.68&   6.15&   5.36&   4.73& 1.20& 1.04& 0.74& 0.78& 428 &439 &10 & 13	\\ 
$[$W71b$]$029--02   & 9.48&   7.28&   5.63&   3.86& 1.06& 0.92& 0.72& 0.52& 470 &    &10 & 19	\\ 
$[$W71b$]$029--04   & 8.62&   6.80&   5.82&   5.01&     &     &     &     &     &    &20:&  4	\\ 
08340--3357         &11.29&   8.14&   5.61&   2.56& 2.32& 1.56& 1.46& 1.16& 590 &    &10 & 18	\\ 
R Pyx               & 4.55&   3.28&   2.53&   1.81&$>$0.74&$>$0.46&$>$0.22&$>$0.34&369&365&13 &11	\\ 
UW Pyx              & 4.87&   3.46&   2.68&   1.85&     &     &     &     & &423 &10 &  5	\\ 
T Cnc               & 3.21&   1.81&   1.07&   0.43&     &     &     &     &     &482 &20 &  1	\\ 
08535--4724         &     &  10.78&   7.58&   4.05&     & 1.84& 1.72& 1.54& 570:&    &13 & 14	\\ 
08534--5055         &     &  11.96&   8.50&   4.55&     & 2.56& 1.92& 1.60& 703 &    &10 & 15	\\ 
IQ Hya              & 5.68&   4.06&   2.89&   1.64& 1.14& 0.94& 0.70& 0.68& 382 &397 &10 & 15	\\ 
CQ Pyx              &     &   9.31&   5.98&   2.09&     & 1.84& 1.82& 1.48& 659 &    &10 & 33	\\ 
09164--5349         & 4.14&   2.73&   2.11&   1.40&     &     &     &     &     &&25$^\dag$ & 12	\\ 
09176--5147         &11.57&   8.59&   6.31&   3.63& 1.50& 1.54& 1.36& 1.02& 431:&    &13 & 16	\\ 
$[$ABC89$]$Vel19    &10.11&   7.98&   6.75&   5.64&     &     &     &     &     &    &21 & 20	\\ 
$[$W71b$]$046--02   & 9.03&   7.27&   6.38&   5.78& 1.00& 0.76& 0.40& 0.34& 265 & &10 & 15	\\ 
$[$ABC89$]$Vel44    & 8.02&   6.25&   5.15&   4.09& 0.74& 0.60& 0.40& 0.32& 413 &    &10 & 16	\\ 
09433--6233         &11.33&   8.94&   7.18&   5.17& 1.06& 1.10& 1.10& 1.10& 590:&    &13 & 12	\\ 
CW Leo              & 7.34&   4.04&   1.19& --2.54& 2.06& 2.14& 2.06& 1.74& 651 &630 &10 & 37	\\ 
W Sex               & 5.11&   3.95&   3.49&   3.06& 0.32& 0.26& 0.16& 0.12& 195 &134 &20 & 17	\\ 
09484--6242         & 7.63&   6.27&   5.56&   4.89& 0.06& 0.10& 0.08& 0.06& 244:&    &20 &  7	\\ 
09513--5324         &10.92&   7.60&   5.07&   2.04& 1.92& 1.70& 1.54& 1.30& 630 &    &10 & 12	\\ 
09529--5506         &10.66&   7.91&   5.82&   3.11& 2.56& 2.08& 1.72& 1.50& 688:&    &10 &  8	\\ 
09533--6021         &12.78&   9.54&   7.03&   4.04& 1.96& 1.90& 1.72& 1.38& 714 &    &10 & 12	\\ 
09521--7508         & 8.11&   5.36&   3.22&   0.70& 1.98& 1.96& 1.74& 1.28& 539 &    &10 &  9	\\ 
09586--6150         &     &  12.02&   9.19&   5.90&     & 1.78& 1.68& 1.38& 506 &    &10 & 11	\\ 
10019--6156         & 7.87&   6.55&   5.99&   5.49&     &     &     &     &     &    &20 &  7\\ 
10023--5946         & 6.94&   5.53&   4.90&   4.20& 0.32& 0.26& 0.18& 0.12& 571:&    &20 & 11\\ 
10026--5849         & 9.80&   7.49&   5.90&   4.22& 1.04& 0.94& 0.68& 0.52& 531 &    &13 & 10\\ 
10052--5906         &     &   7.46&   6.47&   5.67& 0.52& 0.38& 0.20& 0.10& 448 &    &20 & 12\\ 
10098--5742         &     &  10.29&   7.55&   4.24&     & 1.38& 1.26& 1.16& 585 &    &10 & 14\\ 
10109--5958         & 7.26&   5.49&   4.27&   3.03& 0.78& 0.70& 0.58& 0.50& 423:&    &13 & 11\\ 
RW LMi              & 6.18&   3.43&   1.32& --1.21& 1.86& 1.72& 1.50& 1.28& 617 &640 &10 & 14\\ 
10130--5703         & 6.30&   4.07&   3.07&   2.23&     &     &     &     &     & &20 &  5\\ 
10136--5743         & 8.33&   6.40&   5.23&   3.94&     &     &     &     &&&15:$^\dag$ & 13\\ 
10145--6046         & 5.78&   4.35&   3.65&   2.97&     &     &     &     &     &    &20 & 11\\ 
10149--5919         & 6.68&   5.18&   4.53&   3.90&     &     &     &     &     &    &20 &  6\\ 
10151--6008         & 9.28&   7.86&   7.21&   6.51&     &     &     &     &     &    &20 &  6\\ 
10175--5957         & 8.28&   6.61&   5.66&   4.86&     &     &     &     &     &    &20 & 12\\ 
10199--5801         &11.81&   8.51&   6.20&   3.52& 1.66& 1.56& 1.28& 1.02& 675 &    &10 & 14\\ 
10220--5858         & 8.54&   6.62&   5.39&   4.20& 0.86& 0.64& 0.42& 0.32& 585:&    &10 & 11\\ 
CPD--58 2175        &12.33&   9.39&   7.14&   4.54& 1.74& 1.58& 1.42& 1.22& 548 &    &10 & 10\\ 
CZ Hya              & 4.73&   3.32&   2.39&   1.38& 1.42& 1.30& 0.96& 0.92& 444 &442 &10 & 12\\ 
$[$ABC89$]$Car5     & 9.76&   8.28&   7.53&   6.83& 0.20& 0.20& 0.14&     & 168 &    &20 & 12\\ 
$[$ABC89$]$Car11    & 8.50&   6.99&   6.17&   5.41&     &     &     &     &     &    &20:&  4\\ 
TV Vel              & 5.08&   3.82&   3.24&   2.74& 0.60& 0.54& 0.40& 0.40& 404 &365 &10 & 14\\ 
U Ant               & 1.24&   0.01& --0.49& --1.11&     &     &     &     &     &    &20 & 25\\ 
$[$ABC89$]$Car28    & 8.95&   7.29&   6.36&   5.47& 0.56& 0.42& 0.24& 0.20& 495 &    &20 & 16\\ 
$[$ABC89$]$Car32    & 7.67&   5.85&   4.84&   3.90&     &     &     &     &     &    &20 & 15\\ 
FU Car              & 6.41&   4.81&   3.85&   2.89& 0.70& 0.58& 0.32& 0.28& 431:&365:&20 & 13\\ 
$[$ABC89$]$Car54    & 6.98&   5.16&   4.26&   3.44& 0.22& 0.22&     &     & 232:&    &20 &  6\\ 
$[$ABC89$]$Car59    & 7.22&   5.56&   4.62&   3.78&     &     &     &     &     &    &20:&  4\\ 
V Hya               & 1.78&   0.29& --0.70& --1.86&$>$0.52&$>$0.42&$>$0.30&$>$0.28& 532 &531 &22$^\dag$ & 75\\ 
$[$ABC89$]$Car73    &10.69&   8.10&   6.34&   4.41& 1.88& 1.32& 1.06& 0.84& 483 &    &10 & 14\\ 
$[$ABC89$]$Car81    & 9.12&   7.28&   6.28&   5.46&     &     &     &     &     &    &20:&  3\\ 
$[$ABC89$]$Car84    & 8.07&   6.37&   5.34&   4.25& 1.00& 0.80& 0.52& 0.46& 501 &    &10 & 15\\ 
$[$ABC89$]$Car87    & 8.66&   7.10&   6.12&   5.43& 1.08& 0.88& 0.52& 0.38& 473 &    &10 &  9\\ 
$[$ABC89$]$Car93    & 8.64&   6.64&   5.54&   4.56& 0.72& 1.06& 0.86& 0.44& 416 &    &10 & 12\\ 
$[$ABC89$]$Car105   & 7.32&   5.47&   4.24&   2.91& 0.90& 0.70& 0.46& 0.34& 497 &&11:$^*$ & 15\\ 
11145--6534         &10.00&   6.93&   4.49&   1.69& 1.72& 1.72& 1.58& 1.46& 623 &    &10 & 12\\ 
$[$W65$]$ c1        & 9.16&   7.59&   6.81&       &     &     &     &     &     &    &20 &  8\\ 
$[$W65$]$ c2        & 8.03&   6.70&   6.26&       &     &     &     &     &     &    &24$^*$ &  8\\ 
$[$W65$]$ c13       & 8.68&   6.92&   5.85&   4.91& 0.96& 0.74& 0.44& 0.34& 395 &    &10 & 14\\ 
$[$TI98$]$1130--1020& 7.82&   5.68&   4.04&   2.20& 1.38& 1.24& 1.12& 0.94& 443 &    &10 & 20\\ 
11318--7256         & 4.01&   2.13&   0.85& --0.66& 1.38& 1.14& 0.90& 0.68& 526 &535 &10 &  9\\ 
$[$ABC89$]$Cen3     & 9.63&   7.87&   7.00&       &     &     &     &     &     &    &20:&  5\\ 
$[$ABC89$]$Cen4     & 7.12&   5.53&   4.68&   3.83& 0.76& 0.68& 0.46& 0.44& 514 &    &10 & 16\\ 
11463--6320         &10.39&   7.50&   5.44&   2.99& 2.02& 2.06& 1.80& 1.60& 615 &    &10 & 14\\ 
$[$ABC89$]$Cen32    & 8.95&   6.67&   5.05&   3.28& 1.14& 1.04& 0.84& 0.70& 652 &    &10 & 14\\ 
$[$ABC89$]$Cen43    & 9.16&   7.18&   5.82&   4.41& 1.36& 1.10& 0.78& 0.50& 535 &    &10 & 13\\ 
$[$ABC89$]$Cen60    & 8.66&   6.84&   5.71&   4.56& 1.02& 0.86& 0.60& 0.68& 414 &    &10 & 13\\ 
$[$ABC89$]$Cen78    & 9.39&   7.51&   6.48&   5.65& 0.54& 0.34& 0.20&     & 401 &    &20 & 13\\ 
CF Cru              & 8.91&   7.14&   6.21&       & 0.72& 0.62& 0.42&     & 430 &    &10 & 15\\ 
$[$ABC89$]$Cen97    &10.3 &   7.90&   6.57&   5.27& 2.12& 1.32& 0.80& 0.46&     &    &23 & 14\\ 
12194--6007         & 9.59&   7.00&   4.96&   2.57& 1.66& 1.54& 1.26& 0.94& 627 &    &10 & 12\\ 
SS Vir              & 2.93&   1.56&   0.75&   0.01& 0.92& 0.68& 0.36& 0.20& 359 &364 &20 & 55\\ 
12298--5754         & 9.28&   6.65&   4.40&   1.70& 1.56& 1.66& 1.54& 1.24& 580 &    &10 & 10\\ 
CGCS3268            & 5.63&   4.28&   3.43&   2.78&     &     &     &     &     &396 &10 &  1\\ 
12394--4338         & 7.49&   4.89&   2.91&   0.72& 1.68& 1.62& 1.30& 1.24& 551 &    &14$^*$ & 12\\ 
12421--6217         &     &       &   8.47&   4.50&     &     & 2.40& 2.02& 806 &    &10 &  8\\ 
RU Vir              & 4.86&   3.09&   1.80&   0.29& 1.40& 1.24& 0.98& 0.90& 444 &433 &10 & 46\\ 
V Cru               & 4.71&   3.46&   2.88&   2.45& 0.76& 0.68& 0.50& 0.64& 380 &376 &10 & 96\\ 
12540--6845         & 8.20&   5.52&   3.52&   1.15& 2.04& 1.78& 1.56& 1.30& 586 &    &10 & 10\\ 
$[$ABC89$]$Cru17    &10.16&   8.39&   7.45&   6.15&     &     &     &     &     &    &20:&  1\\ 
$[$ABC89$]$Cir1     & 7.91&   5.95&   4.85&   3.56&     &     &     &     &     &    &25 & 11\\ 
13343--5807         & 9.75&   7.04&   4.98&   2.57& 1.86& 1.88& 1.62& 1.42& 556 &    &10 & 13\\ 
13477--6532         &     &   9.67&   6.62&   2.78&     & 1.46& 1.68& 1.36& 690 &    &10 & 13\\ 
13482--6716         & 8.47&   5.79&   3.78&   1.49& 1.66& 1.66& 1.44& 1.18& 500 &    &10 & 12\\ 
13509--6348         &     &       &   5.12&   2.72&     & 1.70& 1.46& 1.36& 678 &    &10 & 16\\ 
$[$ABC89$]$Cir18    &10.04&   8.27&   7.33&   6.5 &     &     &     &     &     &    &20 & 13\\ 
$[$ABC89$]$Cir26    & 9.13&   6.71&   5.12&   3.36& 1.74& 1.36& 0.96& 0.74& 495 &    &10 & 15\\ 
$[$ABC89$]$Cir27    & 9.50&   7.12&   5.27&   3.10& 1.18& 1.22& 1.02& 0.84& 538 &    &13$^*$ & 12\\ 
$[$W71b$]$093--02   &10.5 &   8.67&   7.41&       &     &     &     &     &     &    &10:&  4\\ 
14395--5656         & 9.15&   7.21&   6.01&   4.73& 0.86& 0.66& 0.44& 0.28& 488 &    &10 & 11\\ 
14404--6320         &     &  11.46&   8.32&   4.39&     & 2.08& 2.18& 1.70& 643 &    &10 &  9\\ 
14443--5708         &     &  12.77&   8.93&   4.94&     & 2.98& 2.16& 1.80& 723 &    &10 &  8\\ 
15082--4808         &10.00&   7.00&   4.36&   0.86& 1.76& 1.68& 1.64& 1.48& 632 &    &10 & 13\\ 
15084--5702         &     &  10.85&   7.33&   3.37&     & 2.38& 2.36& 1.94& 948 &    &10 & 10\\ 
II Lup              & 5.92&   3.58&   1.79& --0.33& 1.04& 0.92& 0.82& 0.88& 576 &580 &11$^\dag$ & 28\\ 
15261--5702         &     &   8.45&   5.96&   3.07&     & 2.34& 2.02& 1.40& 716 &    &10 & 15\\ 
15471--5644         &     &       &       &   7.78&     &     &     &     &     &   &10$^*$ &  1\\ 
CGCS3660            & 6.82&   5.67&   5.38&   5.19&     &     &     &     &     &    &20:&  2\\ 
16079--4812         &     &  10.24&   6.85&   2.88&     & 2.02& 2.32& 1.96& 710 &    &10 & 12\\ 
NP Her              & 5.79&   4.36&   3.50&   2.79&     &     &     &     &     &448 &10 &  1\\ 
16171--4759         & 9.58&   6.89&   4.89&   2.70& 1.26& 1.28& 1.06& 0.86& 560 &    &13 & 14\\ 
V Oph               & 3.65&   2.35&   1.61&   0.99& 1.20& 0.86& 0.48& 0.38& 294 &297 &10 & 11\\ 
SU Sco              & 3.08&   1.77&   1.11&   0.56&     &     &     &     &     &414:&20 &  5\\ 
CGCS3721            & 6.56&   5.08&   4.12&   3.28&     &     &     &     &     &353 &10 &  1\\ 
16406--1406         &     &  11.79&   8.74&   5.10&     &     &     &&&&15:$^\dag$ & 28\\ 
16538--4633         & 8.84&   6.07&   4.27&   2.30& 1.18& 1.04& 0.84& 0.72& 527 &    &13 & 12\\ 
16545--4214         & 7.14&   4.50&   2.52&   0.38& 1.62& 1.68& 1.46& 1.28& 534 &    &10 & 12\\ 
T Ara               & 4.53&   3.27&   2.77&   2.27& 0.20& 0.12& 0.10& 0.08& 327 &    &20 & 13\\ 
V901 Sco            & 5.52&   4.01&   3.18&   2.38& 0.70& 0.58& 0.38& 0.34& 439:&    &20 & 10\\ 
17047--2848         & 9.45&   7.09&   5.36&   3.28& 1.88& 1.74& 1.52& 1.38& 531 &    &10 & 12\\ 
V2548 Oph           &     &   8.48&   5.61&   1.98&     & 2.08& 2.06& 1.80& 747 &    &10 & 40\\ 
SZ Ara              & 6.16&   5.00&   4.45&   4.06& 0.94& 0.68& 0.44& 0.42& 222 &220 &10 &  8\\ 
V617 Sco            & 4.67&   3.15&   2.17&   1.23& 1.42& 1.48& 1.14& 1.04&     &524 &10 &  6\\ 
17105--3746         & 9.74&   7.16&   5.14&   2.60& 1.98& 2.18& 2.10& 1.82& 568 &    &10 & 10\\ 
17130--3907         & 8.92&   6.15&   4.23&   2.09& 2.32& 1.82& 1.36& 1.18& 628 &    &13 & 14\\ 
17209--3318E        &12.63&   9.04&   6.24&   3.18&     &     &     &     &     &    &10 &  4\\ 
17217--3916         &     &   9.30&   6.42&   3.14&     & 1.90& 1.94& 1.60& 630 &    &10 & 11\\ 
17222--2328         & 9.67&   6.66&   4.60&   2.30& 1.88& 1.74& 1.58& 1.48& 603 &    &13 &  9\\ 
V833 Her            & 9.58&   6.66&   4.19&   1.00& 2.36& 2.80& 2.72& 2.10& 540 &    &13 & 13\\ 
V Pav               & 2.08&   0.76&   0.19& --0.39&     &     &     &     & 437 &225 &20 & 69\\ 
17446--7809         & 7.22&   4.69&   2.72&   0.55&     &     &     &     &     &    &10 &  4\\ 
17446--4048         & 8.15&   5.54&   3.60&   1.47& 1.34& 1.24& 1.10& 1.18& 545 &    &13$^*$ & 12\\ 
17463--4007         &10.04&   8.37&   7.22&   5.92& 1.58& 1.32& 0.98& 0.90& 399 &    &10 & 14\\ 
V348 Sco            & 7.09&   6.11&   5.66&       & 0.26& 0.30& 0.26& 0.12&     &274 &20 &  6\\ 
17581--1744         & 8.26&   5.77&   4.03&   2.12& 1.70& 1.40& 1.08& 0.94& 628 &    &10 &  8\\ 
18036--2344         &10.71&   7.37&   4.93&   2.17& 2.66& 2.78& 2.22& 1.96& 664 &    &13$^*$ &  9\\ 
FX Ser              & 7.03&   4.45&   2.59&   0.54& 1.70& 1.54& 1.28& 1.02& 519 &    &10 & 26\\ 
V1280 Sgr           & 5.33&   3.58&   2.39&   1.02& 1.10& 1.02& 0.78& 0.70& 532 &523 &13 & 57\\ 
18119--2244         &     &   8.39&   5.60&   2.65&     & 2.24& 1.82& 1.56& 611 &    &10 & 12\\ 
18147--2215         &     &   9.92&   6.50&   2.91&     &     &     &     &     &    &10 &  5\\ 
V5104 Sgr           & 9.05&   5.87&   3.52&   0.78& 1.80& 1.72& 1.52& 1.34& 655 &    &10 & 44\\ 
V2548 Sgr           & 4.47&   2.96&   2.11&   1.22&     &     &     &     &     &159 &23 & 11\\ 
18239--0655         &10.31&   7.23&   4.64&   1.69& 1.58& 1.48& 1.44& 1.24& 635 &    &10 &  9\\ 
18244--0815         &11.4 &   8.58&   5.97&   3.27&     &     &     &     &     &    &10 &  6\\ 
V1076 Her           &     &   9.14&   5.84&   1.95&     & 2.20& 1.98& 1.62& 609 &    &10 & 12\\ 
18248--0839         &12.8 &   9.14&   6.13&   2.88&     &     & 2.04& 2.04& 659:&    &10 &  7\\ 
18269--1257         &13.1 &   8.98&   5.85&   2.53&     &     &     &     &     &  &10$^*$ &  5\\ 
18320--0352         &     &  11.61&   8.31&   4.45&     &     &     &     &     &    &10 &  4\\ 
V627 Oph            & 7.90&   5.97&   4.66&   3.17&     &     &     &     &     &452 &10 & 3\\ 
18367--0452         &     &  10.47&   7.11&   3.17&     &     &     &     &     &    &10:&  2\\ 
V1417 Aql           & 6.23&   3.78&   1.96& --0.08& 1.30& 1.10& 0.92& 0.72& 617 &    &13 & 25\\ 
18400--0704         &     &       &   8.28&   4.64&     &     &     &     &     &    &10 &  3\\ 
V821 Her            & 5.98&   3.68&   1.85& --0.25& 1.88& 1.78& 1.52& 1.22& 524 &511 &10 & 15\\ 
18424+0346          & 9.73&   6.99&   4.87&   2.48&     &     &     &     &     &    &10 &  6\\ 
V874 Aql            & 8.71&   7.61&   7.22&   6.79&     &     &     &     &     &145 &10 &  3\\ 
V2045 Sgr           & 6.21&   4.50&   3.46&   2.42&     &     &     &     &     &451 &10 &  7\\ 
S Sct               & 2.43&   1.13&   0.57&   0.05&     &     &     &     &     &148 &20 & 16\\ 
18475+0926          &     &       &   5.91&   2.36&     &     &     &     &     &    &10 &  3\\ 
AI Sct              & 6.12&   4.50&   3.48&   2.63&     &     &     &     &     &408 &10 &  5\\ 
V1418 Aql           & 8.29&   5.44&   3.14&   0.47& 2.02& 1.74& 1.42& 1.08& 562 &577 &10 & 20\\ 
19029+2017          & 8.17&   5.68&   3.87&   1.86&     &     &     &     &     &    &10 &  5\\ 
19068+0544          & 8.01&   5.41&   3.75&   2.07&     &     &     &     &     &    &10 &  7\\ 
V1420 Aql           & 5.94&   3.68&   2.08&   0.06& 1.68& 1.44& 1.22& 1.18& 694 &676 &12 & 18\\ 
V374 Aql            & 4.91&   3.30&   2.29&   1.26&     &     &     &     &     &456 &20 &  8\\ 
V1965 Cyg           & 7.64&   5.00&   2.95&   0.55& 2.52& 2.08& 1.68& 1.26& 577 &625 &13 & 11\\ 
19358+0917          &     &  11.91&   8.36&   4.46&     &     &     &     &     &    &10:&  1\\ 
19455+0920          &12.1 &   9.32&   6.73&   3.5 &     &     &     &     &     &    &10 &  5\\ 
R Cap               & 5.31&   3.94&   3.05&   2.09& 1.42& 1.18& 0.86& 0.96& 349 &345 &10 & 15\\ 
RT Cap              & 2.48&   1.15&   0.54& --0.06& 0.30& 0.24& 0.14& 0.10& 359 &393 &20 & 23\\ 
BD Vul              & 5.84&   4.30&   3.37&   2.63&     &     &     &     &     &430 &10 &  4\\ 
V442 Vul            & 9.78&   6.84&   4.22&   1.14& 2.14& 2.28& 2.02& 1.56& 661 &    &10 & 12\\ 
RV Aqr              & 4.68&   2.75&   1.39& --0.13& 1.58& 1.32& 1.04& 0.94& 433 &454 &10 &  7\\ 
Y Pav               & 1.89&   0.69&   0.26& --0.08&     &     &     &     &     &233 &20 &  2\\ 
$[$TI98$]$2259+1249 & 6.97&   5.78&   5.24&   4.90& 0.78& 0.74& 0.54& 0.68& 306 &294 &10 & 36\\ 
LL Peg              &     &       &  10.50&   4.27&     &     &     &     &     &696 &10 &  7\\ 
RU Aqr              & 3.10&   2.08&   1.78&   1.51&     &     &     &     &     & 69&20   &  1\\ 
IZ Peg              &     &  10.26&   7.09&   3.04&     & 1.98& 1.94& 1.56& 486 &486 &10 &103\\ 
$[$TI98$]$2223+2548 & 7.19&   5.56&   4.35&   3.06& 1.40& 1.18& 0.90& 0.80& 343 &    &10 & 16\\ 
\\
\multicolumn{12}{l}{CS stars, peculiar and uncertain C stars}\\
R Ori               & 5.82&   4.70&   4.15&   3.76& 0.92& 1.00& 0.78& 0.96& 381 &377 &   & 14\\ 
R CMi               & 4.03&   2.97&   2.48&   2.22& 0.88& 0.82& 0.86& 0.74& 335 &338 &   & 12\\ 
08276--5125         &     &       &  11.69&   6.63&     &     &     &     &     &    &   &  1\\ 
08439--2734         & 6.56&   4.85&   3.79&   2.40& 1.66& 1.52& 1.28& 1.26& 475 &    &  & 14	\\ 
UX Pyx              & 3.93&   2.91&   2.59&   2.26&     &     &     &     &     &423 &   &  1\\ 
MU Vel              &     &       &   9.49&   5.30&     & 1.54& 1.56& 1.48& 597 &    &   & 33\\ 
10226--5229         & 8.89&   6.25&   4.56&   2.60& 2.98& 2.32& 1.86& 1.50& 756 &    &   & 10\\ 
$[$ABC89$]$Car10    & 7.00&   5.59&   4.99&   4.37& 1.14& 1.00& 0.88& 0.88& 397 &    &   & 15\\ 
TU Car              & 7.16&   6.07&   5.52&   4.92& 0.78& 0.86& 0.74& 0.58& 254 &258 &   & 10\\ 
V354 Cen            & 9.22&   8.33&   8.04&   7.6 & 0.38& 0.32& 0.18&     & 150:&150 &   & 11\\ 
$[$ABC89$]$Cen50    & 9.73&   7.54&   6.01&   4.34& 0.92& 0.82& 0.70& 0.60& 512 &    &   & 12\\ 
BH Cru              & 3.21&   2.03&   1.56&   1.23& 0.62& 0.66& 0.58& 0.64& 491 &421 &   & 46\\ 
BH Cru              & 3.15&   1.98&   1.40&   1.01& 0.72& 0.70& 0.50& 0.64& 524 &421 &   & 36\\ 
TT Cen              & 4.35&   2.99&   2.43&   1.91&$>$0.72&$>$0.76&$>$0.68&$>$0.80& 448 &462 &  & 89\\ 
RV Cen              & 3.33&   2.08&   1.47&   1.00& 0.68& 0.56& 0.36& 0.46& 447 &446 &   & 87\\ 
16316--5026         & 4.28&   2.56&   1.82&   1.00& 1.48& 0.96& 0.84& 0.74& 565 &    &   & 13\\ 
VX Aql              & 4.90&   3.59&   3.06&   2.41&     &     &     &     &     &604 &   &  2\\ 
18595--3947         & 3.30&   1.52&   0.51& --0.69& 2.02& 1.50& 1.10& 0.98& 449 &    &   & 22\\ 
V1293 Aql           & 2.10&   1.11&   0.83&   0.59&     &     &     &     &     &    &   &  5\\ 
\end{longtable}\end{center}
\dag These stars are discussed in the section \ref{trends} on long term
trends.\\
$*$These are stars for which the second parameter 
of the variable type depends on the combination of our observations and 
photometry from other sources.
{\bf V718 Tau} 
 Epchtein et al. (1990) and 2MASS data obtained before and after SAAO
observations, at phases which are similar to our faintest measurements (i.e.
not in the gap) suggest that it has been much fainter ($K$=3.94,
3.88 respectively, $\Delta K\sim 0.7$) and redder, than the faintest SAAO
measurements.
{\bf 05418--3224} 
 An observation by Epchtein et al. (1990) predating ours is 3 and 5 mag
brighter at $K$ and $J$ respectively. 2MASS photometry contemporaneous with 
ours and Fouqu\'e et al. (1992) measurements predating ours are within the 
range shown in our light curve. We therefore class this star as showing 
obscuration events, but note that the evidence is very limited.
{\bf V617 Mon}
 Although Noguchi et al. (1981) present measures distinctly different from
ours, a comparison with 2MASS suggests they actually observed BD+08$^{\rm
o}$1312, an M star about one arcmin away from V617 Mon.
{\bf 06531--0216}
 Note that the 2MASS ($K=4.45$) and Epchtein et al. (1990) ($K=2.89$)
observations do not agree with the phasing of the SAAO data.  The period
must therefore be regarded as uncertain.
{\bf [ABC89]Car105}
 The Aaronson et al. (1989) observation ($K=5.41$), which predates the SAAO
photometry, is much fainter than our minimum ($K=4.48$). This may
indicate an obscuration event.
{\bf [W65]~c2}
 While the 2MASS, the 1985 Aaronson et al. (1989) and the SAAO observations
differ by less than 0.1 mag at $K$, the 1988 Aaronson et al. photometry is 
considerably fainter ($J=9.14$, $K=7.21$).
{\bf 12394--4338}
 The Fouqu\'e et al. (1992) observation ($J=8.55$, $K=4.48$) is significantly
fainter at $K$, but not at $J$, than the SAAO minimum ($J=8.60$, $K=3.74$),
while the 2MASS and Epchtein et al. (1990) observations are comparable to
those listed here.
{\bf[ABC89]Cir27}
 The Aaronson et al. (1989) measurement ($K=3.87$) is significantly brighter
than the SAAO maximum ($K=4.57$) and may indicate that the source was
obscured during the SAAO observations.
{\bf 15471--5644} 
Too crowded for measurement at $JHK$, but 2MASS has $K=14.8$ and Groenewegen
et al. (1993) have $K=11.4$, $L=4.6$, so it is certainly a large amplitude
variable.
{\bf 17446--4048}
 The Fouqu\'e et al. (1992) measurement ($K=4.81$) is significantly fainter than
the SAAO minimum ($K=4.21$) and may indicate the star was being obscured at
the time the observation was made. 2MASS is also faint ($K=4.65$).
{\bf 18036--2344}
 The Guglielmo et al. (1993) measurement ($K=6.53$) is significantly fainter
than the SAAO minimum ($K=5.81$) and may indicate that the star was in an
obscuration phase.
{\bf 18269--1257}
2MASS has $K=8.02$ on JD\,2450937, so all our observations are near maximum
and $P\sim 700$ days.
\twocolumn

Some of the photometry discussed in the present paper has already been
published by Whitelock et al. (1994, 1995, 1997, 2000), Olivier et al.
(2001), Feast et al. (1985, 2003), Groenewegen et al. (1998) or by Lloyd
Evans (1997). All of these data are included in the electronic table for
ease of reference.  A small number of measurements were made at SAAO as part
of other programmes by T. Lloyd Evans and/or by S. Bagnulo and I. Short.
These are used in the means quoted and in the diagrams, but the basic data
will be published elsewhere. Individual observations of some objects also
appeared in other papers without dates (e.g. Gaylard \& Whitelock 1988;
Gaylard et al. 1989); these measurements are included in the present
tabulation.

\begin{table}
\begin{center}
\caption{Individual $JHKL$ Observations (Full table available
electronically).}\label{irdata}
\begin{tabular}{rrrrrc}
\hline
\multicolumn{1}{c}{JD}& \multicolumn{1}{c}{$J$}& \multicolumn{1}{c}{$H$}&
\multicolumn{1}{c}{$K$}& \multicolumn{1}{c}{$L$}& Tel.\\
\multicolumn{1}{c}{(day)}& \multicolumn{4}{c}{(mag)}\\
\hline
\multicolumn{6}{l}{\bf R Scl}    \\               
 2443123.5 &  1.75 &  0.60 & --0.03 & --0.62 \\
 2443405.5 &  2.04 &  0.76 &  0.01 & --0.54 \\
 2444187.2 &  1.87 &  0.61 & --0.04 & --0.67 \\
 2446265.8 &  2.17 &  0.74 & --0.05 & --0.79 \\
 2446300.5 &  2.44 &  0.95 &  0.09 & --0.74 \\
 2446303.5 &  2.53 &  1.03 &  0.13 & --0.61 \\
 2446334.5 &  2.55 &  1.07 &  0.15 & --0.64 \\
 2446356.5 &  2.55 &  1.06 &  0.13 & --0.65 \\
 2446373.5 &  2.46 &  0.98 &  0.10 & --0.70 \\
 2446391.2 &  2.43 &  0.96 &  0.10 & --0.69 \\
 2446640.5 &  2.02 &  0.62 & --0.14 & --0.83 \\
 2446655.5 &  2.15 &  0.71 & --0.09 & --0.80 \\
 2446662.5 &  2.18 &  0.75 & --0.07 & --0.84 \\
 2446690.5 &  2.40 &  0.91 &  0.02 & --0.76 \\
 2446695.5 &  2.44 &  0.97 &  0.08 & --0.68 \\
 2446712.5 &  2.53 &  1.02 &  0.11 & --0.67 \\
 2446741.5 &  2.54 &  1.04 &  0.13 & --0.63 \\
 2446749.5 &  2.50 &  1.00 &  0.11 & --0.62 \\
 2446754.2 &  2.45 &  0.98 &  0.09 & --0.67 \\
 2446775.2 &  2.31 &  0.88 &  0.05 & --0.62 \\
 2446782.2 &  2.26 &  0.85 &  0.03 & --0.64 \\
 2446805.2 &  2.14 &  0.77 &  0.02 & --0.56 \\
 2446984.8 &  1.61 &  0.33 & --0.28 & --0.90 \\
 2447014.5 &  1.80 &  0.44 & --0.26 & --0.94 \\
 2447056.5 &  2.15 &  0.69 & --0.13 & --0.88 \\
 2447073.5 &  2.28 &  0.79 & --0.09 & --0.84 \\
 2447113.5 &  2.32 &  0.86 & --0.03 & --0.74 \\
 2447144.2 &  2.19 &  0.75 & --0.07 & --0.79 \\
 2447176.2 &  2.07 &  0.67 & --0.07 & --0.65 \\
 2447191.2 &  2.03 &  0.67 & --0.08 & --0.66 \\
 2447364.8 &  1.62 &  0.32 & --0.35 & --0.98 \\
 2447379.5 &  1.70 &  0.36 & --0.31 & --1.03 \\
 2447394.5 &  1.73 &  0.41 & --0.32 & --1.01 \\
 2447427.5 &  2.02 &  0.61 & --0.19 & --0.94 \\
 2447447.5 &  2.16 &  0.73 & --0.12 & --0.89 \\
 2447497.2 &  2.36 &  0.92 &  0.01 & --0.69 \\
 2447512.2 &  2.30 &  0.88 &  0.01 & --0.73 \\
 2447534.2 &  2.19 &  0.80 & --0.04 & --0.72 \\
 2447732.8 &  1.56 &  0.27 & --0.35 & --0.93 \\
 2447745.5 &  1.55 &  0.25 & --0.38 & --1.03 \\
 2447761.5 &  1.57 &  0.26 & --0.39 & --1.01 \\
 2447779.5 &  1.74 &  0.38 & --0.31 & --1.01 \\
 2447805.5 &  2.03 &  0.61 & --0.18 & --0.88 \\
 2447816.5 &  2.14 &  0.74 & --0.09 & --0.88 \\
 2447821.5 &  2.21 &  0.78 & --0.06 & --0.84 \\
 2447841.5 &  2.42 &  0.96 &  0.05 & --0.77 \\
 2447873.5 &  2.54 &  1.11 &  0.16 & --0.61 \\
 2448073.8 &  1.98 &  0.65 & --0.05 & --0.68 \\
 2448077.8 &  1.96 &  0.64 & --0.05 & --0.73 \\
 2448109.8 &  1.91 &  0.55 & --0.14 & --0.74 \\
 2448141.8 &  2.00 &  0.61 & --0.12 & --0.77 \\
 2448172.5 &  2.13 &  0.70 & --0.08 & --0.78 \\
 2448211.2 &  2.43 &  0.96 &  0.08 & --0.67 \\
 2448224.5 &  2.50 &  1.03 &  0.11 & --0.69 \\
 2448252.2 &  2.53 &  1.06 &  0.14 & --0.61 \\
 2448280.2 &  2.41 &  0.97 &  0.10 & --0.58 \\
 2448492.5 &  1.76 &  0.46 & --0.19 & --0.87 \\
 2448519.5 &  1.92 &  0.55 & --0.16 & --0.85 \\
 2448873.5 &  1.82 &  0.47 & --0.22 & --0.88 \\
 2448900.5 &  2.03 &  0.62 & --0.15 & --0.87 \\
\multicolumn{6}{l}{\it continued in the next column...}\\
\hline
\end{tabular}
\end{center}
\end{table}
\begin{table}
\begin{center}
\setcounter{table}{2}
\caption{continued...}
\begin{tabular}{rrrrrc}
\hline
\multicolumn{1}{c}{JD}& \multicolumn{1}{c}{$J$}& \multicolumn{1}{c}{$H$}&
\multicolumn{1}{c}{$K$}& \multicolumn{1}{c}{$L$}& Tel.\\
\multicolumn{1}{c}{(day)}& \multicolumn{4}{c}{(mag)}\\
\hline
 2448933.5 &  2.29 &  0.81 & --0.04 & --0.82 \\
 2448960.2 &  2.41 &  0.92 &  0.04 & --0.75 \\
 2448990.2 &  2.42 &  0.94 &  0.04 & --0.73 \\
 2449000.2 &  2.41 &  0.94 &  0.04 & --0.70 \\
 2449022.2 &  2.26 &  0.86 &  0.02 & --0.65 \\
 2449146.8 &  1.83 &  0.56 & --0.08 & --0.65 \\
 2449204.8 &  1.61 &  0.35 & --0.27 & --0.84 \\
 2449212.8 &  1.62 &  0.35 & --0.28 & --0.86 \\
 2449271.5 &  1.63 &  0.35 & --0.28 & --0.87 \\
 2449223.5 &  1.65 &  0.35 & --0.28 & --0.88 \\
 2449236.5 &  1.67 &  0.37 & --0.29 & --0.85 \\
 2449263.5 &  1.85 &  0.48 & --0.24 & --0.92 \\
 2449282.5 &  2.01 &  0.60 & --0.18 & --0.88 \\
 2449289.5 &  2.09 &  0.65 & --0.15 & --0.92 \\
 2449296.2 &  2.14 &  0.69 & --0.14 & --0.86 \\
 2449346.2 &  2.37 &  0.90 &  0.01 & --0.76 \\
 2449497.8 &  1.73 &  0.47 & --0.16 & --0.69 \\
 2449518.8 &  1.68 &  0.44 & --0.17 & --0.70 \\
 2449581.5 &  1.46 &  0.24 & --0.34 & --0.90 \\
 2449614.5 &  1.64 &  0.34 & --0.29 & --0.95 \\
 2449637.5 &  1.88 &  0.50 & --0.22 & --0.89 \\
 2449642.5 &  1.93 &  0.56 & --0.19 & --0.96 \\
 2449668.5 &  2.33 &  0.89 &  0.03 & --0.78 \\
 2449672.5 &  2.37 &  0.92 &  0.04 & --0.72 \\
 2449709.2 &  2.57 &  1.11 &  0.16 & --0.62 \\
 2449728.2 &  2.53 &  1.09 &  0.16 & --0.64 \\
 2449772.2 &  2.25 &  0.85 &  0.03 & --0.65 \\
 2449941.8 &  1.71 &  0.43 & --0.16 & --0.70 \\
 2449975.5 &  1.84 &  0.51 & --0.18 & --0.83 \\ 
 2449986.5 &  1.95 &  0.58 & --0.12 & --0.73 \\
 2450019.5 &  2.25 &  0.83 &  0.00 & --0.72 \\
 2450029.2 &  2.34 &  0.90 &  0.04 & --0.66 \\
 2450052.5 &  2.52 &  1.04 &  0.13 & --0.61 \\
 2450057.2 &  2.54 &  1.08 &  0.15 & --0.64 \\ 
 2450062.2 &  2.55 &  1.09 &  0.15 & --0.65 \\ 
 2450082.5 &  2.59 &  1.13 &  0.18 & --0.62 \\
 2450109.2 &  2.50 &  1.03 &  0.14 & --0.59 \\
 2450144.2 &  2.24 &  0.84 &  0.04 & --0.55 \\
 2450260.8 &  1.81 &  0.55 & --0.07 & --0.60 \\ 
 2450274.5 &  1.69 &  0.45 & --0.14 & --0.66 \\ 
 2450299.5 &  1.48 &  0.28 & --0.27 & --0.82 \\
 2450317.5 &  1.43 &  0.24 & --0.30 & --0.84 \\
 2450362.5 &  1.73 &  0.42 & --0.25 & --0.88 \\
 2450398.2 &  2.19 &  0.76 & --0.07 & --0.80 \\
 2450414.5 &  2.37 &  0.91 &  0.03 & --0.71 \\
 2450420.5 &  2.43 &  0.96 &  0.06 & --0.69 \\
 2450437.2 &  2.55 &  1.05 &  0.14 & --0.70 \\
 2450469.2 &  2.52 &  1.04 &  0.10 & --0.66 \\
 2450471.2 &  2.50 &  1.01 &  0.08 & --0.64 \\
 2450503.2 &  2.25 &  0.83 & --0.02 & --0.69 \\
 2450681.5 &  1.61 &  0.35 & --0.26 & --0.77 \\
 2450712.5 &  1.67 &  0.38 & --0.27 & --0.83 \\
 2450721.5 &  1.73 &  0.42 & --0.26 & --0.82 \\
 2450753.5 &  2.03 &  0.63 & --0.13 & --0.80 \\
 2450792.2 &  2.33 &  0.87 &  0.00 & --0.76 \\
 2450830.2 &  2.40 &  0.92 &  0.02 & --0.71 \\
 2451025.2 &  1.64 &  0.40 & --0.20 & --0.76 \\
 2451052.2 &  1.56 &  0.34 & --0.26 & --0.79 \\
 2451115.5 &  1.87 &  0.50 & --0.20 & --0.86 \\
 2451154.2 &  2.26 &  0.82 & --0.01 & --0.82 \\
 2451186.2 &  2.49 &  1.01 &  0.10 & --0.69 \\
 2451417.5 &  1.61 &  0.38 & --0.22 & --0.75 \\
 2451481.5 &  1.64 &  0.31 & --0.33 & --0.95 \\
\multicolumn{6}{l}{\it continued in the next column...}\\
\hline
\end{tabular}
\end{center}
\end{table}
\begin{table}
\begin{center}
\setcounter{table}{2}
\caption{continued...}
\begin{tabular}{rrrrrc}
\hline
\multicolumn{1}{c}{JD}& \multicolumn{1}{c}{$J$}& \multicolumn{1}{c}{$H$}&
\multicolumn{1}{c}{$K$}& \multicolumn{1}{c}{$L$}& Tel.\\
\multicolumn{1}{c}{(day)}& \multicolumn{4}{c}{(mag)}\\
\hline
 2451502.5 &  1.75 &  0.38 & --0.31 & --0.94 \\
 2451575.2 &  2.33 &  0.86 & --0.04 & --0.78 \\
 2451600.2 &  2.39 &  0.91 &  0.01 & --0.71 \\
 2451738.8 &  1.80 &  0.54 & --0.09 & --0.70 \\
 2451743.8 &  1.76 &  0.51 & --0.11 & --0.68 \\
 2451782.5 &  1.55 &  0.33 & --0.24 & --0.83 \\
 2451785.5 &  1.58 &  0.33 & --0.25 & --0.87 \\
 2451859.5 &  1.75 &  0.41 & --0.27 & --0.98 \\
 2451869.5 &  1.81 &  0.46 & --0.25 & --0.93 \\
 2451881.2 &  1.89 &  0.52 & --0.20 & --0.94 \\
 2451809.5 &  1.57 &  0.29 & --0.31 & --0.90 \\
 2451831.5 &  1.62 &  0.32 & --0.31 & --0.91 \\
 2451929.2 &  2.26 &  0.81 & --0.04 & --0.79 \\
 2452182.5 &  1.57 &  0.30 & --0.28 & --0.87 \\
 2452208.5 &  1.55 &  0.26 & --0.35 & --1.00 \\
 2452226.2 &  1.63 &  0.32 & --0.34 & --0.98 \\
 2452241.2 &  1.72 &  0.38 & --0.30 & --0.95 \\
 2452257.2 &  1.88 &  0.50 & --0.25 & --0.94 \\
 2452285.2 &  2.14 &  0.71 & --0.09 & --0.86 \\
 2452321.2 &  2.45 &  0.98 &  0.06 & --0.68 \\
 2452529.5 &  1.76 &  0.49 & --0.15 & --0.78 \\
 2452572.5 &  1.69 &  0.38 & --0.27 & --0.88 \\
 2452602.2 &  1.75 &  0.43 & --0.26 & --0.89 \\
 2452691.2 &  2.32 &  0.87 & --0.01 & --0.75 \\
 2452888.5 &  1.64 &  0.39 & --0.23 & --0.81 \\
 2452932.5 &  1.61 &  0.37 & --0.28 & --0.89 \\
 2452960.5 &  1.69 &  0.38 & --0.28 & --0.96 \\
\hline     
\end{tabular}
\end{center}
Some of these observation of R Scl were published by Lloyd Evans (1997) and
by Whitelock et al. (1995, 1997).
\end{table}

\subsection{IRAS and MSX data}
Because energy distributions of C stars typically peak between 3 and 10
$\mu$m, a measure of the energy output beyond the $L$ band is important for
estimating their bolometric flux. Following earlier work (e.g. Whitelock et
al. 2000, 2003) we use the IRAS 12 and 25 fluxes and, where possible,
supplement these with data from the MSX survey (Egan et al. 2003) using the
 $A$- ($8.28\mu$m), $C$- ($12.13\mu$m) and $D$- ($14.65\mu$m) bands.

IRAS photometry was taken preferentially from the IRAS Faint Source
Catalogue (FSC Moshir et al. 1989) or from the PSC (IRAS Science Team 1989).
The IRAS fluxes for CW~Leo were taken from the PSC rather than the FSC as
these were more consistent with comparable values from the literature
(Gezari, Pitts \& Schmitz 1997). The IRAS photometry was colour corrected
using the prescription from the IRAS explanatory supplement for the purpose
of calculating bolometric magnitudes only. For the discussion of colours
etc. the raw magnitudes were used.


There are 18 stars which have no IRAS fluxes, but 13 of these have been
measured in the MSX $A$-band. The remaining 5 sources all have $K-L<1.0$ and
the long wavelength fluxes will not make a significant contribution to their
bolometric magnitude (in fact one of them, V354 Cen, is probably not a C star
and only two, V874 Aql and [ABC]Car93, are classed as C Miras).

The MSX data were extracted from the complete MSX6C catalogue in the
Galactic Plane ($|b|\leq 6^{\rm o}$) and the high latitude, $|b|> 6^{\rm o}$,
subsections only. A few of our sources have detections in the low
reliability sources ([ABC89]Car5) and singleton source (extracted from a
single scan but with good fluxes: Y~Tau, CL~Mon, 07080--0106, 07220--2324,
FF~Pup, 08340--3357, MU~Vel, 09533--6021, 10145--6046, TV~Vel, 11145--6534)
sections of the catalogue, but a close examination suggested these were
unreliable, e.g. some singleton sources showed unphysical colours.

\begin{figure}
\includegraphics[width=8.5cm]{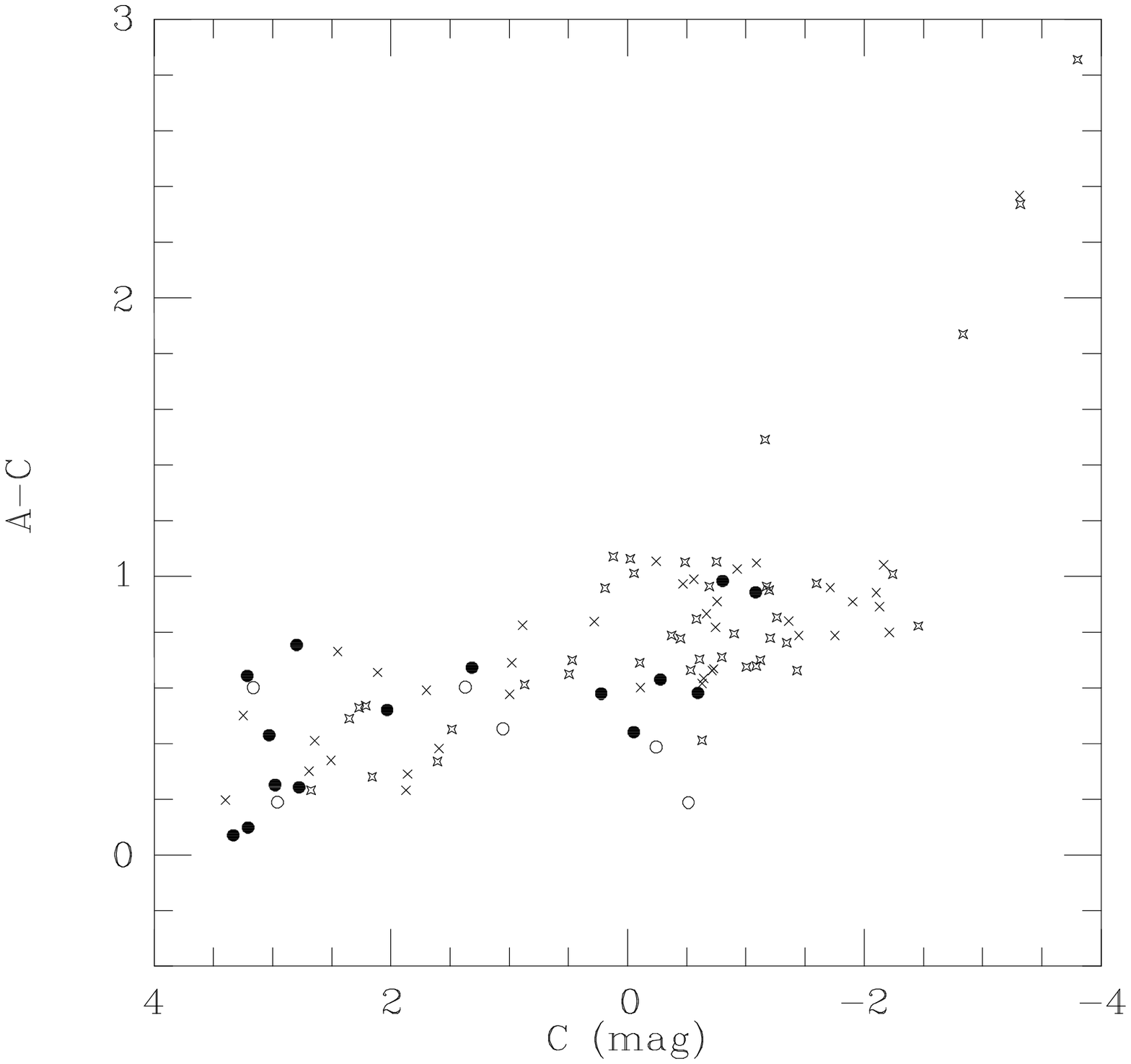}
 \caption{ The MSX $A-C$ colour as a function of the MSX-$C$
magnitude. Symbols: crosses: well observed Miras without obvious
peculiarities; open crosses: other Miras; open circles: well observed small
amplitude variables without obvious peculiarities; close circles: other
non-Miras. The four sources with large $A-C$ colours are saturated in the
$A$ band. Note the difference between the typical $A-C$ colour for the stars
brighter and fainter than $C\sim 1$, particularly the Miras.}
\label{fig_msxcac}
\end{figure}

The very bright sources appear to be saturated in the MSX-$A$ 
band, which is more sensitive than any of the others, as can be seen in the
plot of $A-C$ against $C$ (Fig.~\ref{fig_msxcac}). The four very bright
stars are II Lup, V1417 Aql, V1418 Aql and V1965 Cyg. Their MSX-$A$
magnitudes are inconsistent with other MSX and IRAS magnitudes, although
this is not clear from the flags provided with the catalogue. It appears
that among this type of object anything brighter than 270\,Jy at
$C$ will be saturated at $A$.

Note also from Fig.~\ref{fig_msxcac}, the difference in the $A-C$ colours of
stars, particularly Miras, that are brighter and fainter than $C\sim 1$
mag. This occurs because the bright Miras are intrinsically different from
the faint ones - the stars with large colour indexes, i.e. the ``red''
stars, are all relatively bright. There are various selection effects
contributing to this, but critically we would have been unable to perform
$JHK$ photometry of faint red stars. A similar dichotomy is seen in the IRAS
data, in that most of the Miras with $[12]>0$ have $[12-25]<0.5$ and those
with $[12]<0$ have $[12-25]>0.5$. There are, however, some notable faint
non-Miras with $[12-25]>1$: EV Eri (see Section~\ref{trends}) , 10151--6008,
[ABC89]Cru17 and [ABC89]Cir1.

\begin{figure}
\includegraphics[width=8.4cm]{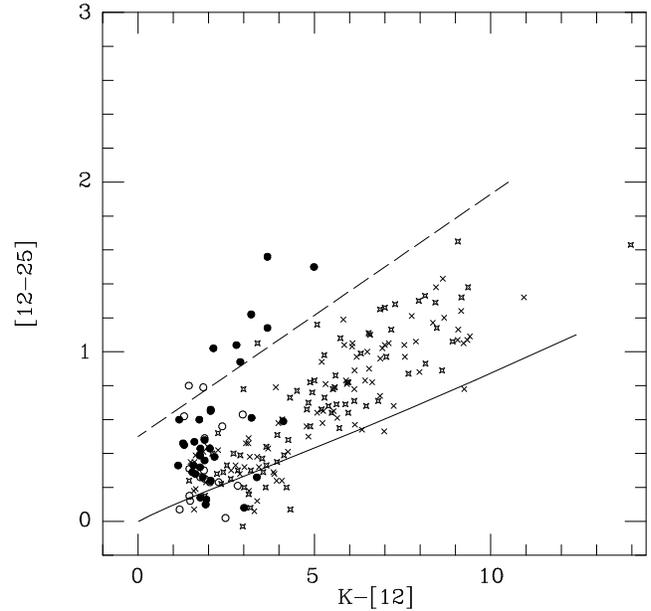}
 \caption{ Combined IRAS near-infrared two-colour diagram;
symbols as in Fig.~\ref{fig_msxcac}. The solid line is the blackbody locus
while the dashed line roughly separates C- from O-rich stars. Note that
sources selected according to the IRAS selection criterion described in
Section \ref{sources} will only find sources with $[12-25]>0.565$. Symbols
are the same as in Fig.~\ref{fig_msxcac}.}
\label{fig_k1225}
\end{figure}

\begin{figure}
\includegraphics[width=8.4cm]{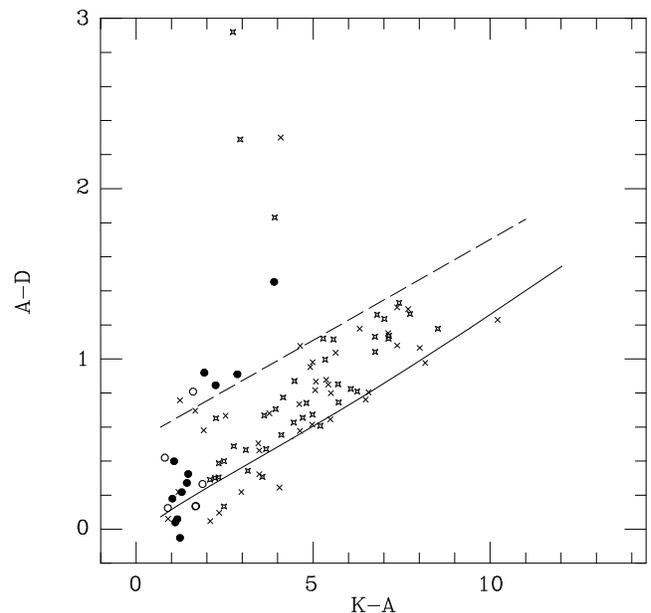}
 \caption{ Combined MSX near-infrared two-colour diagram;
symbols and lines as in Fig.~\ref{fig_k1225}.}
\label{fig_msx1}
\end{figure}

Figs.~\ref{fig_k1225} \& \ref{fig_msx1} illustrate the types of two-colour
diagram which are frequently used to distinguish between O- and C-rich stars
(e.g. Ortiz et al. 2005) using IRAS and MSX data respectively.  The solid
line on both figures is the blackbody locus, while the dashed line provides
a rough division between C- and O-rich stars, which in the case of the MSX
figure has been copied directly from Ortiz et al. The division is far from
being a precise one and stars close to the line must be regarded as having
uncertain chemical type unless spectroscopic information is available.
Nevertheless, most of the programme stars fall in the region you would have
expected given their C-rich nature. A similar separation can be achieved
using slightly different combinations of colours, as was discussed, e.g., by
Guglielmo et al. (1993).

In the MSX diagram the four peculiar non-Miras lying above the upper line
are: 09164--5349 (see Section~\ref{trends}), [ABC89]Cir1, V2548 Sgr and
[ABC89]Cru17, the latter being the only point well away from the line (NB
the differences between Miras and non-Miras and between `peculiar' and
`normal' relate to the variability characteristics and are described in
Section~\ref{var_class}).  The normal non-Mira above the line is
[ABC89]Pup3. The stars in a comparable position on the IRAS diagram are
these same peculiar non-Miras plus EV~Eri and 10151--6008, while the normal
non-Miras are R Scl (which has a detached dust shell,  Bujarrabal \&
Cernicharo 1994) and T Ara (which is super lithium rich, Feast 1954).


 \section{Pulsation Characteristics}\label{pulsation}
\subsection{Pulsation Periods}
 Periods (P$_K$) were determined from a Fourier transform of the $K$ light
curve for all of the stars with SAAO near-infrared observations on 8 or more
dates; these are listed in Table~\ref{jhkl}.  Where a pulsation period has
been published elsewhere it is listed in column 11 (P$_{\rm lit}$) of
Table~\ref{jhkl}. These are taken from the following sources, in order of
preference: GCVS, Le Bertre (1992), Jones et al. (1990), Hipparcos (ESA
1997) or Pojma\'nski (2002). Figure~\ref{fig_pp} shows our period plotted
against the value from the literature. The agreement is generally better
than 5 percent and a brief discussion of the exceptional cases follows:
\begin{figure}
\includegraphics[width=8.4cm]{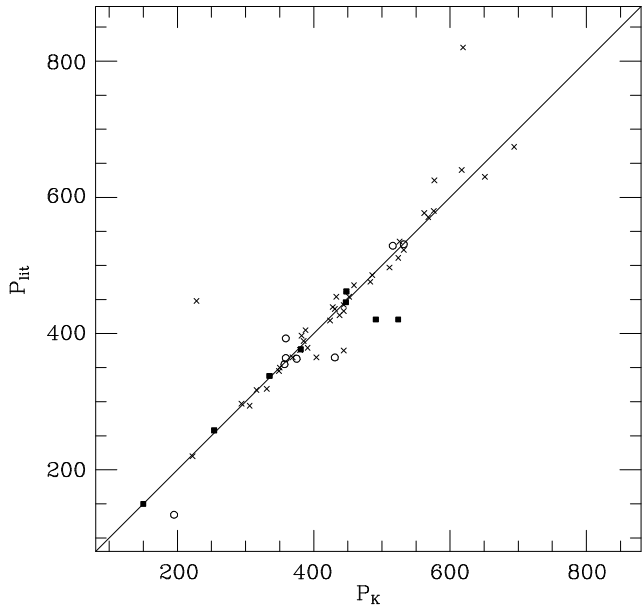} 
 \caption{ Periods, P$_K$, derived from data discussed here plotted
against those from the literature, P$_{\rm lit}$; Symbols: crosses are Mira
C-stars (outstanding points represent V477 Mon and V518 Pup), open circles
are non-Mira C stars; solid squares are various peculiar sources including
non-C stars, SC stars and C-stars with silicate shells (the two outstanding
points represent BH Cru, which has a variable period, at different times).}
\label{fig_pp}
\end{figure}

{\bf V477 Mon} The original period is from Maffei (1966) who indicated it as
``M? P=820:: days''. The near-infrared observations are not consistent with
this value and we use the newly determined 619 days which provides a very
good fit to these data.

{\bf V1965 Cyg} The original period is from Jones et al. (1990) (who
incorrectly associates AFGL 2417 with V1129 Cyg) and there is evidence for
rather erratic behavior in both their and our light curves. Analyzing the
two data sets together (using only the 6 observations actually listed in
their paper) suggests a period of 617 days. 

{\bf BH Cru} This is known to have a lengthening period (Bateson et al.
1988; Walker et al. 1995; Zijlstra et al. 2004) and our data, which were
obtained in two batches with a gap in the middle, suggests a change
from 491 to 524 days, between 1984/9 and 1997/2004.

{\bf V617 Mon} The GCVS period is 375: days with which the near-infrared
observations are not consistent. We use the newly determined 444 days which
provides a very good fit to our data.

{\bf FU Car} The GCVS variability type and period of M: and P=365: days
respectively, come from Luyten (1927) who records FU~Car's variability along
with that of another star and notes ``The available plates are insufficient
for a determination of the light curves, but the possibility is indicated
that both variables have a period of nearly one year, and a range of at
least two magnitudes''. This is consistent with our determination of 431
days and our assertion that it is not a Mira.

{\bf TV Vel} GCVS records this as variability type M with P=365 days and no
indication of any uncertainty. There is no clear source for the period. We
also classify TV~Vel as a Mira, but it is certainly a border-line case. Our
data are inconsistent with the 365 day period.

{\bf W Sex} The difference between the GCVS period of 134 days and the
Hipparcos period of 200 days was noted by Whitelock et al. (2000). Our newly
determined period of 195 days is consistent with the Hipparcos value. It is 
not a Mira.

{\bf RT Cap} This is not a Mira and the GCVS period of 393 days is
inconsistent with the 359 days derived here.

{\bf V518 Pup} The 448 day period comes from the ASAS database of
Pojma\'nski (2002) who also classifies it as a Mira. The 228 day period
derived from the IR data is inconsistent with the ASAS data. While 228 days
is the best fit to the infrared photometry, 448 days is also a possible
solution. In this case 448 days would be preferred and is in fact the value
we adopt, but the Mira classification must be regarded as uncertain.

Figs. \ref{fig_lcmira} and \ref{fig_lcsr}, in the appendix, illustrate for
Miras and other variables, respectively, the $K$ light curves plotted as a
function of phase for the stars with sufficient data to determine a period.

From this it can be seen that the accuracy of the derived periods and
amplitudes varies considerably from one star to the next, depending on the 
number of observations, their distribution in time and the stability of the
light curve over the sampling interval. Most of the illustrations show the
$K$ magnitude phased at the period determined from the $K$ data. Where
satisfactory periods could not be determined from our data they are shown
phased at the GCVS period. For 10220--5858 a period of 585: days is given,
but 290: is also possible. Although we use the 585 day period in the
following discussion the blue colours and low amplitude would fit better 
with the shorter period.

\subsection{Variability Class}\label{var_class}
 In view of the fact that we wish to use the Mira Period-Luminosity (PL)
relation to estimate the distance to the C-stars it is vital that we decide
if particular stars can be classed as Miras or not. According to the
classical GCVS definition Miras have characteristic late-type emission
spectra (Ce in the cases we are discussing) and $V$ light amplitudes greater
than 2.5 mag. Their periodicity is well pronounced, and periods lie in the
range from 80 to 1000 days. For many of the stars of interest very little is
known about the $V$ magnitude and we must make the best estimate we can from
other sources of information. Unfortunately the distinction between Miras
and other types of long period variable is not as clearcut for C-rich stars
as it is for O-rich ones, where the $JHKL$ colours of Miras and non-Miras
are distinctly different (e.g. Whitelock et al. 1995).

 Where there are sufficient observations we determine if a star is a Mira or
not on the basis of the SAAO infrared photometry; if it is clearly periodic
and has a peak-to-peak $K$ amplitude over 0.4 mag we call it a Mira and
assign it to class 1n. If the variations are not periodic or are less than
0.4 mag at $K$ we assign it to class 2n, and call it a non-Mira. If there
are insufficient data to do this we use the GCVS or ASAS classification if
there is one. Otherwise, if our $K$ magnitudes differ from published values
(Aaronson, 2MASS ...) by 0.4 mag or more we assign it to class 1n
else to class 2n. Stars classified in this last way are labeled with a colon
after the classification in Table~\ref{jhkl} as they are clearly less
certain than the others. There are 74 class 2n and 165 class 1n in the
table.

\begin{table}
\begin{center}
\caption{Second parameter, n, of the variability type 1n or 2n.}\label{vclass}
\begin{tabular}{cl}
\hline
n & description of light curve\\
\hline
0 & sinusoidal and reasonably repeatable \\
1 & evidence of obscuration events or a long term trend\\
2 & for pronounced second peak in light curve (Miras only) \\
3 & for erratic behavior, includes large amplitude non-Miras\\
4 & inconsistent with other published photometry\\
5 & star with peculiarities described in text\\
\hline
\end{tabular}
\end{center}
\end{table}

The assignment of the second digit, or sub-class, of the variability
classification is described in Table~\ref{vclass}. The sub-class to which a
variable is assigned obviously depends strongly on how many observations we
have, e.g. many objects of type 13 would probably be classed as type 11
given more data. There is actually only one star, V1420 Aql, catalogued as
type 12, which is illustrated in the last panel of Fig.~\ref{fig_lcmira},
although there are several marginal cases, e.g. [ABC89]Cir27, as can be seen
from the illustrated light curves. Even with V1420 Aql there is some
uncertainty in its period and if it is plotted at the other period given in
Table~\ref{jhkl} (676 days from Le Bertre 1992) then the second peak is less
pronounced.

Our classification agrees reasonably well with the GCVS for most of the 102
stars in common and we briefly discuss the 10 differences here. We classify
the following stars as Miras whereas the GCVS provided the classification
given in parenthesis after the name: V471 Pup (SR:), V518 Pup (SR: also Mira
in ASAS), UW Pyx (Lb: also Mira in ASAS), RW LMi (SRa), CF Cru (Lb), FX Ser
(Lb:). The ASAS classification, based on optical data, is relevant as they
have many more observations than we do and can do a better job with
classifying visually bright variables. However, we agree with the GCVS
classification of V374 Aql (SR) rather than the M: assigned by ASAS. The
following were classed as Miras in the GCVS: FU Car, V354~Cen, V348~Sco and
V503~Mon, but are classed as non-Miras here, because their $K$ amplitudes
are much less than expected from Miras.

The Mira nature of 10220--5858 is uncertain, it has a low amplitude and small
$K-[12]$ for its period and limited data suggest an erratic light curve.

A number of authors have noted that the IRAS colours of [ABC89]Cir1 are
those expected for a silicate shell (e.g. Chen et al. 1993 and references
therein) and it has been investigated on this basis. We note that it has a
slightly larger $K-L$ than normal non-Miras.

Of the 57 C stars in the Aaronson sample, 26 are Miras, and periods in the
range 265 to 652 days have been determined for 24 of them, while the other 31
are erratic or low amplitude variables.

Of the 93 C-stars in the IRAS sample, 86 are Miras and periods ranging from
431 to 948 days were determined for 69 of them, while the other 7 are low
amplitude or erratic variables. Of the 26 C-stars in the faint IRAS sample
12 are Miras, and periods in the range 495 to 714 days were determined for
10 of them, while the other 14 are erratic or low amplitude variables.

\subsection{Mean Magnitudes and Amplitudes}\label{amps}
 Table \ref{jhkl} gives the Fourier-mean magnitudes and amplitudes for all
of the stars discussed here. These were derived by fitting first order sine
curves to the light curves, and the peak-to-peak amplitudes of those curves
are also tabulated. The number of points used in the fit is listed in the
last column.  This may be less than the full number of observations
available, as it is for stars showing obscuration events where we have excluded
cycles showing heavy obscuration (see section \ref{trends}). This is
somewhat subjective and the results dependent on how well any particular
light curve is covered.

\begin{figure}
\includegraphics[width=8.5cm]{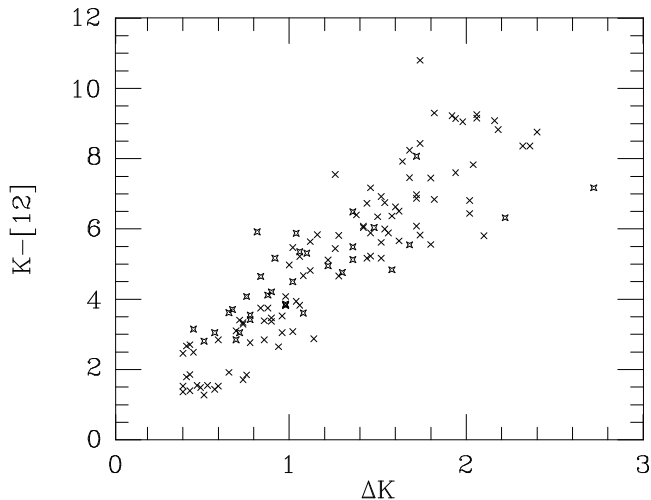}
 \caption{For the Miras, $K-[12]$ colour as a function of the pulsation
amplitude at $K$, $\Delta K$. Symbols as in Fig.~\ref{fig_msxcac}.}
\label{delkk12}
\end{figure}

The pulsation amplitude generally decreases with increasing wavelength,
although there are a few examples of $\Delta J < \Delta H$. Where the source
is faint at $J$ a low amplitude may indicate contamination of the $J$ flux
by another source in the aperture; there are, however, examples, e.g.
CW~Leo, where the measured amplitudes are definitely a true reflection of
the C star variations. Fig.~\ref{delkk12} shows the dependence of pulsation
amplitude on colour. Although there is a great deal of scatter, the
amplitude and colour are clearly correlated. While there will be several
effects contributing to this dependence, the primary one will be the
amplitude's direct or indirect dependence on temperature fluctuations of the
star. We also anticipate a correlation between the stellar pulsation
amplitude, as measured by $\Delta K$, and the mass-loss rate, as measured by
$K-[12]$ which is proportional to the optical depth of the shell (see
section 6 and Whitelock, Pottasch \& Feast (1987). It is not possible to
separate the effects of temperature fluctuations and pulsation amplitude
changes with the available information.

The peaks of the energy distributions for these stars is at wavelengths over
2.5$\mu$m (the combined effect of low temperature and thick
circumstellar dust shells), therefore at $J$ and $H$ and usually also at
$K$ and $L$, we are sampling the Wien part of the energy distribution which
is extremely sensitive to temperature fluctuations. The cooler the star the
larger the flux changes at $JHKL$ that will be caused by small changes of the
stellar temperature. Molecular opacity fluctuations in the $JHKL$ bands,
also in response to temperature changes of the star, will serve to magnify
the amplitude dependence on stellar temperature. Thus to a first
approximation we might expect the $JHKL$ amplitudes to tell us more about
the temperature of the C star, and changes in its temperature around the
pulsation cycle, than about the bolometric amplitude of the pulsations
themselves. The colours that we discuss here, including $K-[12]$, are much
more strongly influenced by reddening of the circumstellar shell than they
are by the temperature of the underlying star (see Section \ref{ir_cols}).
Nevertheless, the cooler stars will tend to have the thicker dust shells, so
we still understand the correlation in Fig.~\ref{delkk12} to be very
considerably a consequence of fluctuations in the stellar temperature, but
with a good deal of scatter as the thickness of the shell is not a simple
function of the temperature of the star.

\section{Mira Period-Luminosity Relation}\label{pl}
 The existence of a PL relationship for Mira variables was discussed in
detail by Feast et al. (1989) for O- and C-rich Miras in the LMC.
Subsequently Whitelock et al. (2003) discussed the PL relation for
longer-period, thick-shelled Miras also in the LMC, including photometry
from IRAS and ISO. The studies by Feast et al. and by Whitelock et al.
encompassed stars with multiple near-IR observations and therefore well
defined mean magnitudes. Groenewegen \& Whitelock (1996) used data for
spectroscopically confirmed C stars only, but included those with single
observations to provide a larger sample of LMC stars. All of these papers
found rather similar bolometric PL relations for the C stars.

\begin{figure}
\includegraphics[width=8.5cm]{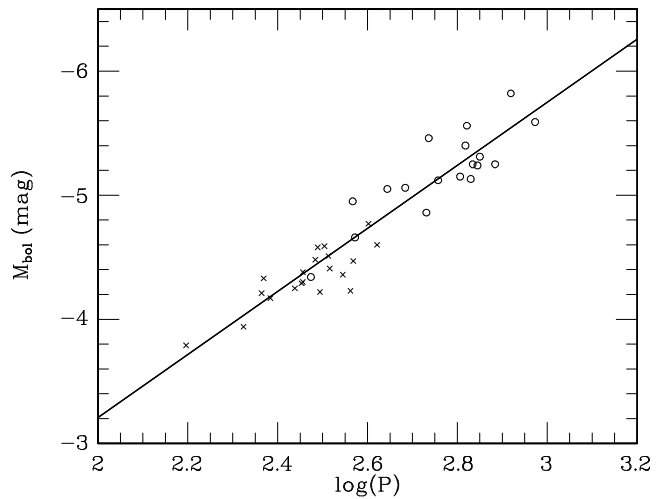}
 \caption{The PL relation for C-rich Miras in the LMC. The
crosses and circles represent stars from Feast et al. (1989) and from
Whitelock et al. (2003), respectively. The straight line is the locus given by
equation \ref{lmc_pl}.}
\label{fig_pl} 
\end{figure}

 Here we combine the data from Feast et al. (1989) and Whitelock et al.
(2003) to derive a PL relation for the C stars that covers the period range
of interest for the Galactic C stars under discussion.  We omit four stars
classed by Whitelock et al. as C-rich: WBP14 for which the data were
uncertain; 04496--6958 and SHV\,05210--6904, which lie above the PL relation
(Whitelock et al. suggested that this may be the result of extra energy
from hot bottom burning, although that conclusion is controversial for a
carbon star and more detailed studies are required to investigate these
luminous stars); 05128--6455 which Matsuura et al. (2005) have shown to be
O-rich. Thus we have 38 C-rich Miras which are illustrated in a PL diagram
(Fig.~\ref{fig_pl}), for which a least squares fit gives:
\begin{equation} M_{bol}= -2.54 \log P +1.87, \ \ (\sigma=0.17) \label{lmc_pl}
\end{equation} assuming that the distance modulus of the LMC is 18.50 mag.
This is close to the relationships given in the various references cited
above and is what we use in Section \ref{is_ext} to derive distances. Some
of the 0.17 mag dispersion will be introduced by the limited temporal
coverage of IRAS (the satellite did not observe long enough to provide mean
magnitudes for these long period stars).

This PL is distinctly different from that derived for O-rich Miras (Feast et
al. 1989; Whitelock et al. 2003). Due to the differences in slope the
relations, which are close at short periods, diverge at long period. At a
period of 500 days the C-Miras are 27 percent fainter than their O-rich
counterparts at the same period. Part of this difference may be due to the
different energy distributions of the O- and C-rich stars which can lead to
different systematic errors affecting the estimates of total luminosity
(e.g. the strong water features which are present in O-rich, but not C-rich
Miras). However, the large differences at long periods suggest that there
may be real luminosity differences between the two types of Mira at a given
period.

The PL relationship is revisited in Paper III of this series, where the
kinematics of the Galactic C Miras are used to derive a zero-point.

\section{Infrared Colours}\label{ir_cols}
 In the following analysis we compare various data on Galactic C-rich Miras
with comparable measurements of LMC objects.  The LMC samples are taken from
Feast et al. (1989) (with updated periods from Glass \& Lloyd Evans (2003))
and Whitelock et al. (2003). Note that the Feast et al. stars, which were
optically selected, do not have $L$ or IRAS observations; many of the
Whitelock et al. sample, which were selected from IRAS sources, do not have
$J$ measurements; thus not all the stars appear in all the diagrams.
Two bright LMC C  stars with distinctly blue colours for their period are
always distinguished in the illustrations. These stars, which also lie above
the bolometric period-luminosity relation, are thought to be undergoing hot
bottom burning (Whitelock et al. 2003). 

\begin{figure}
\includegraphics[width=8.4cm]{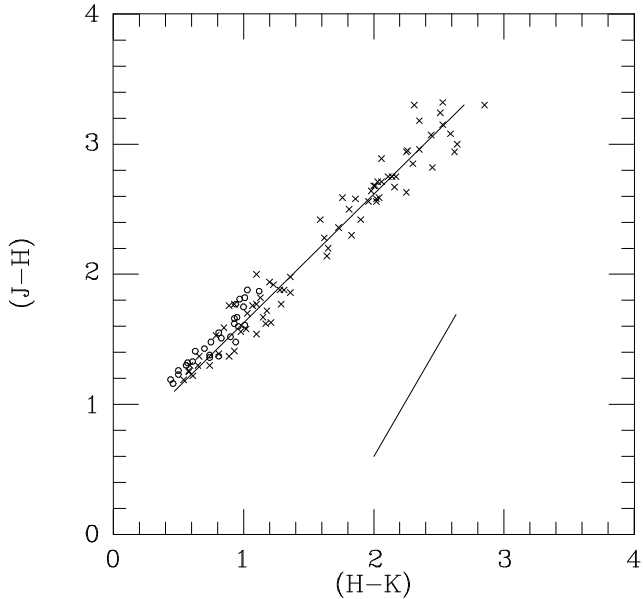}
 \caption{A two-colour diagram comparing Miras (crosses)
having known periods, with non-Miras (open circles); stars from both groups
are well observed (at least 9 measurements) and lack obvious
peculiarities. The locus for the Miras, given in equation \ref{jhhk}, is
illustrated as is a reddening vector for $A_V=10$ mag.}
\label{fig_jhhku}
\end{figure}

 Figs.~\ref{fig_jhhku} and \ref{fig_hkklu} illustrate the colours prior to
correction for reddening. The stars illustrated here are those with 10 and
20 classifications in Table~\ref{jhkl} and with at least 9 observations
contributing to the mean. Thus they represent well-characterized {\it
normal} Miras and non-Miras as far as it is possible to define them. 

 It is clear that the Miras spread to much redder colours than the non-Miras
as one might expect given their higher mass-loss rates and resultant
circumstellar shells. At the blue extreme the non-Miras and Miras follow
slightly different loci in the two-colour diagrams, but the differences 
are subtle and it is not possible to distinguish between individuals in the 
two groups simply on the basis of their colours. This is in marked contrast
to the situation for O-rich stars (e.g. Whitelock et al. 1995).

 The lines illustrated in these two figures represent a maximum likelihood
fit to the Mira data and are given in equations \ref{jhhk} and \ref{hkkl},
which are discussed later.  

\begin{figure}
\includegraphics[width=8.4cm]{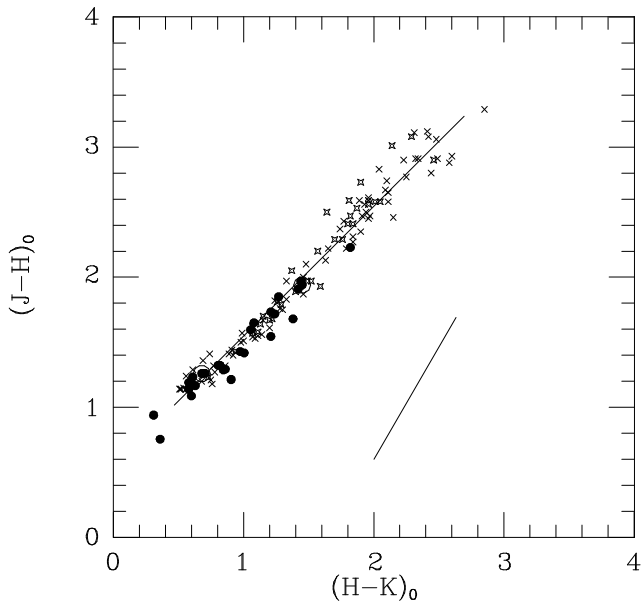}
 \caption{As Fig.~\ref{fig_jhhku}, but after correcting 
for interstellar extinction as described in section \ref{is_ext} and showing
only the Miras, but including those with peculiarities (open crosses). The
filled circles are the LMC Miras, with the two large open circles
representing LMC stars suspected of undergoing hot bottom burning.}
\label{fig_jhhk} 
\end{figure}

\begin{figure}
\includegraphics[width=8.4cm]{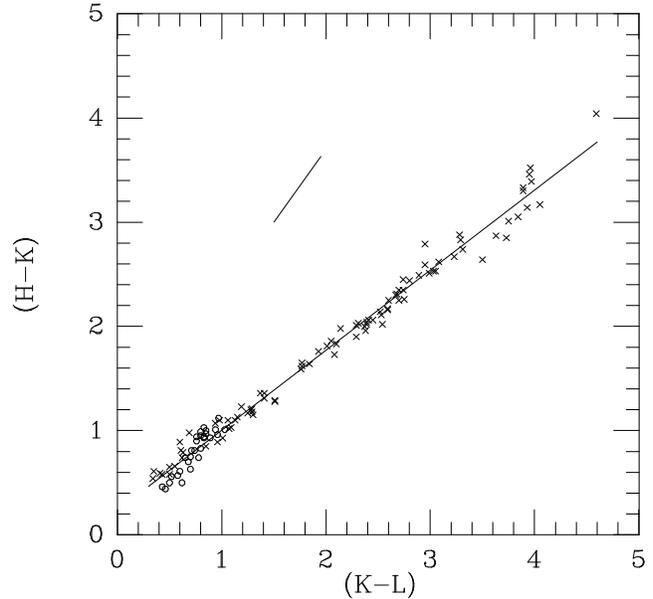}
 \caption{As Fig.~\ref{fig_jhhku}, but for alternative
colours. The locus described by equation \ref{hkkl} is shown.}
\label{fig_hkklu}
\end{figure}
\begin{figure}
\includegraphics[width=8.4cm]{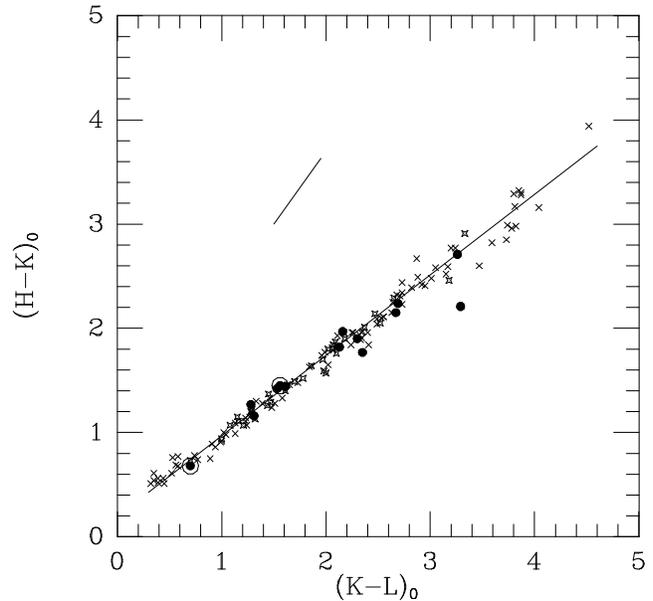}
\caption{ As Fig.~\ref{fig_hkklu}, but after correcting for interstellar
extinction as described in section \ref{is_ext} and showing only the Miras;
the symbols are as in Fig.~\ref{fig_jhhk}.} 
\label{fig_hkkl}
\end{figure}

\begin{figure}
\includegraphics[width=8.4cm]{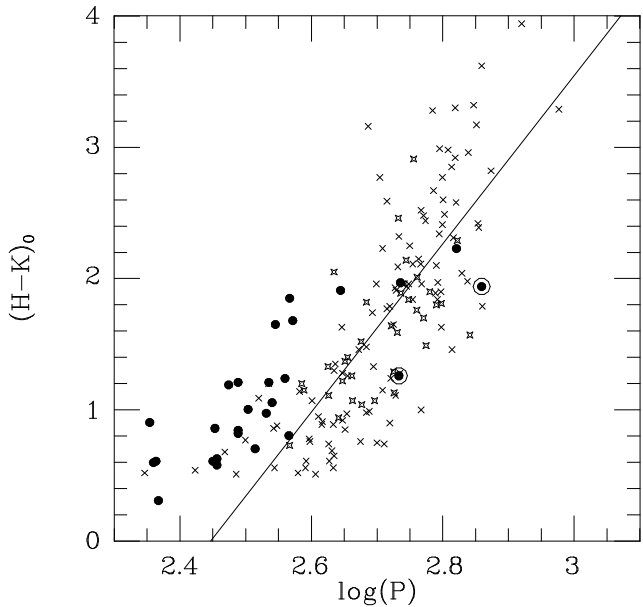}
 \caption{ Period $(H-K)$ colour relationship for Galactic
(crosses) and LMC (closed circles) C stars. The open crosses represent
Galactic Miras with 1n (n$>$1) classifications.  The two LMC points with
circles around them are luminous Miras thought to be undergoing hot bottom
burning. The locus described by equation 4 is shown.}
\label{fig_hkp}
\end{figure}

\begin{figure}
\includegraphics[width=8.4cm]{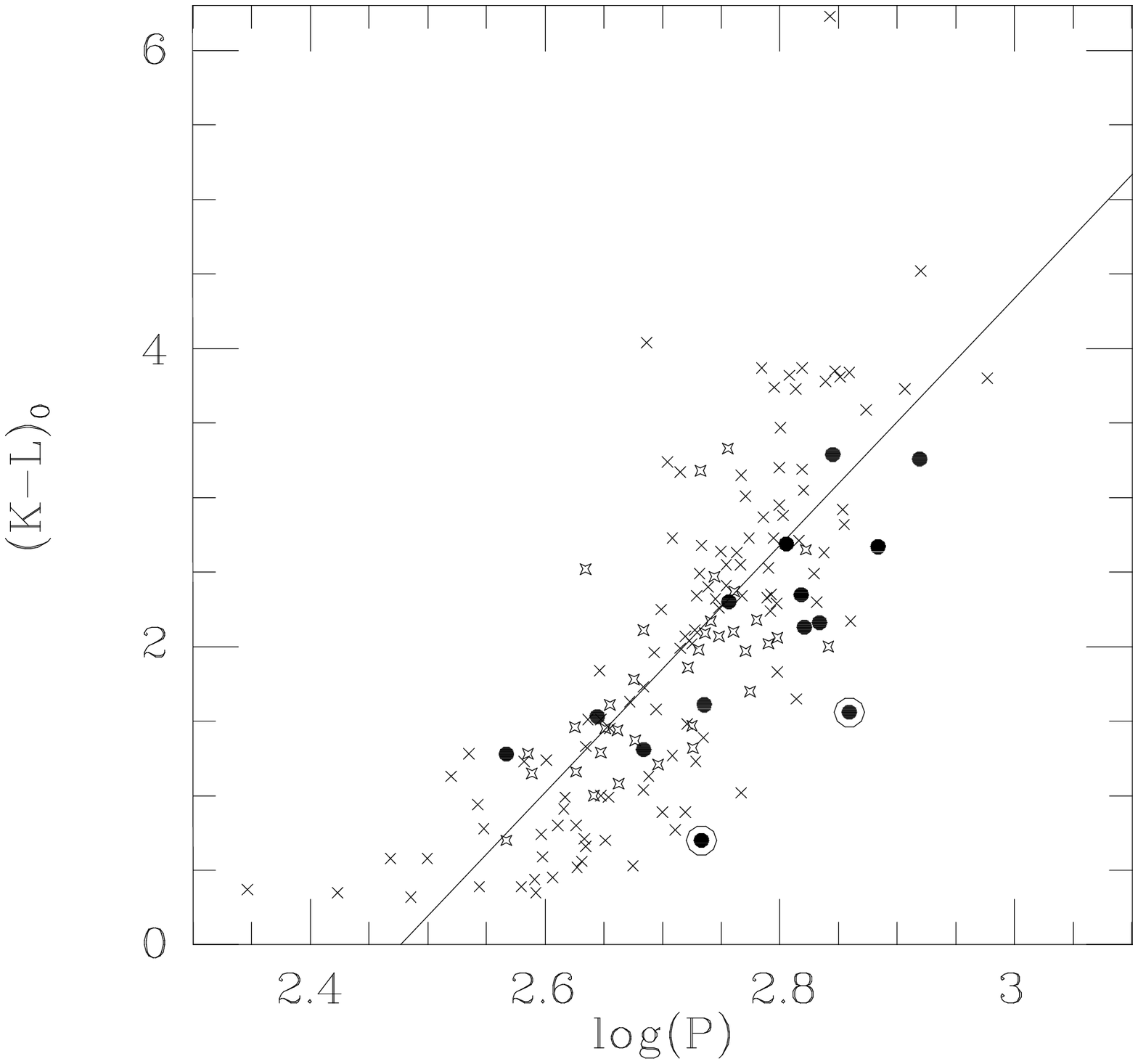}
 \caption{ Period $(K-L)$ colour relationship with symbols
 as in Fig~\ref{fig_hkp}. The locus described by equation 5 is shown.} 
\label{fig_klp}
\end{figure}

The reddening-corrected colours for the Miras only are illustrated in Figs.
\ref{fig_jhhk} and \ref{fig_hkkl} where they are compared to the colours of Miras in
the LMC (the derivation of the reddening corrections is given in Section
\ref{is_ext}, below). The colours
of the Miras with peculiarities (class 1n, n=1,5), which are illustrated as
open crosses, do not differ significantly from the class 10 Miras. The 
loci illustrated in the two figures were fitted by maximum likelihood to the
class 10 objects only:
\begin{equation} (H-K)_0=-0.549+1.002(J-H)_0, \ \ (\sigma=0.010),
\end{equation} 
\begin{equation} (K-L)_0=-0.252+1.295(H-K)_0, \ \ (\sigma=0.014),
\end{equation} 

Note that the $J$
values for some of the faint LMC sources in Fig.~\ref{fig_jhhk} are
uncertain by up to 0.2 mag, as are the $H$ values for the faintest LMC
sources in Fig~\ref{fig_hkkl}. There appears to be a small shift between
the LMC and the Galactic $JHK$ colours in that $(J-H)$ for the LMC sources
is on average $\leq 0.1$ mag less for the LMC stars than for the Galactic
ones at the same $(K-L)$. Cohen et al. (1981) discuss a colour difference
between Galactic and LMC C-stars in the same sense, but few of their sample
are Miras and their stars are in the colour range $0.4<(H-K)_0<0.8$. For
$(H-K)_0<0.8$ we find that LMC and Galactic C-rich Miras occupy the same
part of the two-colour diagram. This diagram is obviously rather sensitive
to reddening corrections and to errors in transformation between different
photometric systems, which tend to be largest for $J$.

The Galactic $(H-K)_0$ and $(K-L)_0$ period-colour relationships are
compared to those for Miras in the LMC in Figs.~\ref{fig_hkp} and
\ref{fig_klp}. There is no obvious difference between the class 10 and class
1n (n$>$1) Miras.  There appear to be differences between the distributions
of LMC and Galactic Miras in these figures, but there is a great deal of
overlap between the two samples and the differences could plausibly be
attributed to the very different selection effects in the Galactic and LMC
samples. In particular it is possible that we would have been unable to
measure the $H$ flux for any LMC Miras with extremely red colours
($H-K>2.8$).  It is also likely that our selection criteria for the
Aaronson and for the IRAS samples would have excluded most short period
Miras. The relatively blue $(H-K)$ and $(K-L)$ colours of the Galactic
Miras, compared to the LMC ones, with periods less than 500 days is
notable.

Although neither $H-K$ nor $K-L$ shows a linear dependence on period there
is certainly a trend to larger colours at longer period.
The least-squares fit to the Galactic data in Figs.~\ref{fig_hkp} and
\ref{fig_klp} (omitting the points at the shortest period, SZ Ara, in both
figures and the one with the largest $K-L$, LL Peg, in Fig.~\ref{fig_klp})
gives the following relationships:
\begin{equation} (H-K)_0=-15.69+6.412\log P \ \ (\sigma=0.48),
\end{equation}
and 
\begin{equation} (K-L)_0=-20.56+8.300\log P \ \ (\sigma=0.61).
\end{equation}

\begin{figure}
\includegraphics[width=8.4cm]{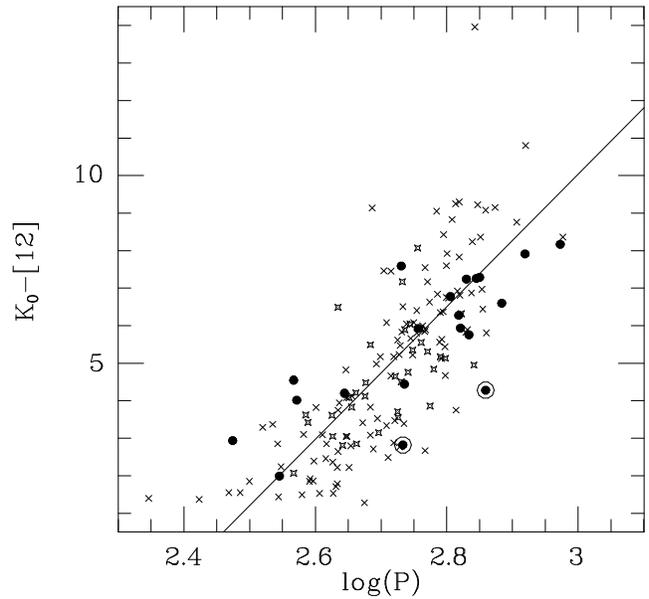}
 \caption{ $K-[12]$ as a function of period for all of the
Miras (class 1n) discussed here, compared with C stars from the   
LMC (closed circles) C stars. The two LMC points with rings around them
are luminous Miras thought to be undergoing hot bottom burning. The locus
described by equation \ref{k12p} is shown.}
\label{fig_k12p1}
\end{figure}

\begin{figure}
\includegraphics[width=8.4cm]{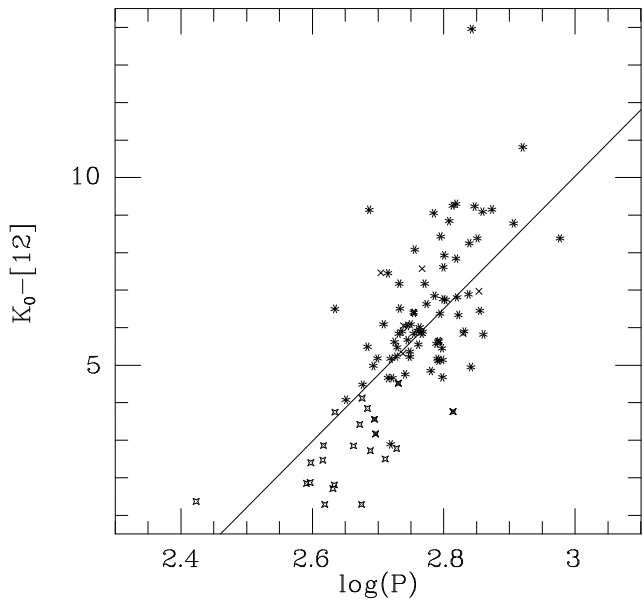}
 \caption{ As Fig.~\ref{fig_k12p1} but showing only 
stars from the IRAS sample (asterisks), the faint IRAS sample (crosses) and
the Aaronson sample (open crosses); note that four of the Aaronson
sample are also faint IRAS sources. The line is equation \ref{k12p}.} 
\label{fig_k12p2}
\end{figure}

Fig.~\ref{fig_k12p1} shows $K-[12]$ as a function of period and compares
it to LMC data. 
The straight line is a least squares fit to the Galactic data, which
yields:
\begin{equation} K_0-[12]=-43.0 +17.7 \log P \ \ (\sigma=1.5) 
\label{k12p} \end{equation}

$K-[12]$ is a very good indicator of  the optical depth of the dust shell 
and hence of the dust mass-loss rate (e.g. Whitelock et al. 1994). There
has been some discussion in the literature of differences in mass-loss rates
between the Magellanic Clouds and the Galaxy.  Such differences might be
expected if Magellanic Cloud Miras have lower metallicity than their
Galactic counterparts and therefore form dust with lower efficiency. If,
however, the pulsation period of the Mira is a function of its metallicity,
as it is for O-rich Miras (Feast \& Whitelock 2000), differences between the
two systems will be more subtle.   A recent analysis by van Loon (2006)
suggests that, while the situation is very complex, there is no evidence for
a mass-loss rate dependence on metallicity. The differences seen in
Fig.~\ref{fig_k12p1} seem to be relatively minor and may be due to selection
effects in the various samples rather than to fundamental properties of the
two galaxies.

The very large effect of selection criteria is well illustrated by Fig.
\ref{fig_k12p2} which shows only the Galactic C stars, but distinguished
according to the original sample from which they were drawn. Note that there
is almost no overlap between the bright IRAS selected stars, most of which
have $K-[12]\geq 5$ mag, and the optically selected Aaronson sample which
all have $K-[12]\leq 5$ mag. It is clear that the C Miras show a large
range of colour at a given period and that selection criteria will
determine how this range is sampled.

\section{Apparent Bolometric Magnitudes}\label{bol_mag} 
 The bolometric magnitude was calculated by spline fitting to the $JHKL$, 12
and 25 $\mu$m flux as a function of frequency, as described in Section 6 of
Whitelock et al. (1994) and/or by spline fitting to $JHKL$ MSX-$A$ flux in
the same way. 

Some preliminary tests were done to compare the results of using spline fits
to $JHKL$ and MSX-$A$, -$C$ and -$D$ fluxes with those on $JHKL$ and
MSX-$A$. This because there were only 67 stars with the full data set. These
tests indicated that a small colour correction was necessary if the
integrations using only MSX-$A$ were to give the same result as the full
dataset. Thus the bolometric magnitudes derived by spline fitting $JHKL$ and
MSX-$A$ were corrected by $0.0119-0.0148\times(K-L)$; its largest effect is to
change the bolometric magnitude by only --0.055 mag.

There were 91 Miras with both MSX-$A$ and IRAS data and the mean difference in
the bolometric magnitudes derived from the two sets (IRAS--MSX) was
$-0.02\pm0.03$ mag, with extreme values of --0.73 and +0.35. The  
corrections with the largest absolute values were always for the redder
sources, as the mid-infrared is less important for the bluer sources.

Where there is no value for $J$ (24 stars), $H$ (3
stars) or $L$ (1 star) an estimate is made from one of the following
relations, for the purpose of calculating the bolometric magnitude:
\begin{equation} (H-K)=-0.642 + 1.011(J-H), \label{jhhk}\end{equation}
\begin{equation} (H-K)=+0.233 + 0.769(K-L). \label{hkkl}\end{equation}
These loci are illustrated in Figs.~\ref{fig_jhhku} and
\ref{fig_hkklu}. If the 12 $\mu$m flux has been measured, but the 
25 $\mu$m flux has not, the colour corrected [25] is estimated from:
\begin{equation} ([12]-[25])_{cc}=0.0337+0.109(K-[12]). \end{equation}

 Because it is only for very red stars that $J$ or $H$ were not measured,
and only for blue ones that [25] was not measured, these estimates do not
compromise the quality of the final bolometric magnitude. The $L$ flux is
much more important, but only one Mira with a known period, CF~Cru, does not
have a measured value for $L$, which is estimated from $H$ via equation
\ref{hkkl} above.

Given that we chose to reject the faint cycles for the very well observed
stars we examine briefly the difference it would make to the bolometric
magnitudes if we used faint rather than bright cycles, assuming that the
cause of the change is line of site obscuration and that the IRAS and/or MSX
fluxes remain constant. R For would go from $m_{bol}=4.54$ to 4.97, R Lep
from $m_{bol}=3.52$ to 4.13 and II Lup from $m_{bol}=3.89$ to 4.07, the
size of the change being strongly dependent on the relative contributions
to the total flux from $JHKL$ and from the mid-infrared. 

\section{Bolometric Corrections}\label{bol_cor}
 The main interest in bolometric corrections is their use when limited
data are available. With this in mind we provide, in Table~\ref{bc},
the coefficients of the best fit least squares fourth order polynomials to
the bolometric correction at $K$, BC$_K$ as a function of the colour
$(x-y)$: 
\begin{equation} BC_K= c_0+c_1 (x-y)+c_2(x-y)^2+c_3(x-y)^3+c_4(x-y)^4, 
\end{equation}
for a variety of colours. The table also includes the maximum (max) and
minimum (min) values of the colour for which the fit was determined (this
type of fit should not be extrapolated), the error ($\sigma$) associated
with an individual observation and the number of stars used in the fit (No).

\begin{table*}
\begin{center}
\caption[Coefficients for calculating the bolometric correction as a function 
of colour according to equation 10.]{Coefficients for calculating the bolometric correction as a
function of colour according to equation 10.}\label{bc} 
\begin{tabular}{cccccc}
\hline
 & $(J-K)_0$ & $(H-K)_0$ & $(K-L)_0$ & $ (K-[12])_0$ & $(K-A)_0$\\
\hline
$c_0$ & 0.972 &2.360 & 3.228 &2.801&3.130\\
$c_1$ & 2.9292 &3.1729& 0.8720&0.7101&0.4807\\
$c_2$ & --1.1144 &--2.5747&--0.7042&--0.1958&--0.1655\\
$c_3$ & 0.1595 &0.5462 &0.06350&0.01032&0.002241\\
$c_4$ &$-9.5689\,10^{-3}$
&$-0.043014$&$-1.6341\,10^{-3}$&$-2.2054\,10^{-4}$&$-2.2405\,10^{-4}$\\
max   & 6.5 & 4.0& 6.2&14& 10\\
min   & 1.5 & 0.5 & 0.3& 1.2& 0.5\\
$\sigma$& 0.23 & 0.27 &0.17& 0.11& 0.13\\
No.   & 123 & 142 & 144 & 144 &70\\
\hline
\end{tabular}
\end{center}
\end{table*}

\begin{figure}
\includegraphics[width=8.4cm]{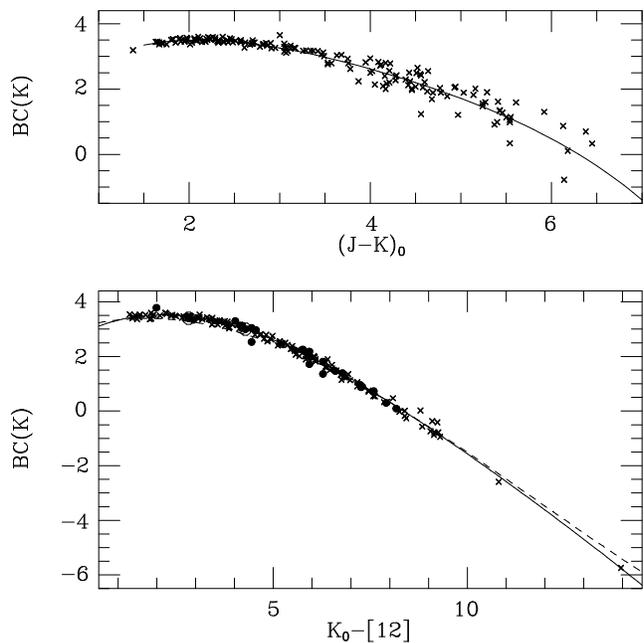}
 \caption{ The bolometric correction at $K$ as a function
of (top) $(J-K)_0$ and (bottom) $K_0-[12]$. The curves are fourth order
polynomials with the parameters given in Table~\ref{bc}. The lower figure
also shows LMC values (solid circles) from Whitelock et al. (2003). The
dashed curve is the relation for $K-[12.5]$ from Guandalini et al. (2005)
fig.~5.}
\label{fig_k12bc}
\end{figure}

Two examples of these fits are shown in Fig.~\ref{fig_k12bc} for $(J-K)_0$
and $K_0-[12]$.  It is clearly
possible to derive an accurate bolometric correction if observations in
widely separated bands, such as $K$ and $[12]$, are available. In practice
it will often only be possible to use something like $(J-K)$ when dealing
with photometry from e.g. the 2MASS survey or typical measurements that are
now being done on extragalactic C stars with large telescopes.

The curves listed in Table~\ref{bc} provide excellent fits to the LMC C-star
bolometric corrections discussed by Whitelock et al. (2003) and illustrated
in their fig.~13, although the LMC data do show more scatter because the
stars are fainter. There is no evidence of any difference between C stars in
the Galaxy and the LMC in respect of the bolometric corrections as a
function of colour. 

The lack of scatter in Fig.~\ref{fig_k12bc} is largely a consequence of the
way the bolometric magnitudes are calculated. Thus, for example, if we
examine II Lup during a bright cycle ($\bar{K}=1.79$ mag) we derive 
 $K-[12]=5.92$ and $BC_K=2.15$, while during a faint cycle ($\bar{K}=3.0$ mag)
we obtain $K-[12]=7.12$ and $BC_K=1.2$; both of these points fit on the
curve in Fig.~\ref{fig_k12bc}. The bolometric magnitude derived from these
data, $m_{bol}=0.48$ and 0.52, are not very different because the $JHKL$
flux is only a small part of the total. In contrast, for R Lep we obtain 
$K-[12]=2.81$ and $BC_K=3.47$ when it is bright ($\bar{K}=0.07$ mag) and
$K-[12]=3.54$ and $BC_K=3.36$ when it is faint ($\bar{K}=0.8$ mag). Again the
points fall close to the curve in Fig.~\ref{fig_k12bc}, but the resulting
bolometric magnitudes,  $m_{bol}=3.52$ and 4.13, differ considerably
because the $JHKL$ flux is a major contributor to the total.

Fig.~\ref{fig_k12bc} also shows a comparison with the bolometric correction
derived by Guandalini et al. (2005) for a slightly different colour
($K-[12.5]$). The two relations differ by at most 0.19 mag in the region in
which there are many points, $3<K-[12]<4$, and this difference goes up to
0.36 mag at $K-[12]=14$.

\section{Interstellar Reddening and Distances}\label{is_ext}
 We assume that the Galactic C-rich Miras obey the LMC bolometric PL
relation as discussed above (Section \ref{pl}). Thus a first estimate can be
made of the distance to the star by comparing the measured and absolute
bolometric magnitudes. The extinction is then estimated using the Drimmel et
al. (2003) three dimensional Galactic extinction model, including the
rescaling factors that correct the dust column density to account for small
scale structure seen in the DIRBE data, but not described explicitly by the
model. The measured mean $JHKL$ magnitudes are corrected for extinction
following the reddening law given by Glass (1999) and the bolometric flux is
recalculated. This process of calculating distance, extinction and
bolometric magnitude is then iterated as necessary, typically two to five
times, until the distance modulus changes by less than 0.01 mag. The
extinction values, $A_V$, derived in this way are listed in
Table~\ref{derived}. They range up to $A_V\sim 5.8$ and therefore have a
significant effect on the $JHKL$ colours and on the derived distances.

We have 70 stars in common with those of Groenewegen et al. (2002) for which
the derived distances can be compared. Groenewegen et al. calculate
distances either from the PL relation of Groenewegen \& Whitelock (1996),
for stars with known periods longer than 390 days, or from the 12 $\mu$m
flux, with a bolometric correction that depends on the 25 to 12 $\mu$m flux
ratio, for the others. The statistical agreement is good, with the average
distance calculated by Groenewegen et al. for the 70 stars in common being
only 8 percent larger than our value.

%
\begin{center}
\onecolumn
\begin{longtable}{lrcrcccrrcr}
\caption[Derived data for Miras with pulsation periods.]
{Derived data for Miras with pulsation periods.}\label{derived} \\
\hline
 name & dist & $A_V$ & $m_{bol}$ & $(J-H)_0$ & $(H-K)_0$ & $(K-L)_0$ &
 $K_0-[12]$ & $K_0$ & $\log \dot{M}$ & \multicolumn{1}{c}{BC$_K$}\\
& (kpc) & \multicolumn{7}{c}{(mag)}& ($\rm M_{\odot} yr^{-1}$)& \multicolumn{1}{c}{(mag)}\\
\hline
\endfirsthead

\hline
 name & dist & $A_V$ & $m_{bol}$ & $(J-H)_0$ & $(H-K)_0$ & $(K-L)_0$ &
 $K_0-[12]$ & $K_0$ & $\log \dot{M}$ & BC$_K$\\
& (kpc) & \multicolumn{7}{c}{(mag)}& ($M_{\odot} yr^{-1}$)& (mag)\\
\hline
\endhead
  \multicolumn{10}{l}{{Continued on Next Page\ldots}} \\
\endfoot    

  \\ \hline 
\endlastfoot
YY Tri          & 2.25 & 0.24 &  6.53 &      & 2.99 & 3.74 &  8.43 &  6.79 & --4.45 &--0.26 \\
R For           & 0.70 & 0.04 &  4.54 & 1.67 & 1.20 & 1.28 &  3.62 &  1.21 & --5.87 &  3.33 \\
$[$TI98$]$0418+0122 & 6.42 & 0.37 &  9.24 & 1.83 & 1.33 & 1.46 & 3.61 &5.98 &--5.73 &  3.26 \\
V718 Tau        & 1.46 & 1.35 &  6.12 & 1.70 & 1.15 & 1.15 &  3.42 &  2.72 & --5.85 &  3.41 \\
R Lep           & 0.47 & 0.25 &  3.52 & 1.43 & 0.94 & 1.00 &  2.81 &  0.05 & --6.03 &  3.47 \\
QS Ori          & 2.65 & 0.69 &  7.17 & 1.50 & 0.98 & 1.04 &  3.08 &  3.74 & --5.89 &  3.43 \\
05418--3224      & 2.73 & 0.04 &  7.24 & 2.47 & 1.82 & 2.11 & 5.49 &  4.86 & --5.19 &  2.38 \\
06088+1909      & 2.37 & 1.18 &  6.91 & 2.37 & 1.74 & 1.96 &  4.98 &  4.15 & --5.08 &  2.76 \\
ZZ Gem          & 1.76 & 0.61 &  6.75 & 1.32 & 0.77 & 0.58 &  1.85 &  3.15 & --6.65 &  3.60 \\
V617 Mon        & 2.94 & 1.01 &  7.48 & 1.68 & 1.22 & 1.29 &  3.05 &  4.08 & --5.96 &  3.41 \\
V636 Mon        & 1.09 & 0.45 &  5.11 & 1.83 & 1.28 & 1.39 &  3.39 &  1.78 & --5.76 &  3.34 \\
V477 Mon        & 2.61 & 1.01 &  6.86 & 2.31 & 1.84 & 2.24 &  5.12 &  4.46 & --4.83 &  2.40 \\
RT Gem          & 3.43 & 0.54 &  8.08 & 1.24 & 0.56 & 0.39 &  1.44 &  4.57 &        &  3.51 \\
06487+0551      & 2.40 & 0.80 &  6.84 & 2.47 & 1.91 & 2.34 &  5.47 &  4.55 & --5.36 &  2.29 \\
CG Mon          & 1.99 & 0.75 &  6.69 & 1.29 & 0.61 & 0.52 &  1.53 &  3.23 &        &  3.46 \\
CL Mon          & 1.11 & 0.47 &  5.22 & 1.67 & 1.15 & 1.27 &  3.34 &  1.84 & --5.85 &  3.38 \\
06531--0216      & 2.01 & 0.86 &  6.35 & 1.97 & 1.49 & 1.70 &  3.86 &  3.27&        & 3.07 \\
06564+0342      & 2.99 & 0.83 &  7.24 & 2.58 & 2.11 & 2.55 &  5.84 &  5.35 & --5.27 &  1.89 \\
07080--0106      & 3.76 & 0.16 &  7.70 & 2.80 & 2.44 & 2.73 &  6.63 & 6.22 & --4.71 &  1.48 \\
VX Gem          & 1.88 & 0.03 &  6.66 & 1.22 & 0.61 & 0.35 &  1.92 &  3.13 & --6.54 &  3.53 \\
R Vol           & 0.88 & 0.49 &  4.84 & 1.89 & 1.40 & 1.61 &  3.83 &  1.67 & --5.70 &  3.18 \\
HX CMa          & 1.68 & 0.74 &  5.73 & 2.38 & 1.79 & 2.17 &  5.81 &  3.55 & --4.90 &  2.18 \\
07217--1246      & 2.18 & 0.84 &  6.47 & 2.47 & 1.97 & 2.35 &  5.64 & 4.22 & --4.98 &  2.24 \\
07220--2324      & 2.75 & 1.13 &  7.09 & 2.59 & 1.96 & 2.26 &  5.22 & 4.67 & --5.22 &  2.42 \\
$[$W71b$]$007--02    & 4.34 & 1.33 &  8.29 & 1.62 & 1.07 & 1.08 &  2.85 & 4.81 &    &  3.49 \\
07373--4021      & 1.09 & 0.54 &  5.30 & 1.81 & 1.26 & 1.44 &  4.21 &  2.18 &--5.51 &  3.12 \\
V471 Pup        & 4.17 & 1.43 &  8.39 & 1.14 & 0.56 & 0.44 &  1.85 &  5.00 &        &  3.39 \\
07454--7112      & 0.83 & 0.45 &  4.60 & 2.90 & 2.23 & 2.73 &  6.08 &  2.78 &--5.14 &  1.82 \\
V831 Mon        & 2.33 & 0.09 &  7.31 & 1.53 & 1.09 & 1.13 &  3.29 &  3.88 & --6.05 &  3.43 \\
07576--4054      & 2.86 & 1.25 &  7.26 &      & 2.59 & 3.17 &  7.45 &  6.51& --5.18 &  0.75 \\
07582--1933      & 2.49 & 0.44 &  6.91 & 2.91 & 2.32 & 2.68 &  6.51 &  5.35 &--5.02 &  1.56 \\
$[$ABC89$]$Pup38    &12.30 & 0.98 & 10.63 & 1.75 & 1.30 & 1.33 &  3.75 &  7.34 &    &  3.29 \\
FF Pup          & 3.77 & 0.58 &  8.06 & 1.20 & 0.65 & 0.66 &  2.65 &  4.64 &        &  3.42 \\
V518 Pup        & 2.03 & 0.10 &  6.67 & 2.05 & 1.37 & 1.45 &  4.08 &  3.51 & --5.68 &  3.16 \\
08050--2838      & 2.63 & 0.97 &  7.01 & 3.01 & 2.14 & 2.47 &  6.04 &  4.99& --5.10 &  2.01 \\
08074--3615      & 2.40 & 1.52 &  6.37 &      & 3.94 & 4.52 & 10.80 &  8.95& --4.32 &--2.59 \\
$[$ABC89$]$Ppx19    & 5.89 & 2.19 &  8.93 & 1.97 & 1.52 & 1.78 &  4.12 &  5.91 &    &  3.01 \\
V346 Pup        & 1.36 & 0.88 &  5.57 & 2.65 & 2.11 & 2.55 &  6.40 &  3.67 & --4.85 &  1.90 \\
$[$ABC89$]$Ppx40    & 5.13 & 1.54 &  8.74 & 1.36 & 0.69 & 0.56 &  1.71 &  5.22 &    &  3.52 \\
$[$W71b$]$029--02    & 4.88 & 3.04 &  8.53 & 1.87 & 1.46 & 1.63 &  3.41 &  5.35 &   &  3.17 \\
08340--3357      & 2.26 & 0.83 &  6.60 & 3.06 & 2.48 & 3.01 &  7.17 &  5.53 &--4.84 &  1.07 \\
R Pyx           & 1.35 & 0.38 &  6.00 & 1.23 & 0.73 & 0.70 &  2.07 &  2.50 & --6.38 &  3.50 \\
UW Pyx          & 1.50 & 0.67 &  6.11 & 1.34 & 0.74 & 0.80 &  2.36 &  2.62 & --6.28 &  3.49 \\
08535--4724      & 3.51 & 4.53 &  7.64 &      & 2.91 & 3.33 &  8.07 &  7.17 &--4.53 &  0.47 \\
08534--5055      & 4.29 & 2.20 &  7.90 &      & 3.32 & 3.85 &  9.22 & 8.30 & --4.24 &--0.40 \\
IQ Hya          & 1.55 & 0.53 &  6.26 & 1.56 & 1.14 & 1.23 &  3.10 &  2.84 & --6.05 &  3.42 \\
CQ Pyx          & 1.14 & 0.49 &  4.99 &      & 3.30 & 3.87 &  9.30 &  5.94 & --4.76 &--0.94 \\
09176--5147      & 3.17 & 3.64 &  7.69 & 2.58 & 2.05 & 2.52 &  6.49 & 5.98 & --4.77 &  1.71 \\
$[$W71b$]$046--02    & 5.12 & 5.63 &  9.26 & 1.15 & 0.54 & 0.35 &  1.37 &  5.87 &   &  3.40 \\
$[$ABC89$]$Vel44    & 4.18 & 3.34 &  8.33 & 1.41 & 0.89 & 0.91 &  2.46 &  4.85 &    &  3.48 \\
09433--6233      & 8.66 & 0.90 &  9.60 & 2.29 & 1.70 & 1.97 &  5.31 &  7.10 &--5.02 &  2.50 \\
CW Leo          & 0.14 & 0.07 &  0.40 & 3.29 & 2.85 & 3.73 &  9.25 &  1.18 & --4.65 &--0.78 \\
09513--5324      & 1.79 & 1.87 &  6.05 & 3.12 & 2.41 & 2.95 &  6.75 & 4.90 & --4.82 &  1.15 \\
09529--5506      & 2.80 & 1.82 &  6.90 & 2.55 & 1.98 & 2.63 &  6.87 & 5.65 & --4.69 &  1.25 \\
09533--6021      & 4.81 & 1.48 &  8.03 & 3.08 & 2.42 & 2.92 & 6.97 &  6.90 & --4.74 &  1.14 \\
09521--7508      & 1.12 & 0.72 &  5.19 & 2.67 & 2.09 & 2.49 & 5.83 &  3.15 & --5.19 &  2.03 \\
09586--6150      & 9.13 & 1.03 &  9.80 &      & 2.77 & 3.24 & 7.46 &  9.10 & --4.75 &  0.71 \\
10026--5849      & 5.68 & 4.73 &  8.72 & 1.79 & 1.29 & 1.47 &  3.71 &  5.47 &       &  3.25 \\
10098--5742      & 3.91 & 3.54 &  7.81 &      & 2.52 & 3.15 &  7.55 &  7.23 &       &  0.58 \\
10109--5958      & 2.95 & 1.70 &  7.55 & 1.58 & 1.11 & 1.16 &  3.05 &  4.12 &       &  3.43 \\
RW LMi          & 0.46 & 0.09 &  3.10 & 2.74 & 2.10 & 2.53 &  6.35 &  1.31 & --4.94 &  1.78 \\
10199--5801      & 4.39 & 4.33 &  7.90 & 2.83 & 2.04 & 2.49 & 5.82 &  5.81 & --4.10 &  2.09 \\
10220--5858      & 5.45 & 3.73 &  8.52 & 1.51 & 1.00 & 1.02 &  2.67 &  5.05 &       &  3.47 \\
CPD--58 2175     & 5.76 & 4.53 &  8.72 & 2.45 & 1.96 & 2.40 &  6.04 &  6.73 &       &  1.99 \\
CZ Hya          & 1.35 & 0.13 &  5.80 & 1.40 & 0.92 & 1.00 &  3.05 &  2.38 & --6.06 &  3.42 \\
TV Vel          & 1.84 & 1.08 &  6.57 & 1.14 & 0.51 & 0.45 &  1.53 &  3.14 &        &  3.43 \\
$[$ABC89$]$Car73    & 6.05 & 4.52 &  8.97 & 2.10 & 1.48 & 1.73 &  3.83 &  5.93 &    &  3.04 \\
$[$ABC89$]$Car84    & 4.54 & 4.45 &  8.30 & 1.21 & 0.75 & 0.89 &       &  4.93 &    &  3.36 \\
$[$ABC89$]$Car87    & 7.09 & 3.46 &  9.33 & 1.18 & 0.76 & 0.53 &  1.28 &  5.81 &    &  3.53 \\
$[$ABC89$]$Car105   & 2.95 & 2.57 &  7.37 & 1.57 & 1.07 & 1.21 &  3.15 &  4.01 &    &  3.36 \\
11145--6534      & 1.73 & 1.51 &  5.96 & 2.91 & 2.34 & 2.73 &  6.37 &  4.35 &--4.82 &  1.61 \\
$[$W65$]$ c13       & 5.38 & 4.53 &  8.93 & 1.27 & 0.78 & 0.74 &  1.86 &  5.44 &    &  3.49 \\
$[$TI98$]$1130--1020 & 2.16 & 0.09 &  6.82 & 2.13 & 1.63 & 1.84 &4.82&4.03 & --5.56 &  2.79 \\
11318--7256      & 0.66 & 0.56 &  4.07 & 1.82 & 1.24 & 1.48 &  3.47 & 0.80 & --5.66 &  3.27 \\
$[$ABC89$]$Cen4     & 4.00 & 1.68 &  8.00 & 1.41 & 0.74 & 0.77 &  2.49 &  4.53 &    &  3.47    \\
11463--6320      & 3.13 & 2.77 &  7.27 & 2.59 & 1.89 & 2.33 & 5.56 &  5.19 & --4.84 &  2.08 \\
$[$ABC89$]$Cen32    & 4.40 & 2.58 &  7.94 & 2.00 & 1.46 & 1.65&3.75&  4.82 & --5.45 &  3.12 \\
$[$ABC89$]$Cen43    & 6.06 & 3.93 &  8.85 & 1.55 & 1.11 & 1.23 &  2.77 &  5.46 &    &  3.39    \\
$[$ABC89$]$Cen60    & 5.25 & 3.50 &  8.82 & 1.44 & 0.91 & 0.99 &  2.85 &  5.39 &    &  3.43    \\
CF Cru          & 5.88 & 5.80 &  9.03 & 1.14 & 0.56 &      &  1.79 &  5.68 &        &  3.35 \\
12194--6007      & 2.84 & 2.21 &  7.04 & 2.35 & 1.90 & 2.29 & 5.44 &  4.76 & --5.21 &  2.28 \\
12298--5754      & 1.85 & 1.58 &  6.19 & 2.46 & 2.15 & 2.63 & 6.00 &  4.26 & --4.86 &  1.93 \\
CGCS3268        & 2.05 & 1.36 &  6.83 & 1.20 & 0.76 & 0.59 &   2.39 &  3.31 &       &  3.52 \\
12394--4338      & 1.33 & 0.34 &  5.53 & 2.56 & 1.96 & 2.17 & 4.76 &  2.88 & --5.41 &  2.65 \\
12421--6217      & 4.87 & 5.42 &  8.03 &      &      & 3.73 & 8.76 &  7.98 & --4.24 &  0.06 \\
RU Vir          & 0.91 & 0.08 &  4.94 & 1.76 & 1.28 & 1.51 &  4.08 &  1.79 & --5.69 &  3.15 \\
V Cru           & 1.47 & 0.98 &  6.16 & 1.14 & 0.52 & 0.39 &   1.49 &  2.79 &       &  3.37 \\
12540--6845      & 1.37 & 0.65 &  5.52 & 2.61 & 1.96 & 2.34 & 5.89 &  3.46 & --4.71 &  2.06 \\
13343--5807      & 2.40 & 1.94 &  6.82 & 2.50 & 1.94 & 2.32 & 5.66 &  4.80 & --5.14 &  2.01 \\
13477--6532      & 2.32 & 1.36 &  6.49 &      & 2.96 & 3.78 & 8.24 &  6.50 & --4.52 &--0.01 \\
13482--6716      & 1.70 & 0.87 &  6.17 & 2.59 & 1.96 & 2.25 & 5.17 &  3.70 & --5.20 &  2.47 \\
13509--6348      & 2.98 & 2.30 &  7.06 &      &      & 2.30 & 5.89 &  4.91 & --5.16 &  2.15 \\
$[$ABC89$]$Cir26    & 3.81 & 4.11 &  7.93 & 1.97 & 1.33 & 1.58 &  3.53 &  4.75 &    &  3.18 \\
$[$ABC89$]$Cir27    & 3.57 & 4.14 &  7.70 & 1.93 & 1.59 &1.98 &4.50&  4.89 & --5.35 &  2.80 \\
14395--5656      & 6.58 & 3.41 &  9.13 & 1.57 & 0.99 & 1.13 &  2.71 &  5.70 &       &  3.43 \\
14404--6320      & 3.62 & 2.55 &  7.53 &      & 2.98 & 3.82 &  8.83&  8.09 & --4.75 &--0.55 \\
14443--5708      & 5.17 & 3.43 &  8.27 &      & 3.62 & 3.84 &  9.08&  8.62 & --4.15 &--0.35 \\
15082--4808      & 0.95 & 0.60 &  4.65 & 2.93 & 2.60 & 3.47 &  7.92&  4.31 & --4.67 &  0.35 \\
15084--5702      & 3.48 & 3.58 &  7.05 &      & 3.29 & 3.80 &  8.36&  7.00 & --4.10 &  0.04 \\
II Lup          & 0.64 & 0.48 &  3.89 & 2.29 & 1.76 & 2.10 &  5.92 &  1.75 & --4.82 &  2.15 \\
15261--5702      & 3.31 & 1.51 &  7.23 &      & 2.40 & 2.82 &  6.44&  5.82 & --4.72 &  1.41 \\
16079--4812      & 2.13 & 3.36 &  6.28 &      & 3.18 & 3.82 &  8.36&  6.54 & --4.69 &--0.26 \\
NP Her          & 2.44 & 0.17 &  7.07 & 1.41 & 0.85 & 0.70 &  2.22 &  3.48 & --6.37 &  3.59 \\
16171--4759      & 2.84 & 2.70 &  7.16 & 2.40 & 1.83 & 2.07 &  5.35 &  4.64 &       &  2.52 \\
V Oph           & 0.78 & 0.90 &  5.06 & 1.20 & 0.68 & 0.58 &  1.55 &  1.53 & --6.95 &  3.53 \\
CGCS3721        & 2.71 & 1.29 &  7.56 & 1.34 & 0.88 & 0.78 &  2.24  &  4.00 &       &  3.56 \\
16538--4633      & 2.42 & 2.57 &  6.88 & 2.49 & 1.64 & 1.85 &  4.65 & 4.04 & --4.67 &  2.84 \\
16545--4214      & 1.03 & 0.75 &  5.01 & 2.56 & 1.93 & 2.11 &  5.23 & 2.45 & --5.02 &  2.56 \\
17047--2848      & 3.22 & 1.31 &  7.49 & 2.22 & 1.65 & 2.02 &  5.62 & 5.24 & --5.07 &  2.25 \\
V2548 Oph       & 1.09 & 0.87 &  4.75 &      & 2.82 & 3.59 &  9.15 &  5.53 & --4.47 &--0.78 \\
SZ Ara          & 2.44 & 0.47 &  7.84 & 1.11 & 0.52 & 0.37 &  1.40 &  4.41 & --6.48 &  3.44 \\
V617 Sco        & 1.29 & 1.19 &  5.52 & 1.39 & 0.91 & 0.89 &  2.88 &   2.06 &       &  3.46 \\
17105--3746      & 2.73 & 3.20 &  7.07 & 2.23 & 1.82 & 2.40 &  5.81 & 4.85 & --4.75 &  2.22 \\
17130--3907      & 2.24 & 1.62 &  6.52 & 2.59 & 1.82 & 2.07 &  5.13 & 4.08 & --4.90 &  2.44 \\
17217--3916      & 2.88 & 2.03 &  7.10 &      & 2.75 & 3.19 &  7.60 &  6.24 &       &  0.86 \\
17222--2328      & 2.64 & 2.58 &  6.92 & 2.73 & 1.90 & 2.18 &  4.84 & 4.37 & --5.14 &  2.55 \\
V833 Her        & 1.07 & 0.14 &  5.08 & 2.90 & 2.46 & 3.18 &  7.17 &  4.18 & --4.68 &  0.91 \\
17446--4048      & 1.40 & 0.80 &  5.66 & 2.52 & 1.89 & 2.09 &  5.89 & 3.53 & --5.06 &  2.13 \\
17463--4007      &10.55 & 1.22 & 10.38 & 1.54 & 1.07 & 1.24 &  3.82 &  7.11 &       &  3.27 \\
17581--1744      & 2.34 & 1.72 &  6.62 & 2.30 & 1.63 & 1.83 &  4.67 & 3.87 & --5.25 &  2.75 \\
18036--2344      & 2.25 & 2.61 &  6.47 & 3.06 & 2.28 & 2.64 &  6.32 & 4.69 & --4.59 &  1.78 \\
FX Ser          & 1.13 & 1.39 &  5.25 & 2.43 & 1.77 & 1.99 &  4.66 &  2.46 & --5.12 &  2.78 \\
V1280 Sgr       & 1.34 & 1.03 &  5.59 & 1.64 & 1.13 & 1.32 &  3.55 &  2.30 & --5.71 &  3.29 \\
18119--2244      & 2.21 & 1.67 &  6.54 &      & 2.68 & 2.87 &  6.84 & 5.45 & --4.78 &  1.09 \\
V5104 Sgr       & 1.05 & 0.63 &  4.82 & 3.11 & 2.31 & 2.71 &  6.92 &  3.46 & --4.70 &  1.36 \\
18239--0655      & 1.74 & 1.64 &  5.96 & 2.90 & 2.49 & 2.88 &  6.73 & 4.49 & --4.53 &  1.47 \\
V1076 Her       & 1.14 & 0.36 &  5.07 &      & 3.28 & 3.87 &  9.05 &  5.81 & --4.64 &--0.73 \\
18248--0839      & 2.29 & 1.58 &  6.53 & 3.49 & 2.91 & 3.18 &  7.83 & 5.99 & --4.56 &  0.54 \\
V627 Oph        & 3.34 & 0.85 &  7.75 & 1.84 & 1.26 & 1.45 &  4.13 &  4.58 & --5.59 &  3.16 \\
V1417 Aql       & 0.87 & 0.36 &  4.49 & 2.41 & 1.80 & 2.02 &  5.17 &  1.93 & --4.85 &  2.56 \\
V821 Her        & 0.75 & 0.69 &  4.34 & 2.22 & 1.79 & 2.07 &  5.17 &  1.79 & --5.36 &  2.55 \\
V2045 Sgr       & 2.22 & 1.07 &  6.86 & 1.59 & 0.97 & 0.99 &  2.80 &  3.36 & --5.98 &  3.50 \\
AI Sct          & 2.15 & 1.06 &  6.90 & 1.50 & 0.95 & 0.80 &  3.09 &  3.38 & --5.64 &  3.52 \\
V1418 Aql       & 1.04 & 0.77 &  4.96 & 2.77 & 2.25 & 2.64 &  6.08 &  3.07 & --4.90 &  1.89 \\
V1420 Aql       & 1.00 & 0.52 &  4.66 & 2.20 & 1.57 & 2.00 &  4.95 &  2.03 & --5.04 &  2.62 \\
V1965 Cyg       & 1.13 & 0.56 &  5.11 & 2.58 & 2.01 & 2.37 &  5.55 &  2.90 & --4.89 &  2.21 \\
R Cap           & 1.62 & 0.46 &  6.45 & 1.32 & 0.86 & 0.94 &  2.85 &  3.01 & --6.57 &  3.45 \\
BD Vul          & 2.22 & 0.62 &  6.92 & 1.47 & 0.89 & 0.71 &  2.22 &  3.31 & --6.14 &  3.60 \\
V442 Vul        & 1.41 & 0.58 &  5.45 & 2.88 & 2.58 & 3.05 &  6.81 &  4.17 & --4.82 &  1.29 \\
RV Aqr          & 0.75 & 0.18 &  4.54 & 1.91 & 1.35 & 1.51 &  3.94 &  1.37 & --5.71 &  3.17 \\
$[$TI98$]$2259+1249 & 4.15 & 0.49 &  8.65 & 1.14 & 0.51 & 0.32 &  1.55&5.20& --5.88 &  3.45 \\
LL Peg          & 1.05 & 0.09 &  4.75 &  &      & 6.23 & 13.96 & 10.49 &     --4.28 &--5.74 \\
IZ Peg          & 1.70 & 0.19 &  6.20 &      & 3.16 & 4.04 &  9.14 &  7.07 & --4.86 &--0.87 \\
$[$TI98$]$2223+2548 & 2.84 & 0.16 &  7.69 & 1.61 & 1.20 & 1.28 & 3.37&4.34 & --5.94 &  3.36 \\
\end{longtable}                                                                    
\end{center}                                          
\twocolumn                                            

There are, however, some individually large differences and some systematic
trends. For the bluer sources ($K-[12]<6$) our distances tend to be smaller
because the apparent bolometric flux we derive is larger and there is a
clear trend with colour. The Groenewegen et al. results for the 34 stars
with $K-[12]<5.9$ are 24 percent more distant than we find. The most
extreme case is R Cap (20084--1425), $K-[12]=2.85$, for which we find
1.52 kpc while Groenewegen et al. get 3.45 kpc (Le Bertre et al. (2001) got
1.33 kpc; see below). The main source of this difference is in the apparent
bolometric magnitudes, which differ by 1.25 mag, although the absolute
magnitudes also differ in such a way as to increase the difference in
derived distance. Our results suggest that Groenewegen et al. systematically
underestimated the contribution from $HKL$ to the total flux, thereby
underestimating the apparent bolometric flux for blue sources. This will
have had very little effect on the red sources that make up the bulk of
their C-star sample.

For the redder sources our distances tend to be slightly larger than the
Groenewegen et al. ones, but there is no systematic trend with colour for
stars with $K-[12]>6$ and the effect is not large. For the 36 stars with
$K-[12]>5.9$ Groenewegen et al. get distances 6.5 percent less than ours. The
most extreme examples are RW LMi, $K-[12]=6.35$, for which we and
Groenewegen get 0.32 and 0.46 kpc respectively, and 08534--5055,
$K-[12]=9.22$, for which the distances are 3.10 and 4.29 respectively. For
RW LMi the difference is entirely in the apparent bolometric magnitude and
for 08534--5055 it is largely so.

Le Bertre et al. (2001) calculate distances using a [2.20-3.77]$\mu$m
dependent bolometric correction to the [2.20] mag, applied to observations
obtained with the Japanese IRTS. For this purpose they assume
$M_{bol}=-5.01$ mag for all their sources (this is the bolometric magnitude
we would associate with a Mira with a period of 530 days) indicating a
factor of 1.8 uncertainty for the distances of individual sources owing to
this latter assumption. They have only 3 sources in common with us, R Cap,
HX CMa and 11318--7256, for which they estimate distances 9 to 30 percent
smaller than we do.

CW Leo is a very well studied, thick-shelled, C star for which numerous
distance estimates, in the range 0.11 to 0.17 kpc, have been made. It is
also one of the few stars for which a geometric distance, of 0.145 kpc, has
been measured using the dust outflow speed combined with the proper motion
of the shell (Tuthill et al. 2000). This value is in good agreement with our
estimate of 0.14 kpc. 

Guandalini et al. (2005) tabulate distances (their tables 1 and 2) for
various C stars including some Miras. For the four stars in common in their
table 1 (which they describe as having ``astrometric, or reliable, distance
estimates") their distances are, in the mean, 7 percent larger than ours.
For the 8 stars listed in both our Table 5 and in their table 2, the mean
difference is zero. The distances in table 1 of Guandalini et al.
(2005), which are described as ``astrometric" are taken from Bergeat \&
Chevallier (2005) and are based on a procedure of Knapik et al. (1998) who
corrected the Hipparcos parallaxes by a semi-empirical statistical method of
their own.  Contrary to a standard Lutz-Kelker type approach to this
problem, which would have resulted in negative corrections to the parallaxes
and absolute magnitudes, both positive and negative corrections result from
their method.  They do not discuss the uncertainties in the corrections
applied.  However, it is clear from table 1 of Bergeat et al. (1998) that
these corrections are often large.  Knapik et al. (1998) also say (their
section 5), ``: individual values [of the corrections] may occasionally be
wrong and a few such cases were detected from data at hand...... In such
cases the catalogue value can be kept if positive or the star is abandoned". 

So far as C-Miras are concerned, the possibility of using Hipparcos
parallaxes to calibrate the zero-point of the PL($K$) relation was
investigated by Whitelock \& Feast (2000). A definitive result was not
obtained due primarily to the small number of C-Miras with parallaxes of
significant weight. It may be possible to re-investigate this matter when
the revision of the Hipparcos catalogue is completed (see van Leeuwen 2005,
van Leeuwen \& Fantino 2005).

The distances tabulated here probably represent the best currently available
for carbon Miras, particularly for those where the light curve has been well
characterized over many cycles. Noting the discussion in the last paragraph
of Section \ref{bol_mag} there is a potential problem in calculating
accurate apparent bolometric luminosities during obscuration events (see
also Section \ref{trends}). If we had made $JHKL$ observations of a C star
with a thin dust shell (where most of the bolometric flux is emitted at
near-infrared wavelengths) only during an obscuration event, then we would
have underestimated its luminosity and overestimated its distance. It also
remains possible that we have failed to identify a small number of bright
hot-bottom-burning stars for which we will have underestimated the
distances.

\section{Mass-loss Rates}\label{mdot}
 Mass-loss rates can be derived for many of these stars using the expression
given by Jura (1987):
\begin{equation} \dot{M}=1.7 \times 10^{-7} v_{15}
d^2_{kpc}L_4^{-1/2}F_{\nu,60}\bar{\lambda}^{1/2}_{10}\ \rm
M_{\odot}yr^{-1},\label{e_mdot} \end{equation}
where $v_{15}$ is the outflow velocity in units of 15 km\,s$^{-1}$, $d$ is the
distance in kpc, $L_4$ is the luminosity in units of $10^4 \rm L_{\odot}$,
$F_{\nu,60}$ is the flux from the dust at $60\mu$m and $\bar{\lambda}$ is
the mean wavelength of the light emerging from the star and its shell in
units of 10 $\mu$m.  Note that this equation assumes a constant
dust-to-gas ratio. The results are given in Table~\ref{derived}.

For the outflow velocity ($v_{15}$) we use the expansion velocities
tabulated by Groenewegen et al. (2002) while noting that those authors, in
their own calculation of mass loss add an additional drift velocity (of the
order of 2 or 3 km\,s$^{-1}$) to their expansion velocities to determine the
outflow velocity. For the stars in our sample which have no measured
expansion velocity we assume 19 km\,s$^{-1}$, this being the mean for the 68
stars in our sample which have all the other parameters necessary to
calculate $\dot{M}$. We find that $\bar{\lambda}$ varies from 0.23 to 1.42,
i.e. it is generally somewhat larger than the 0.3 assumed by Groenewegen et
al. particularly for the very red stars which constitute the bulk of their
sample. 

For $F_{\nu,60}$ we use the IRAS flux at $60\mu$m, and where there is no
IRAS flux at this wavelength we do not attempt to calculate mass-loss rates.
Note that this approach will result in an overestimate of the mass-loss rate
if the shell is very thin and a significant fraction of the $60\mu$m flux
actually originates from the underlying star.  We can
estimate the effect by assuming the stars radiate as blackbodies at the
temperatures given by Bergeat et al. (2002) and using the $K$ mag given in
Table~\ref{derived} to estimate the stellar contribution to the measured
$60\mu$m flux. Bergeat et al. tabulate temperatures for 4 of the 5 stars
(none for R Cap) listed with
$\log \dot{M} < -6.4 \, \rm M_{\odot} yr^{-1}$ in Table~\ref{derived}. For V~Oph,
ZZ~Gem, VX~Gem and SZ~Ara the stellar contribution to the $60\mu$m flux will
be approximately 39, 32, 23 and 12 percent, respectively. For these 5 stars
(including R Cap, where the contribution from the star is estimated at 20
percent) the mass-loss rates given in the table have been adjusted by the
amount indicated; for all the others the effect will be negligible.

For very close stars the $60\mu$m flux will have been spatially
resolved and therefore not entirely included in the IRAS PSC estimate. In
which case the mass-loss rates would be underestimated.  The 3 stars with
distances under 500\,pc are the only ones where the extended nature of the
source is likely to be significant; these are CW Leo, R Lep and RW LMi. Two
of these stars (CW Leo and R Lep) were examined by Young, Phillips \& Knapp
(1993) who found extended contributions from both, amounting to about 10
percent of the PSC flux. The mass-loss values tabulated for these three
stars have therefore been increased by 10 percent. For all the other stars
the effect will probably be insignificant, but certainly less than 10
percent. 

A number of authors have calculated mass-loss rates for C-rich Miras and the
results differ quite significantly from one paper to another. There are 
many factors which contribute to these differences, but uncertainty in the 
distance, which appears squared in equation \ref{e_mdot}, is always a major
factor. This aspect has already been discussed in Section \ref{is_ext}.

Whitelock et al. (1987)  showed that 
mass-loss rates of
relatively thin shelled O-rich Miras depended on their pulsation amplitudes,
providing strong support for the role of pulsation in driving mass loss.
While it would be very interesting to do a similar exercise for the C Miras
under discussion, it is unfortunately not practical.  As shown in
Fig.~\ref{delkk12} and discussed in Section \ref{amps} the $JHKL$ amplitudes
tell us little about the pulsation amplitude of the star. Ideally we should
measure the bolometric amplitude, but this must await monitoring at
mid-infrared wavelengths.

\begin{figure} 
\begin{center}
\includegraphics[width=8.0cm]{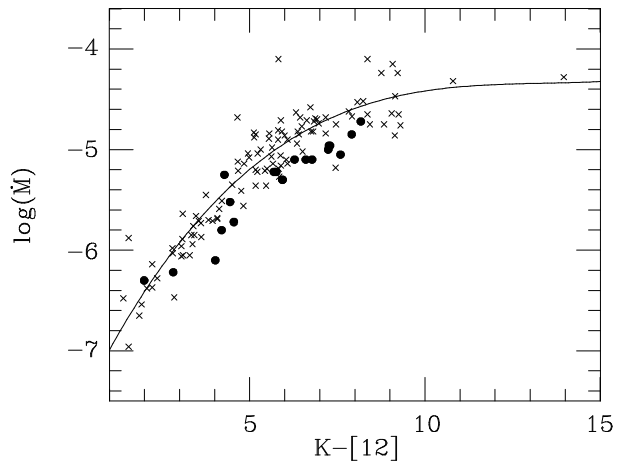}
 \caption{Mass-loss rate as a function of $K-[12]$ colour, for Galactic C
Miras (crosses) and LMC C Miras (closed circles, van Loon et al. 1999), the
line is equation \ref{mdotk12}.}
\label{fig_mdotk12}
\end{center}
\end{figure}

Fig. \ref{fig_mdotk12} shows how the mass-loss rates depend on colour.
The line is a polynomial fit to the Galactic C Miras: 
\begin{eqnarray} 
\log \dot{M}=-7.668+0.7305(K-[12])  \nonumber \\
-5.398\times10^{-2}(K-[12])^2 +1.343\times10^{-3}(K-[12])^3. \label{mdotk12} \end{eqnarray}
The LMC Miras discussed by Whitelock et al. (2003) are shown for comparison.
The two groups follow the same trend and the slight displacement of the LMC
points with respect to the Galactic ones should not be seen as significant
in view of the different assumptions that went into the mass-loss rate
calculations (see van Loon et al. 1999 for the LMC data).  The relationship
is also qualitatively similar to that found for O-rich stars (Whitelock et
al. 1994 fig.~21) and covers the transition from an optically thin dust
shell, $K-[12]<5$, to optically thick one $K-[12]>7$.

\section{Long Term Trends, Obscuration Events and the RCB
phenomenon}\label{trends}

The very extended atmospheres of Miras are intrinsically unstable and all
their light curves show some level of variability from one cycle to another.
However, some Miras show much greater variability (typically around $\Delta
K\sim 1$ mag on top of the normal pulsation), which can be understood as the
result of dramatically changing obscuration from dust. Detailed descriptions
have been given for R For (Whitelock et al. 1997) and II Lup (Feast et al.
2003) where the obscurations are attributed to the ejection of dust puffs in
our line of sight. It is clear from these references that the dust ejection
cannot be in the form of a spherically symmetric shell. A similar phenomenon
has been noted in the photographic red magnitudes of northern C stars,
measured over a period of 30 years (Alksnis 2003).

These obscuration events in C-rich Miras are phenomenologically similar to
those observed in the H-deficient RCB C stars. The RCB stars are
characterized by apparently random declines in brightness of 7 mag or more
in visual light. The brightness variations are a consequence of the ejection
of puffs of material at random times and in essentially random directions
(e.g. Feast 1996). C-rich dust condensing in these puffs of material is
responsible for the observed extinction.  There is as yet no consensus on
the evolutionary status of, or the mechanism for mass-loss from, RCB stars.

As discussed by Feast et al. (2003) there are differences in the details of
obscuration events in Miras and RCB stars, but these are to be expected
given the differences in the sizes, outflow velocities and temperatures of
the stars involved.

In this section we look at the additional information provided on this
phenomenon by the data presented here. The frequency of occurrence is
obviously important; we see clear obscuration events in 5 out of 18 Miras
for which we have at least 25 observations. This should probably be regarded
as a lower limit as some of those with photometry over a limited time may
eventually show obscuration events if observed for long enough. We therefore
estimate the fraction of C-rich Miras exhibiting obscuration events at very
roughly one third. Furthermore, we demonstrated above (see
Figs.~\ref{fig_jhhk} to
\ref{fig_k12p1}) that there is no difference in the infrared properties of
the Miras in which obscuration events have been observed and those in which
they have not. It is therefore possible that all Miras will be seen to do
this if monitored for long enough.

Although, as we discuss below, the phenomenon is also observed among
non-Miras, the statistics for this group are not reliable. Our primary
interest in this work has been the Miras and we generally stopped observing
other stars when we had sufficient observations to establish that they were
not large amplitude variables. 

In the following some individual examples are briefly discussed.
Given in brackets after each star name is the variability type (M or SR) and
the number of SAAO near-infrared observations that are available.
Although more data are presented here we do not reconsider the behaviour
of {\bf R For} (M 209) or {\bf II Lup} (M 264) mentioned at the start of
this section.

\begin{figure} 
\begin{center}
\includegraphics[width=8.0cm]{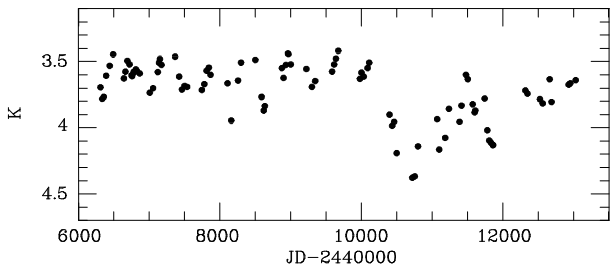}
 \caption{EV Eri at $K$; note the obscuration event at around JD 2450700.}
\label{fig_everi}
\end{center}
\end{figure}

\begin{figure} 
\begin{center} 
\includegraphics[width=8.0cm]{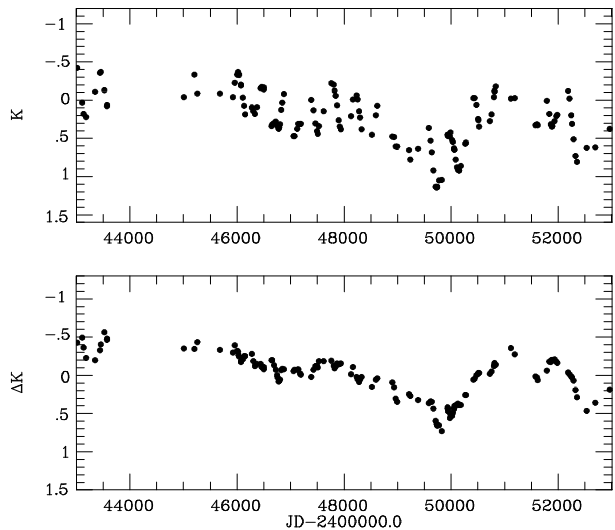}
\caption{(top) R Lep at $K$. (bottom) R Lep at $K$ after removal of the 438
day pulsations. }\label{fig_rlep} 
\end{center} 
\end{figure}

{\bf EV~Eri} (SR 92) was discussed by Whitelock et al. (1997) who estimated
its period at 228 days, not significantly different from the 226 days
derived here from a slightly larger dataset. Since then it has undergone an
obscuration phase (Fig.~\ref{fig_everi}) similar to those shown by the Miras
R~For (Whitelock et al. 1997) and R~Lep (see below and
Fig.~\ref{fig_rlep}). At its dimmest EV~Eri was fainter than usual by
$\Delta J\sim 1.1$, $\Delta H\sim 0.9$, $\Delta K\sim 0.8$ and $\Delta L\sim
0.7$ mag. Gigoyan et al. (1998) classified it as C-type (R or N) on an
objective prism spectrum and identify it with CGCS611 (Ste85-21), presumably
incorrectly as the coordinates differ significantly (EV Eri: 04 09 07.48 -09
14 12.0; CGCS611: 04 03 39.20 -09 13 17.3 (Equinox 2000)). In view of the
similarity of the light curve of EV~Eri to those of RCB stars it would be
interesting to see what a higher resolution spectrum revealed. As discussed
above this star shows IRAS colours (see Fig.~\ref{fig_k1225}, where
$K-[12]=2.14$ and $[12]-[25]=1.02$) that are unusual among non-Mira C-rich
variables. RCB stars show an excess at near-infrared wavelengths (e.g. Feast
et al. 1997; Feast 1997).

{\bf R Lep} (M 154) was discussed by Whitelock et al. (1997). The
light curve from the larger dataset is illustrated in the top part of
Fig.~\ref{fig_rlep}, while the bottom part shows the $K$ curve after
removing the 438 day periodic term (modelled as a sine curve plus its
harmonic). The smoothness of this residual is a measure of the regularity of
the underlying 438 day pulsation.  An analysis of data from the AAVSO
archive gives a period of 436 days and shows an earlier deep minimum around
JD\,2443700 in addition to the one illustrated here.

{\bf R Vol} (M 88). There is little to add to the discussion by Whitelock et
al. (1997), but to note that R Vol is brighter now, at $JHK$ and $L$ than it
has ever been in the 24 years we have been observing it. The most recent
maximum recorded by the AAVSO, around JD\,2453000, is the brightest in
almost 100 years. The star may be emerging from a prolonged obscuration
event.

\begin{figure} 
\begin{center}
\includegraphics[width=8.4cm]{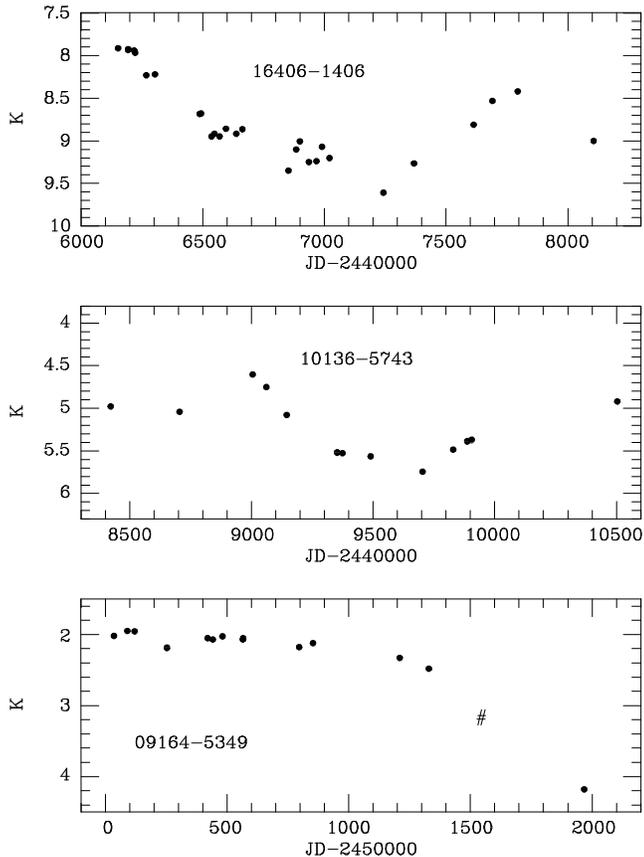}
\caption{09176--5147, 10136--5743 and 16406--1406 at $K$; the \# in the
bottom plot is an observation from 2MASS.}
\label{fig_long}
\end{center}
\end{figure}

{\bf 09164--5349} (SR 14) has a rather remarkable light curve, illustrated
in Fig.~\ref{fig_long}, which was typical of a very low amplitude SR
variable for more than 1000 days before it started a slow decline, changing
in brightness by $\Delta K > 2.0$ mag and $\Delta J > 3.5 $ mag over the
next 800 days.  Epchtein et al. (1990) reported $JHKLM$ photometry from
January 1986, i.e. almost 10 years before our first observations, at the
same bright quiescent level as our early measurements with $K=2.14$. Because
of its unusual IRAS colours (e.g. it lies above the dashed line which
separates most O- and C-rich stars in Fig.~\ref{fig_k1225}) 09176--5147 has
been identified by various authors as a potential C star with silicate dust
shell (e.g. Chen et al. 1993) although no evidence was found for anything
other than a normal C-rich shell. The colour changes during the fading event
suggest an increase in dust absorption. This particular object differs from
most others showing dust obscuration events in that it does not show
evidence for pulsation or other types of large amplitude variability.
Groenewegen et al. (2002) did not detect CO emission.

If 09164--5349 is indeed exhibiting the same type of obscuration event as
the other stars discussed in this section it is particularly important
because its existence proves that the phenomenon is not necessarily
associated with large amplitude pulsation (10136--5743 and 16406--1406 may
not be pulsating, but that is not proved beyond any doubt).

{\bf 10136--5743} (M: 13) has a large amplitude, $\Delta K > 1$ mag, but is
not obviously periodic (Fig.~\ref{fig_long}), although a period of the order
of 1000 days is possible if the light curve is erratic. There has been very
little published on this object beyond confirmation that it is a carbon
star. It may be similar to 16406--1406.

{\bf 16406--1406} (M: 28) is one of the most peculiar stars in the survey.
It is also one of the few objects for which we have no spectroscopic
evidence for its C-type classification, and this must therefore be regarded
as uncertain. There has been very little published about it. Kwok et al.
(1997) describe the IRAS spectrum as type-F, i.e. showing a featureless
continuum.  It is very red $3.3 <K-L< 4.0$, and has large amplitude
variations, $\Delta K > 1.5$ mag, with no evidence for periodicity unless
it is with a period of about 1800 days and very erratic. Its colours, e.g.
$K-[12]=8.62$ and $[12-25]=0.89$, are certainly typical of a C star with a
moderately thick shell. It is in the Galactic Centre quadrant, but well out
of the Galactic plane ($\ell=4.1$, $b=20.2$).

16406--1406 has similarities to 10136--5743. It is also possible that it is
like R For and that the periodicity has been totally disrupted by
an obscuration event. In Table~\ref{jhkl} we tentatively classified it as a
Mira because of its large amplitude variability, but the light curve of
Fig.~\ref{fig_long} does not suggest Mira-like periodicity.

In summary, these obscuration events occur in very roughly one third of
C-rich Miras and in an unknown fraction of other C-rich variable stars.  It
is possible that they are related to the RCB phenomenon as discussed above
and it would certainly be worth making a more detailed study of the
abundances and other properties of these stars. It is also possible that
these stars are in binary systems with low level interactions. It may even
be that the RCB-like mass loss is triggered by binary-related effects. 
Although obscuration events are not seen among solitary O-rich Miras they
are very common among symbiotic Miras (Whitelock 1987), where they are
thought to be a consequence of binary star interaction. It is also worth
noting that the well studied C-rich binary SR variable V~Hya shows colour
changes (e.g. Olivier et al. 2001) which are modulated at its orbital
period but which otherwise look very similar to the obscuration phenomenon
under discussion. At the same time we do have enough data to be certain that
the obscuration events in R~For are not periodic (Whitelock et al. 1997).

Finally we note that Woitke \& Niccolini's (2005) results offer at least a
partial explanation of the obscuration events. In their model for dust
driven winds in AGB stars, instabilities (hydrodynamical, radiative or
thermal) allow the occasional formation of dust clouds close to the star in
temporarily shielded areas; these clouds are then accelerated outward by
radiation pressure. At the same time, but elsewhere, thinner dust-free
matter falls back towards the star. In this way a turbulent and dynamical
environment is created close to the star, which can be expected to produce
the strongly inhomogeneous dust distribution which we observe in the C stars
discussed here.

\section*{Acknowledgments} 
 We thank the following people for their contribution to the SAAO
observations reported here: Brian Carter, Robin Catchpole, Ian Glass, Dave
Laney, Lerothodi Leeuw, Karen Pollard, Greg Roberts, Jonathan Spencer-Jones,
Garry Van Vuren, Hartmut Winkler and Albert Zijlstra. We are grateful to Tom
Lloyd Evans, Steven Bagnulo and Ian Short for allowing us to use their data
in advance of publication. We also thank Luis Balona for the use of his STAR
Fourier analysis package and John Menzies for a critical reading of the
manuscript. This research has made use of the SIMBAD database, operated at
CDS, Strasbourg, France. We acknowledge with thanks the variable star
observations from the AAVSO International Database contributed by observers
worldwide and used in this research.  We are grateful to the referee,
Jacco van Loon for some helpful suggestions.

\appendix
 \renewcommand{\thefigure}{A-\arabic{figure}}

\begin{figure} 
\includegraphics[width=7.0cm]{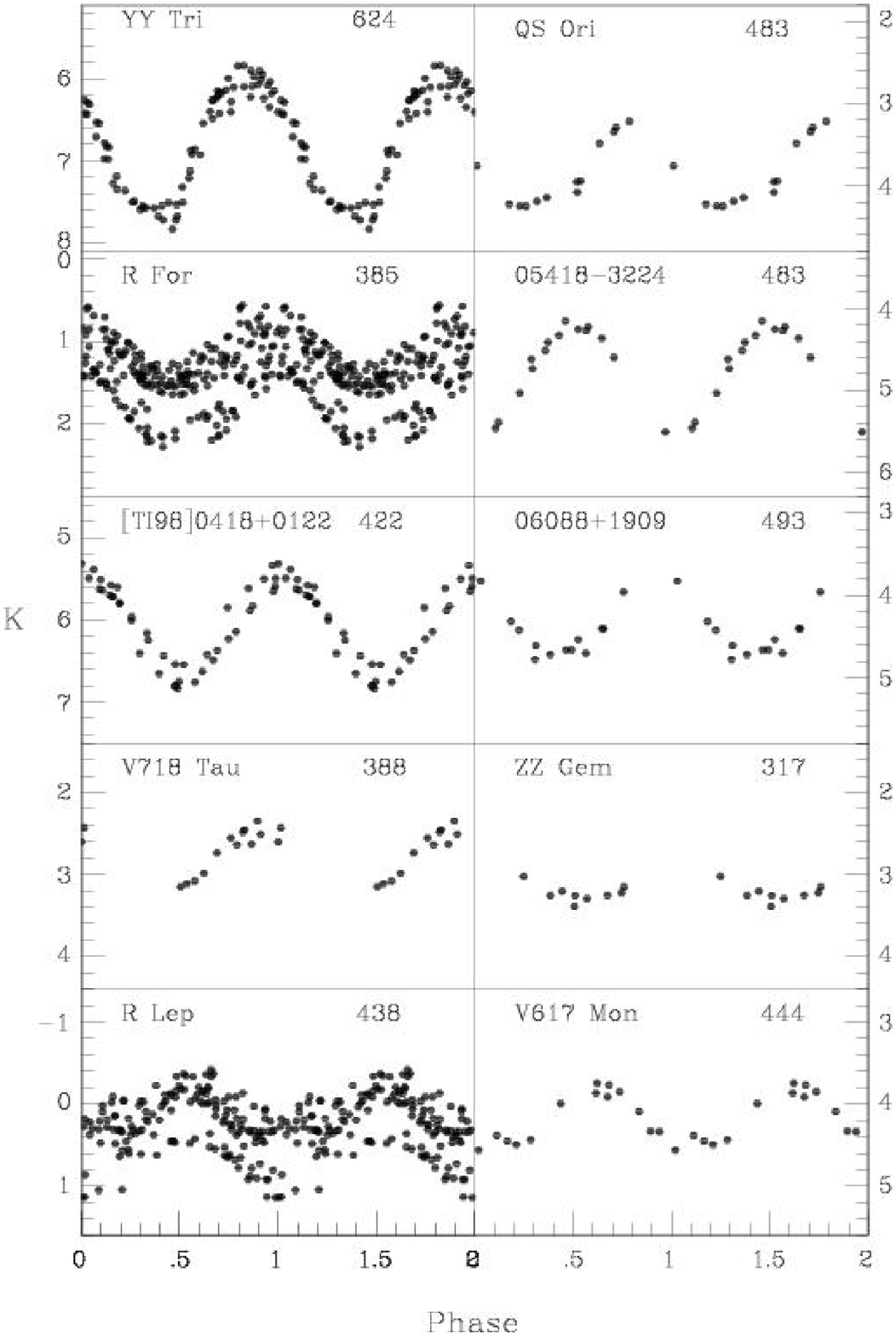}
 \caption{\label{fig_lcmira} $K$ light curves for the Mira variables; each
point is plotted twice to emphasize the periodicity. }
\end{figure}
\setcounter{figure}{0}
\begin{figure} 
\includegraphics[width=7.0cm]{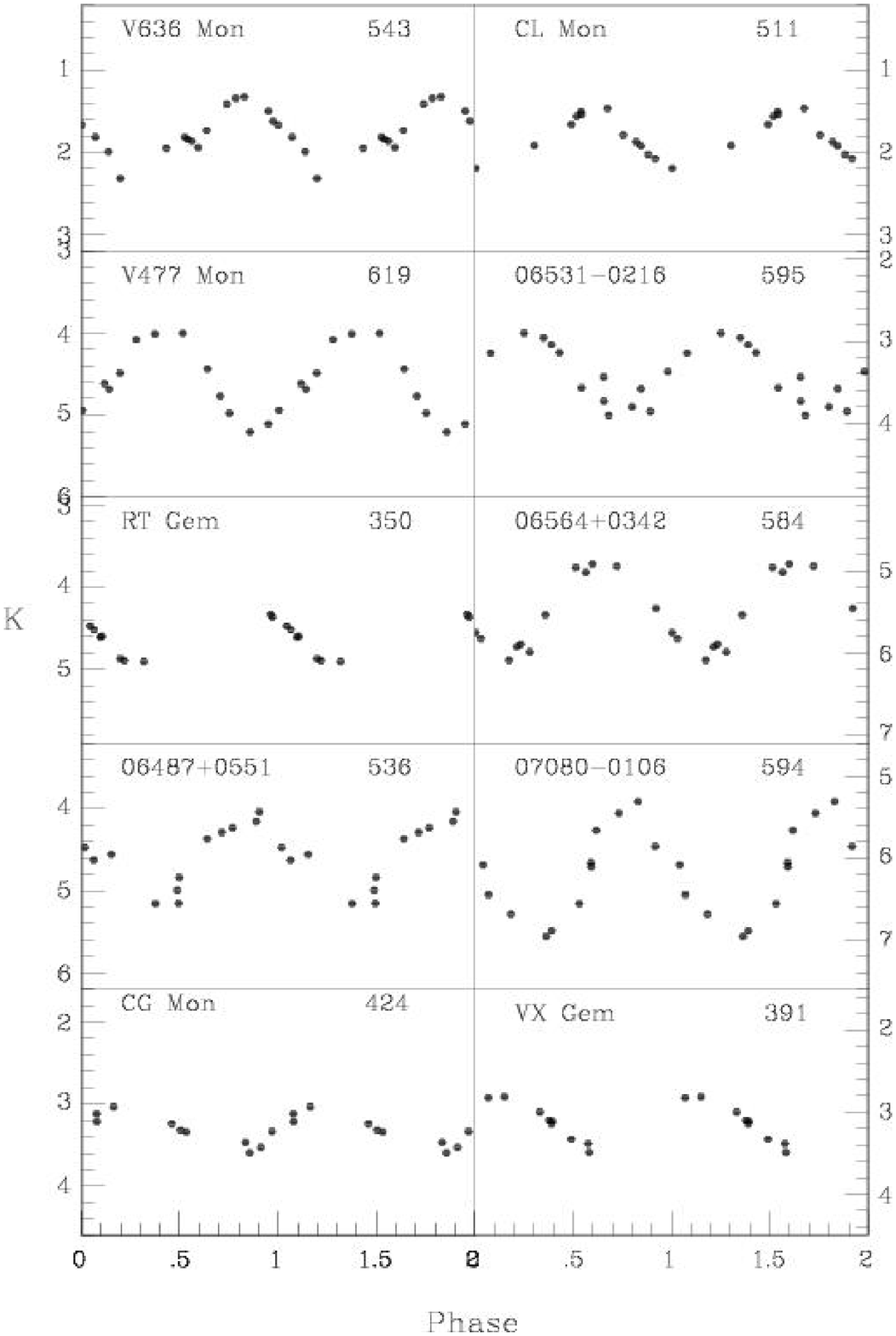}
 \caption{continued $K$ Mira light curves.}
\end{figure}
\setcounter{figure}{0}
\begin{figure} 
\includegraphics[width=7.0cm]{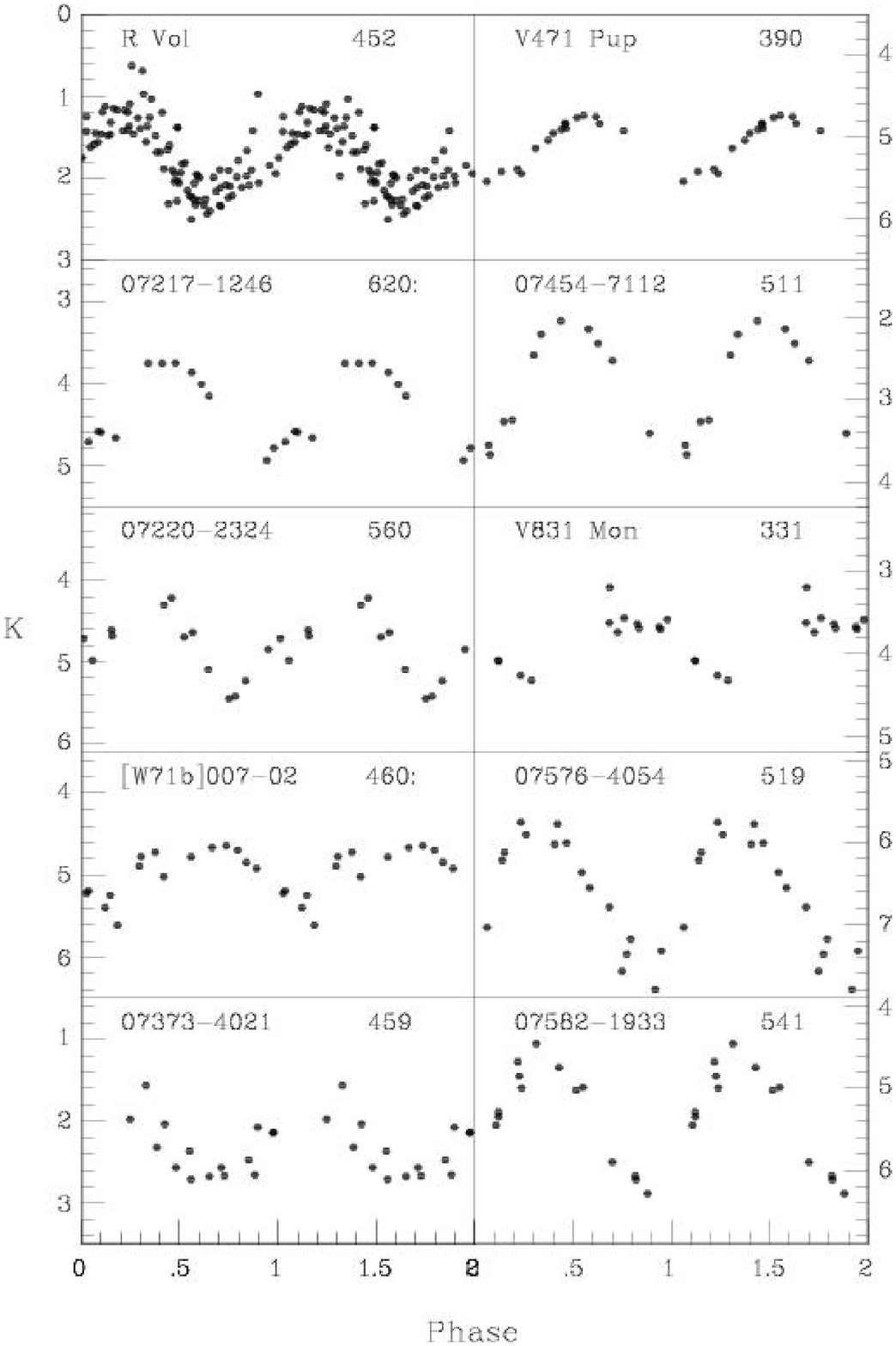}
 \caption{continued $K$ Mira light curves.}
\end{figure}
\setcounter{figure}{0}
\begin{figure} 
\includegraphics[width=7.0cm]{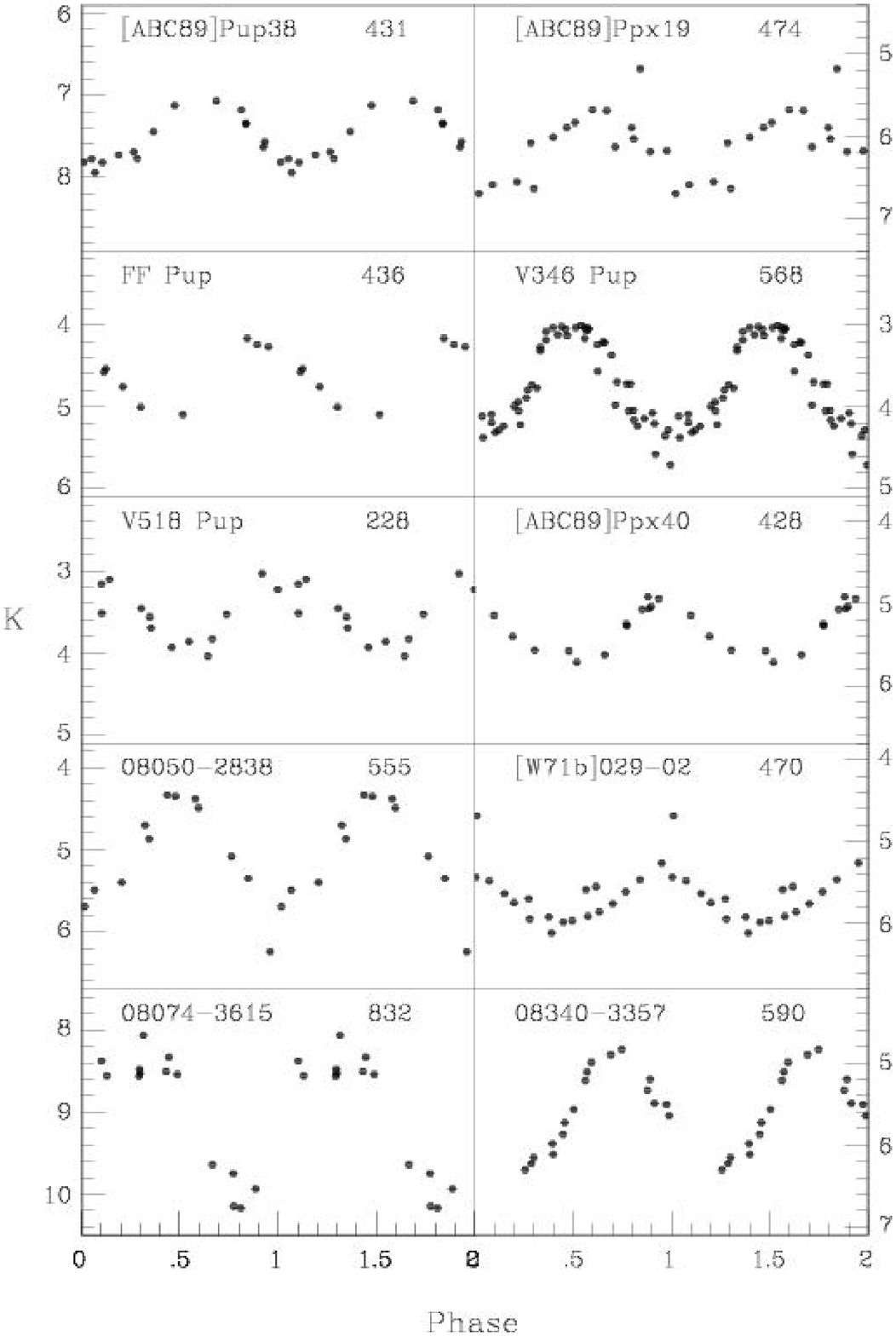}
 \caption{continued $K$ Mira light curves.}
\end{figure}

\setcounter{figure}{0}
\begin{figure} 
\includegraphics[width=7.2cm]{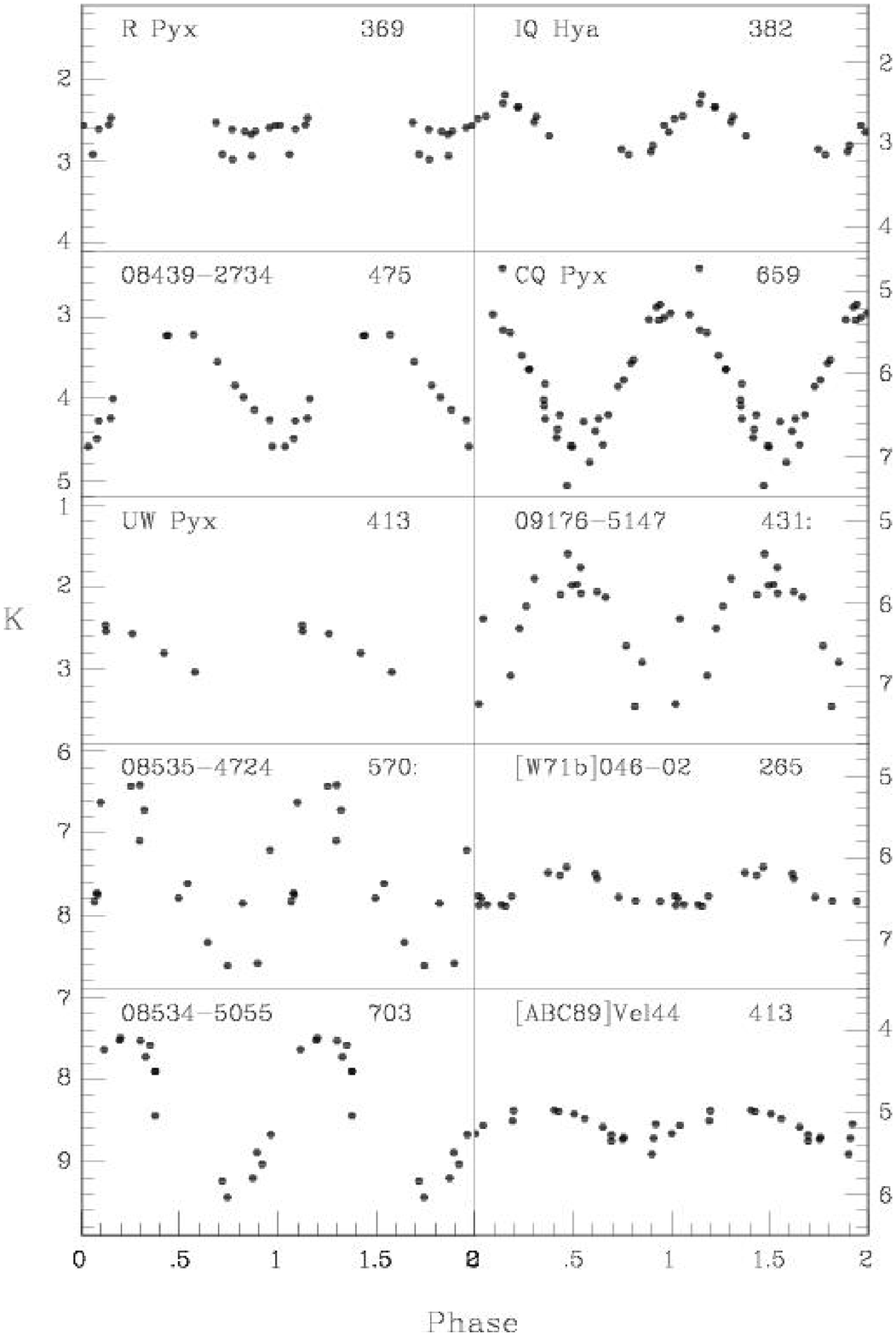}
 \caption{continued $K$ Mira light curves.}
\end{figure}
\setcounter{figure}{0}
\begin{figure} 
\includegraphics[width=7.2cm]{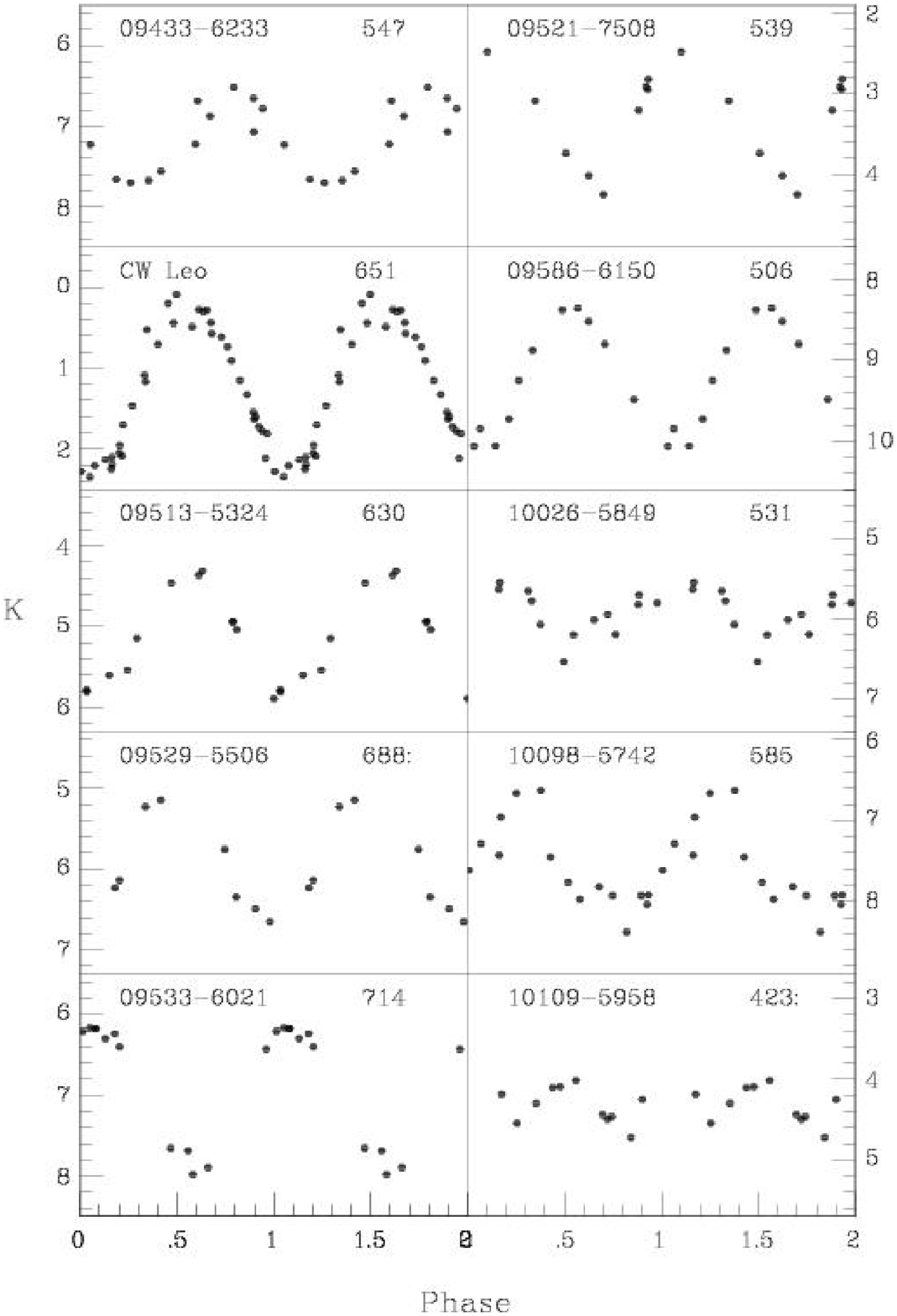}
 \caption{continued $K$ Mira light curves.}
\end{figure}
\setcounter{figure}{0}
\begin{figure} 
\includegraphics[width=7.2cm]{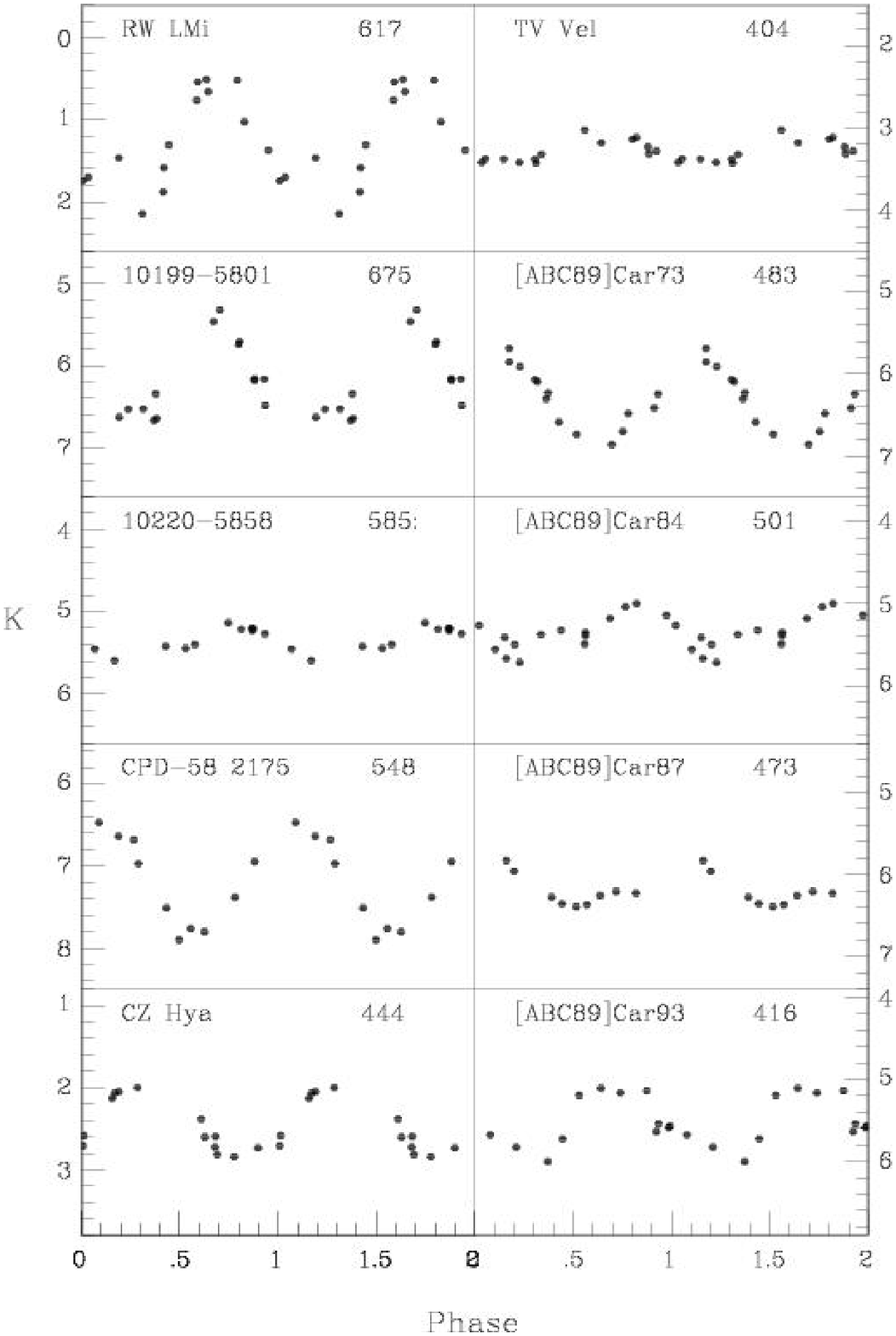}
 \caption{continued $K$ Mira light curves.}
\end{figure}
\setcounter{figure}{0}
\begin{figure} 
\includegraphics[width=7.2cm]{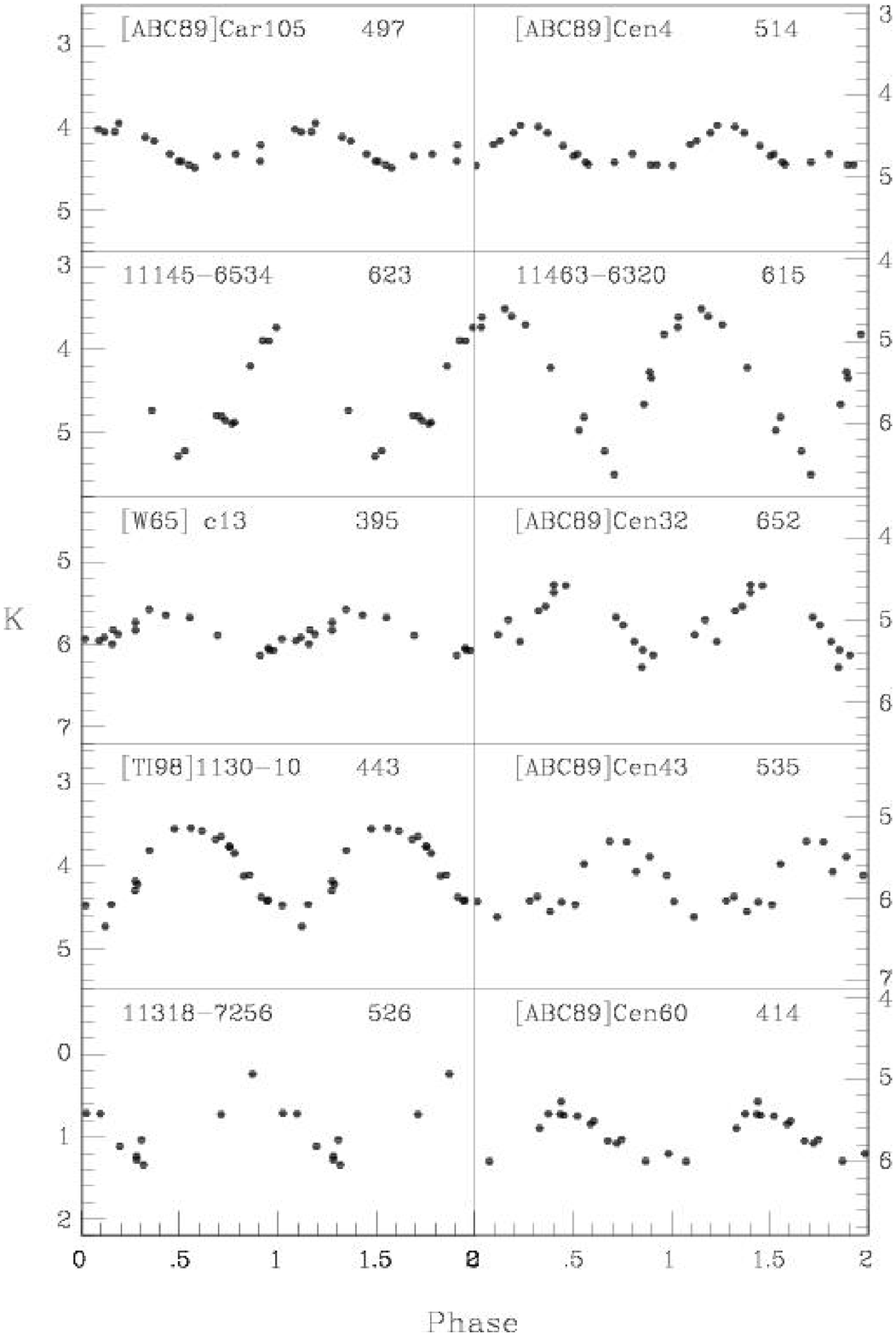}
 \caption{continued $K$ Mira light curves.}
\end{figure}
\setcounter{figure}{0}
\begin{figure} 
\includegraphics[width=7.2cm]{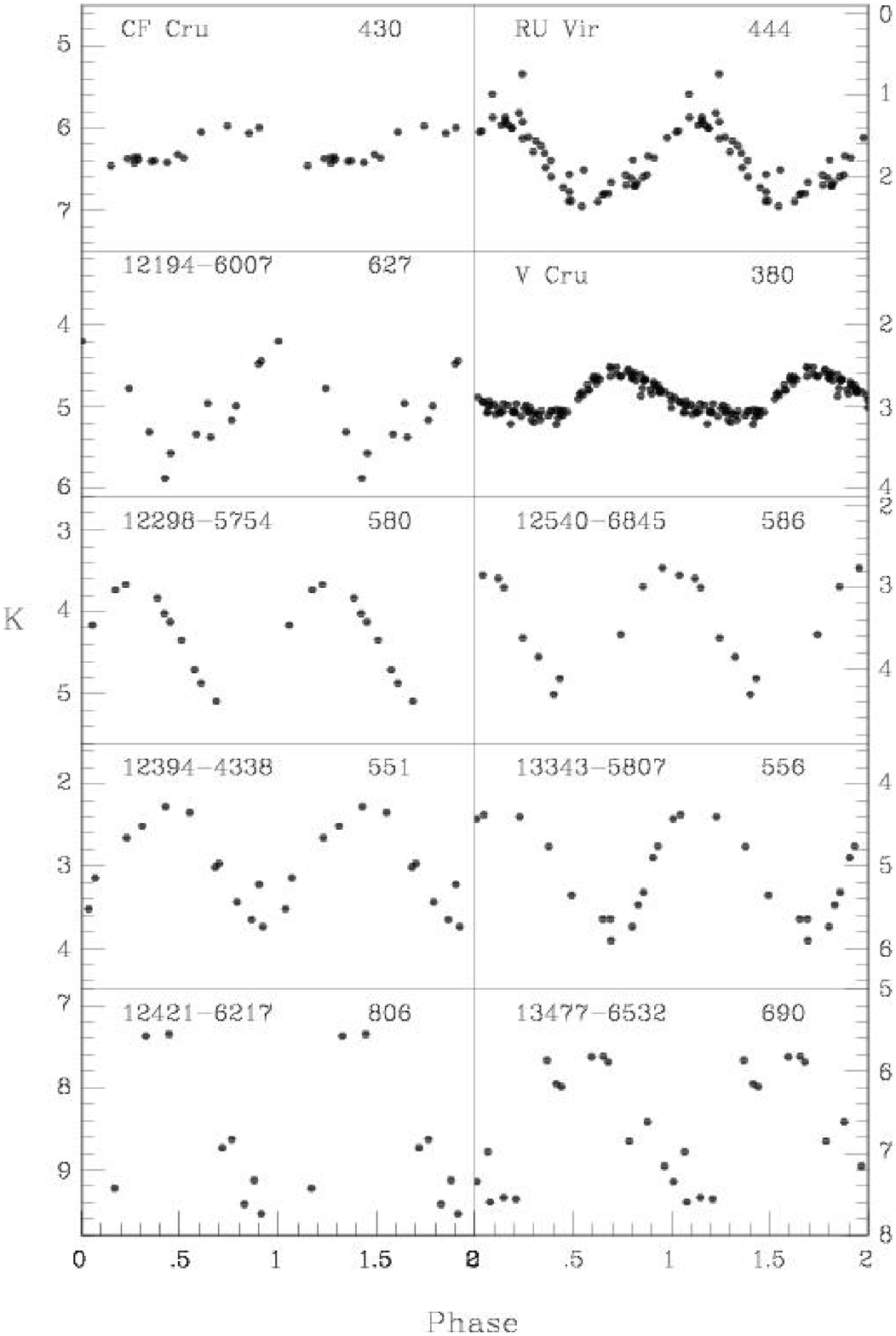}
 \caption{continued $K$ Mira light curves.}
\end{figure}
\setcounter{figure}{0}
\begin{figure} 
\includegraphics[width=7.2cm]{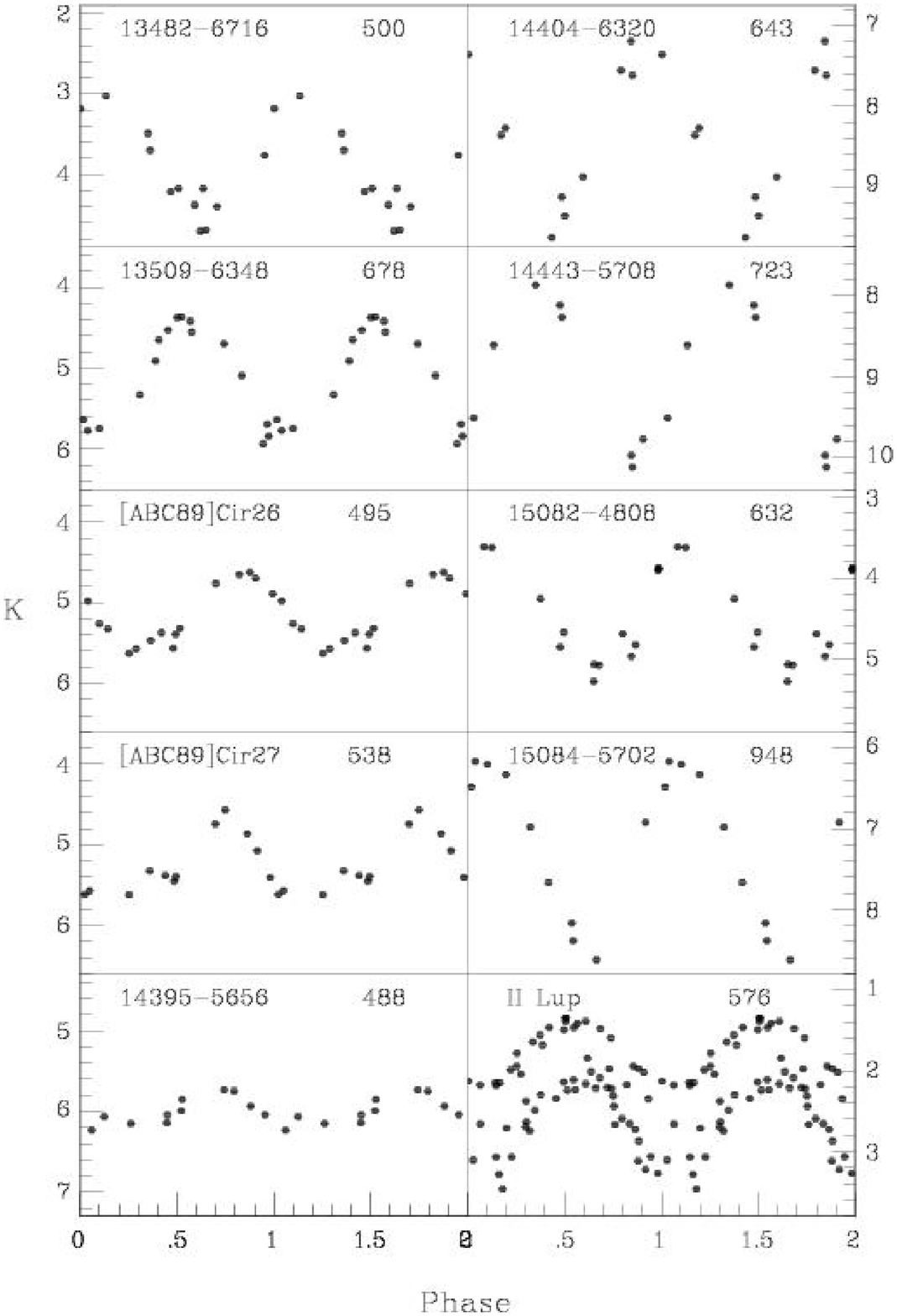}
 \caption{continued $K$ Mira light curves.}
\end{figure}
\setcounter{figure}{0}
\begin{figure} 
\includegraphics[width=7.2cm]{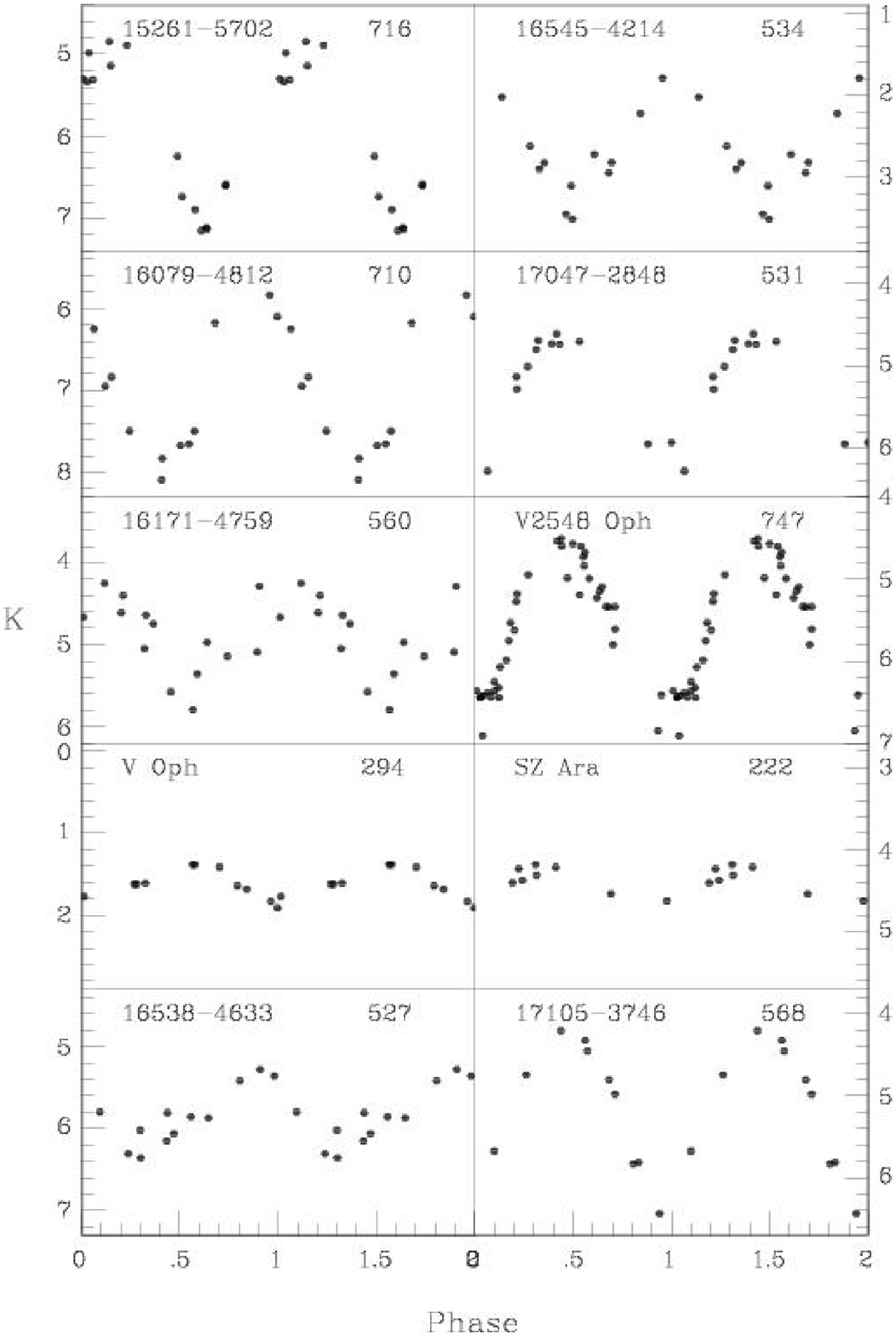}
 \caption{continued $K$ Mira light curves.}
\end{figure}
\setcounter{figure}{0}
\begin{figure} 
\includegraphics[width=7.1cm]{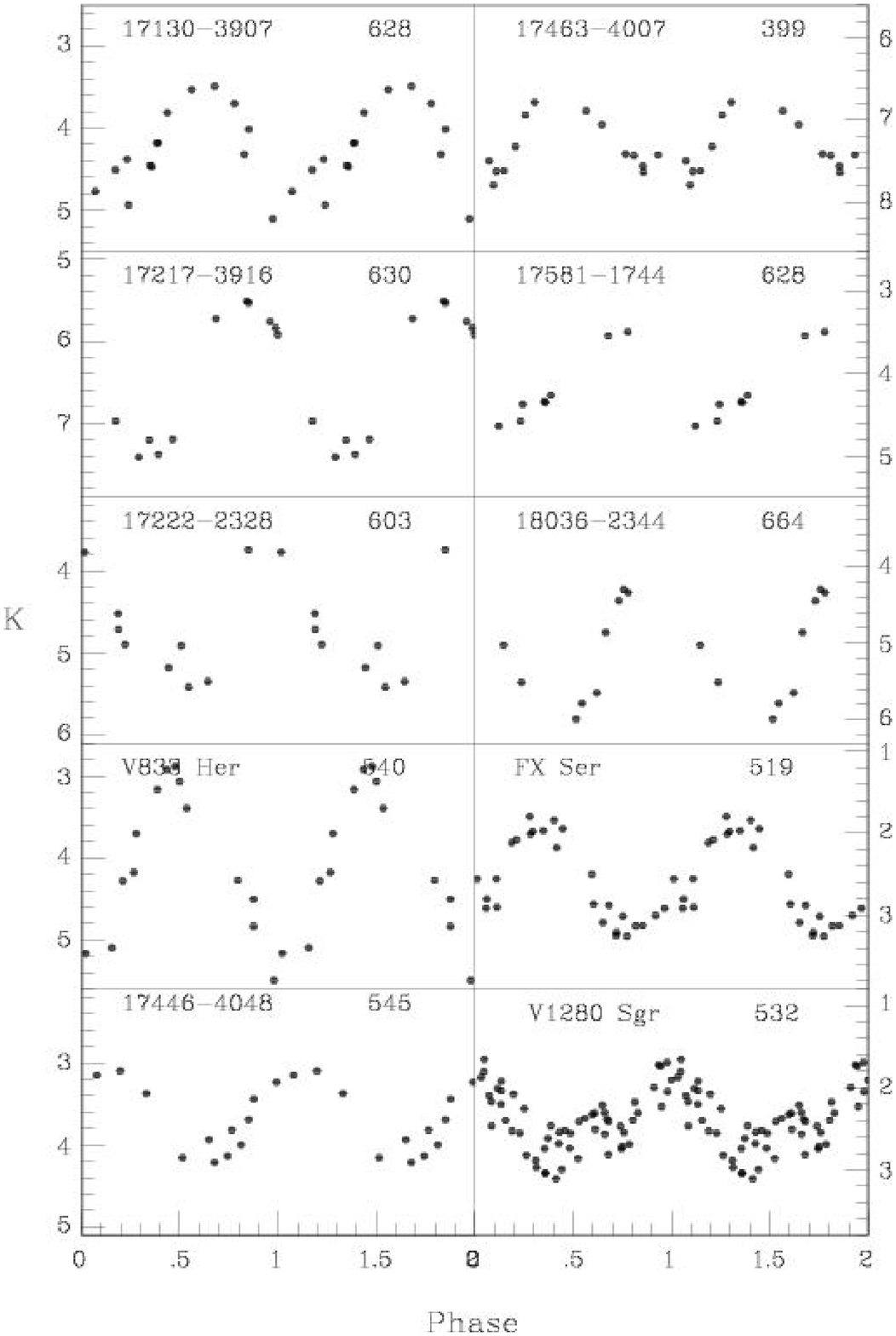}
 \caption{continued $K$ Mira light curves. The curve for V833~Her 
extends beyond the range shown here.}
\end{figure}

\setcounter{figure}{0}
\begin{figure} 
\includegraphics[width=7.2cm]{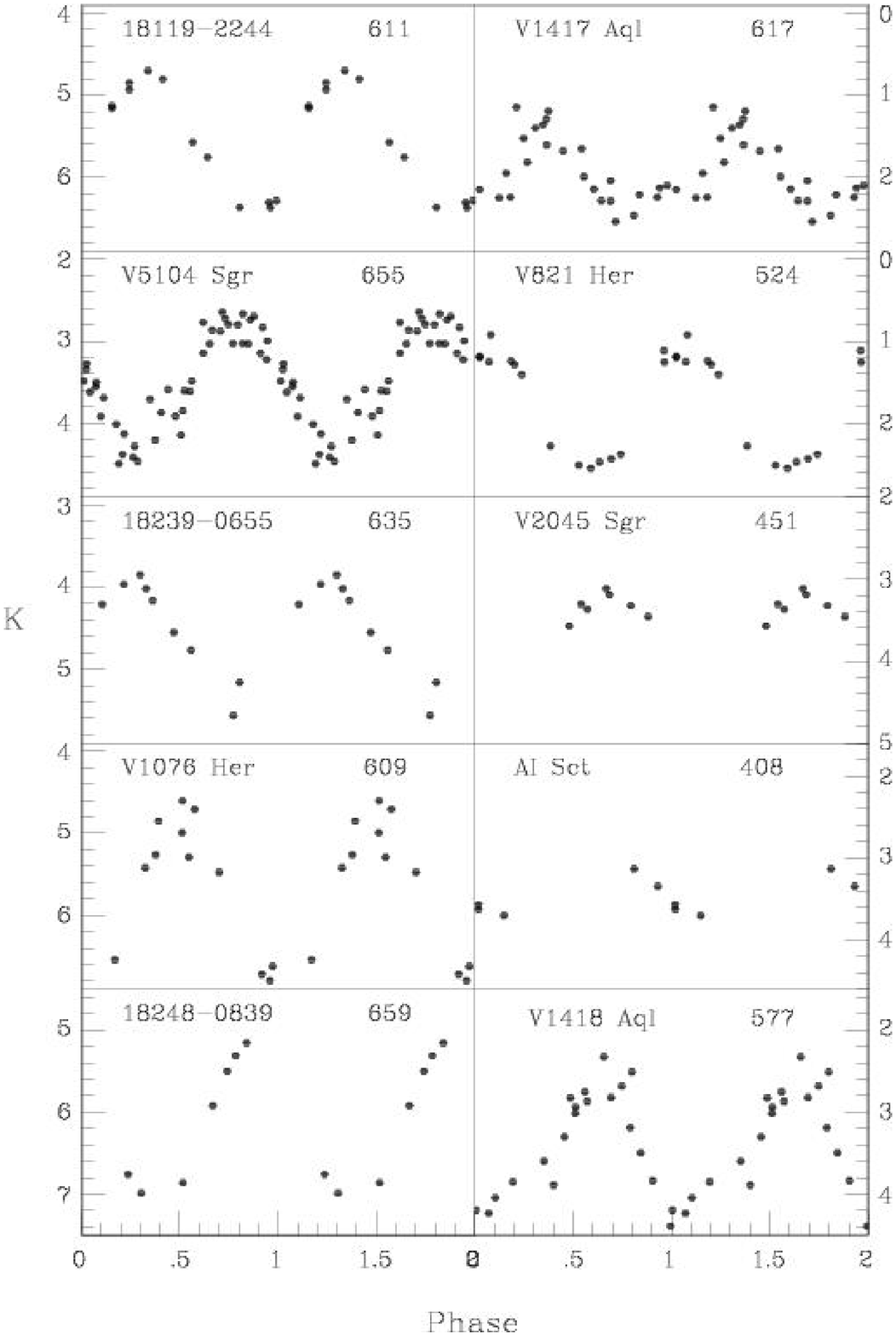}
 \caption{continued $K$ Mira light curves.}
\end{figure}
\setcounter{figure}{0}
\begin{figure} 
\includegraphics[width=7.2cm]{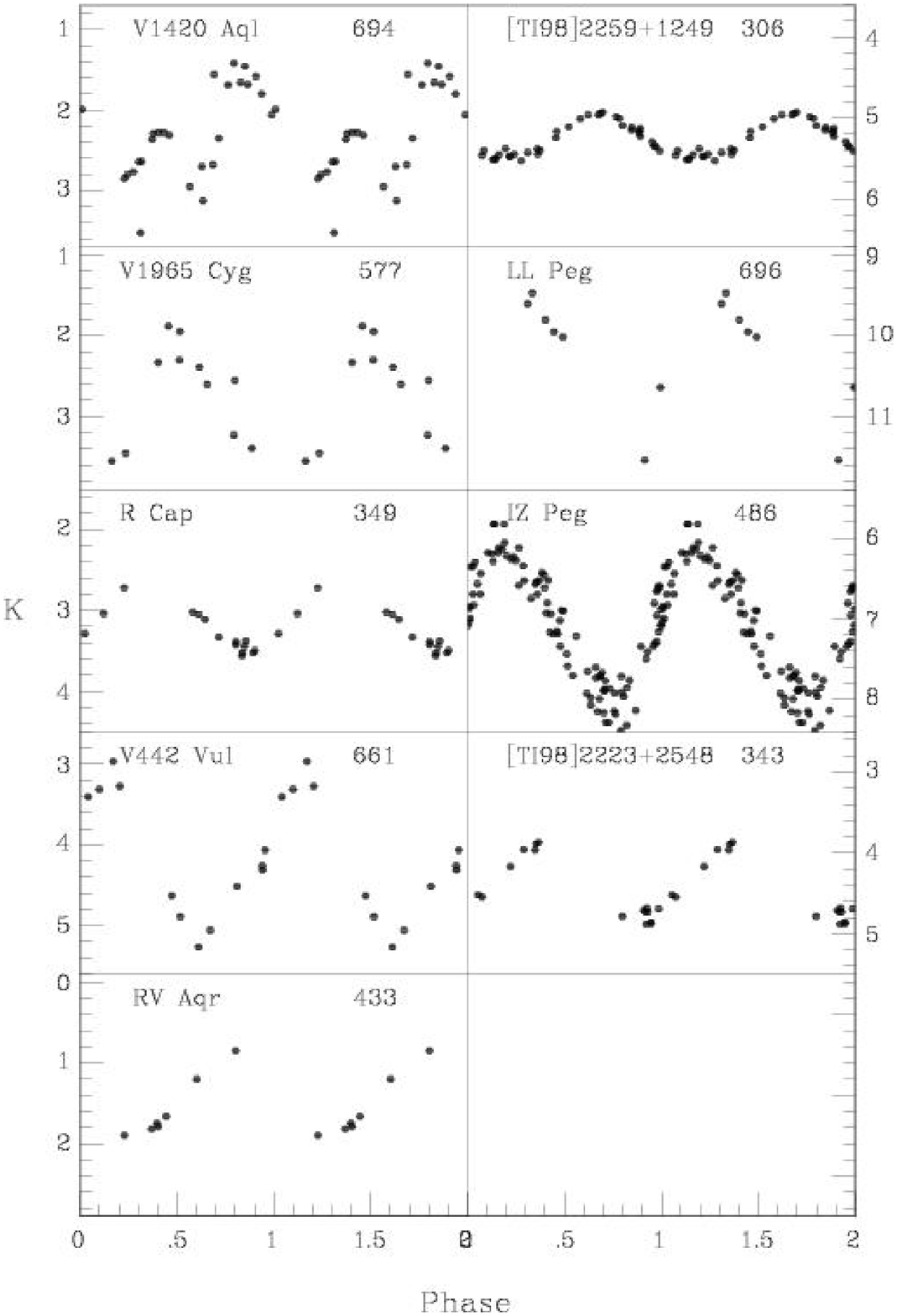}
 \caption{continued $K$ Mira light curves.}
\end{figure}

\begin{figure} 
\includegraphics[width=7.2cm]{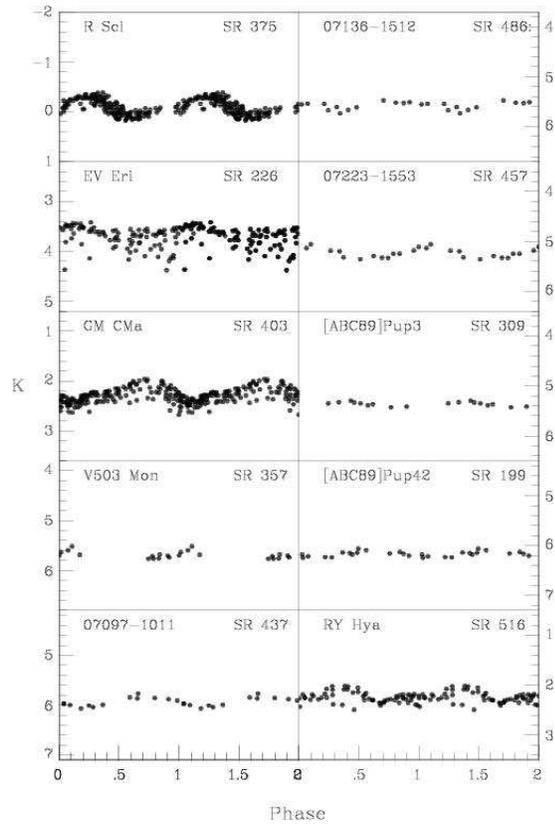}
 \caption{\label{fig_lcsr}$K$ light curves for semi-regular variables 
on the same scale as the Miras.}
\end{figure}
\setcounter{figure}{1}
\begin{figure} 
\includegraphics[width=7.2cm]{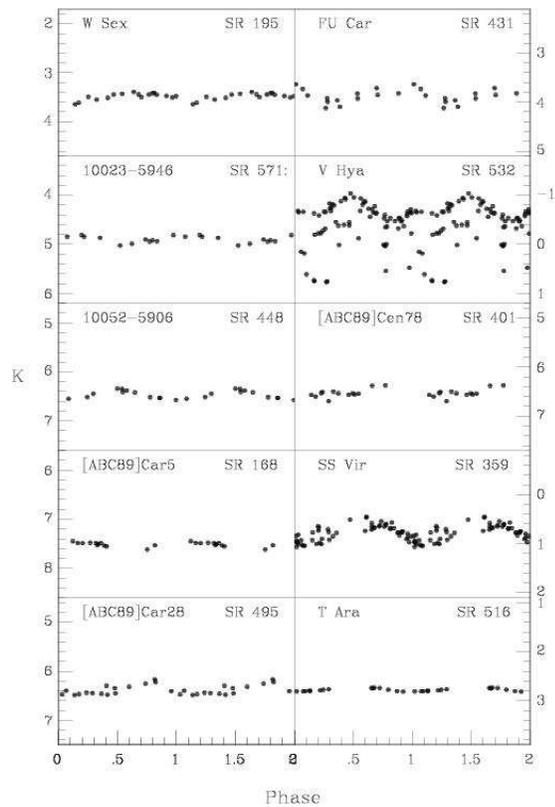}
 \caption{continued $K$ SR light curves.}
\end{figure}
\setcounter{figure}{1}
\begin{figure} 
\includegraphics[width=3.6cm]{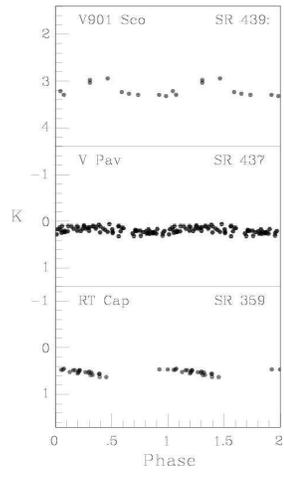}
 \caption{continued $K$ SR light curves.}
\end{figure}

\end{document}